\documentclass[11pt]{article}

\usepackage[margin=1in]{geometry}
\usepackage{amsmath,amssymb,amsfonts}
\usepackage{bm}
\usepackage{mathtools}
\usepackage{tensor}
\usepackage[hidelinks]{hyperref}
\usepackage[numbers,sort&compress]{natbib}

\title{\textbf{Imaginary Time as a Complex Representation of Relativity}}

\author{
William McConville\thanks{Corresponding author:
\texttt{wfmcconville@widener.edu}}
\\[0.5em]
Department of Physics, Widener University\\
Chester, PA, USA\\[0.5em]
ORCID: \href{https://orcid.org/0009-0000-3847-2423}
{0009-0000-3847-2423}
}

\date{}

\hypersetup{
    pdftitle={Imaginary Time as a Complex Representation of Relativity},
    pdfauthor={William McConville}
}

\begin{document}
% ============================================================

\maketitle

% ------------------------------------------------------------
\begin{abstract}
We develop a systematic complex representation of Lorentzian relativity based on the imaginary temporal coordinate \(X^0=ict\). Rather than treating this substitution as a Wick rotation to Euclidean spacetime, we formulate it as an invertible change of representation in which physical spacetime is identified with a real subspace of a complex vector space. This construction introduces two related metric structures: a complex-bilinear form and a Hermitian form, both of which reproduce the original Lorentzian interval when restricted to physical spacetime. We extend the representation to relativistic dynamics, including proper time, four-momentum, and the Hamilton–Jacobi formulation, and show that the metric, connection, curvature tensors, Einstein equations, and conservation laws retain their conventional content in curved spacetime. Explicit applications to the Kerr and Friedmann–Lemaître–Robertson–Walker geometries demonstrate consistency for rotating and time-dependent spacetimes. We further examine relativistic wave equations and their nonrelativistic limit, showing how the Schrödinger equation and its relation to Euclidean-time evolution emerge within the same complex temporal framework. The resulting formulation introduces no new physical predictions, but clarifies the geometric role of imaginary temporal coordinates and reveals structural connections among relativistic geometry, action, wave propagation, and analytic continuation.
\end{abstract}

\medskip

% ============================================================

\section{Introduction}
% ============================================================

Complex numbers have played strikingly different roles in the mathematical formulations of relativity and quantum mechanics. In quantum mechanics, complex state spaces, Hermitian inner products, and complex phases are fundamental to the standard formalism. In relativity, by contrast, the geometry of spacetime is conventionally formulated on a real Lorentzian manifold. This distinction was less pronounced in some of the earliest formulations of special relativity. Poincaré and Minkowski introduced an imaginary temporal coordinate, conventionally written \(ict\), which allowed the invariant spacetime interval to be expressed in a formally Euclidean form and Lorentz transformations to be represented as rotations through complex angles \cite{Poincare1906,Minkowski1909}. Although this notation provided an appealing formal symmetry between space and time, it was largely abandoned as the geometric interpretation of relativity matured.

The limitations of the \(ict\) notation are well known. In their discussion ``Farewell to \(ict\),'' Misner, Thorne, and Wheeler emphasized that the notation suppresses the underlying Lorentzian metric structure, obscures the distinction between covariant and contravariant quantities and the causal structure associated with the light cone, and does not naturally generalize to curved spacetime \cite[Box~2.1, p.~51]{MTW1973}. These objections are substantive if the replacement \(ct\rightarrow ict\) is interpreted simply as a device for treating spacetime as an ordinary Euclidean vector space. In particular, the expression of the Minkowski interval as a sum of four squares does not by itself define the appropriate inner-product structure on a complex vector space, nor does it specify how vectors, covectors, metrics, or derivatives should transform once the construction is extended beyond flat spacetime.

More recently, Gopalan has revisited Euclidean representations of
special relativity through a renormalized spacetime construction in
which Lorentz boosts can be represented as Euclidean rotations
\cite{Gopalan2021}. Modern treatments of complex spacetime geometry
have approached related questions from several different directions.
Wick rotation, for example, is ubiquitous in quantum field theory
and Euclidean approaches to gravity, but a naive continuation
$t\rightarrow it$ becomes ambiguous in a generic curved spacetime.
Samuel has formulated Wick rotation directly in the tangent space
using the tetrad formalism, while Visser has argued that
curved-spacetime Wick rotation is more appropriately regarded as a
complex deformation of the metric than as an analytic continuation
of a coordinate \cite{Samuel2016,Visser2017}.

Helleland and Hervik have examined Wick rotations between
pseudo-Riemannian geometries by treating them as real slices
of a common holomorphic complexification
\cite{HellelandHervik2018}. Their work demonstrates that a
complex-bilinear form with a formally Euclidean appearance
can induce an indefinite metric when restricted to an
appropriate real slice, and investigates the conditions
under which Wick rotations between such geometries exist.
Complexified tangent bundles and the distinction between
complex-bilinear and Hermitian extensions of a real metric
are standard constructions in complex differential geometry
\cite{Huybrechts2005}. An earlier application of imaginary temporal coordinates in
gravitational theory appears in the work of Ogievetskii and
Polubarinov, who explicitly adopted the convention
\(x_4=it\) in their formulation of spinor interactions
with gravity~\cite{OgievetskiiPolubarinov1965}. Their work
focused on the generally covariant transformation properties
of spinors rather than on developing a systematic complex
representation of Lorentzian spacetime geometry.

Other approaches have employed complex spacetime coordinates,
including imaginary proper-time parameters, in attempts to
reformulate gravitational or quantum dynamics \cite{Perko2019}.
Related work in Lorentz-covariant quantum mechanics has also
employed imaginary temporal components and indefinite inner-product
structures. Bedi\'c and Kong used such a representation in their
treatment of the covariant harmonic oscillator, while subsequent
work with Ting developed a pseudo-Hermitian, group-theoretical
formulation using real Minkowski four-vectors
\cite{BedicKong2020,BedicKongTing2021}. 
Together, these approaches illustrate the variety of roles that
complex structures can play in relativistic physics. 

The present work takes a different approach, asking whether the
historical $ict$ notation can be developed into a systematic and
exactly equivalent complex representation of Lorentzian relativity.
Rather than interpreting $ict$ as a Wick rotation to Euclidean
spacetime, we formulate spacetime events as complex vectors related
invertibly to their conventional real-coordinate representation.
Beginning with real coordinates \(x^\mu=(ct,\mathbf{x})\), we define
\[
X^\mu=S^\mu{}_\nu x^\nu,
\qquad
S=\operatorname{diag}(i,1,1,1),
\]
so that \(X^\mu=(ict,\mathbf{x})\). The physical spacetime is therefore not analytically continued to a Euclidean geometry; rather, the same Lorentzian events are represented within a complex vector space. This viewpoint naturally introduces two related metric structures on the complex vector space: a complex-bilinear form
\begin{equation}
    h=(S^{-1})^TgS^{-1},
\end{equation}
and a Hermitian form
\begin{equation}
    G=(S^\dagger)^{-1}gS^{-1}.
\end{equation}
Both reproduce the original spacetime interval for complex vectors corresponding to physical spacetime displacements, while defining distinct extensions of that geometry within the surrounding complex vector space. In flat spacetime these become \(h=I\) and \(G=\eta\), respectively. Thus the Euclidean appearance of the bilinear line element does not imply that physical spacetime has become a positive-definite Euclidean space. The relationship between these two structures, and the restrictions that distinguish the physical spacetime vectors within the larger complex vector space, will be developed below.

This distinction also produces an interesting structure in relativistic dynamics. Introducing the imaginary proper-time parameter
\[
\Sigma=ic\tau
\]
leads naturally to the complementary momentum vector
\[
\Pi^\mu
=
mc\frac{dX^\mu}{d\Sigma}
=
\left(\frac{E}{c},-i\mathbf{p}\right),
\]
rather than the conventional imaginary-time four-momentum in which the energy component carries the factor of \(i\). The difference follows from transforming both the spacetime coordinates and the affine parameter. Together with the Hermitian form \(G\), this complementary position--momentum structure yields an invariant pairing related directly to the relativistic action. These features provide a suggestive mathematical connection to structures familiar from quantum mechanics, where complex vector spaces, Hermitian pairings, and the phase \(e^{i\mathcal{S}/\hbar}\), with \(\mathcal{S}\) denoting the action, play fundamental roles. We do not interpret these correspondences as a derivation of quantum mechanics or as evidence that imaginary time is physically fundamental; rather, they motivate examining how completely the complex representation can encode relativistic dynamics.

The principal aim of this paper is therefore to determine whether this representation can be formulated consistently through both special and general relativity while preserving the original Lorentzian physics. We first develop Lorentz transformations, four-velocity and momentum, and uniformly accelerated motion in the complex representation. We then extend the construction to curved spacetime through the complexified tangent and cotangent spaces, establishing the transformation properties of the metric, Levi-Civita connection, geodesics, curvature, Einstein equations, and stress-energy tensor. The construction is tested explicitly in two complementary geometries. For the Kerr spacetime, the formalism must accommodate rotation, an off-diagonal temporal--azimuthal metric sector, and the vacuum Einstein equations. In the cosmological example, it must instead accommodate explicit time dependence, nonzero Ricci curvature, a perfect-fluid stress-energy tensor, and covariant conservation. In both cases the standard relativistic equations are recovered. Finally, we examine relativistic wave equations and their nonrelativistic limit, exploring how the complex temporal representation connects the Hamilton--Jacobi formulation, Schrödinger evolution, and Euclidean continuation. The resulting formulation is therefore not proposed as an alternative gravitational theory, but as an exactly equivalent complex representation of Lorentzian relativity whose mathematical structure may prove useful in settings where relativistic and intrinsically complex descriptions meet.

% ============================================================
\section{Complex Representation of Minkowski Spacetime}
\label{sec:minkowski}
% ============================================================

\subsection{Complex spacetime coordinates}
\label{sec:complex_coordinates}

We begin with Minkowski spacetime in standard inertial coordinates
\begin{equation}
    x^\mu=(x^0,x^1,x^2,x^3)=(ct,\mathbf{x}),
\end{equation}
with metric signature $(-,+,+,+)$. We introduce the constant complex
transformation
\begin{equation}
    S^\mu{}_\nu
    =
    \operatorname{diag}(i,1,1,1),
\end{equation}
and define the complex spacetime vector
\begin{equation}
    X^\mu
    =
    S^\mu{}_\nu x^\nu
    =
    (ict,\mathbf{x}).
    \label{eq:complex_coordinates}
\end{equation}
The transformation is invertible, with
\begin{equation}
    (S^{-1})^\mu{}_\nu
    =
    \operatorname{diag}(-i,1,1,1),
\end{equation}
so that
\begin{equation}
    x^\mu=(S^{-1})^\mu{}_\nu X^\nu.
\end{equation}

Equation~\eqref{eq:complex_coordinates} defines the representation of a
spacetime event used throughout this work. An event represented
conventionally by the real coordinate vector $x^\mu$ is represented here by
the complex vector $X^\mu$. Since $S$ is invertible, the two representations
contain the same information and correspond uniquely to the same physical
event. The transformation is therefore not an analytic continuation from
a Lorentzian spacetime to a distinct Euclidean spacetime. Rather, it
provides a complex-vector representation of the original Lorentzian
spacetime.

The corresponding differentials transform as
\begin{equation}
    dX^\mu=S^\mu{}_\nu\,dx^\nu,
    \qquad
    dx^\mu=(S^{-1})^\mu{}_\nu\,dX^\nu.
    \label{eq:differential_transformation}
\end{equation}
Because $S$ is constant, derivatives transform by the inverse relation
\begin{equation}
    \frac{\partial}{\partial X^\mu}
    =
    (S^{-1})^\nu{}_\mu
    \frac{\partial}{\partial x^\nu},
    \label{eq:derivative_transformation}
\end{equation}
so that, explicitly,
\begin{equation}
    \frac{\partial}{\partial X^0}
    =
    -\frac{i}{c}\frac{\partial}{\partial t},
    \qquad
    \frac{\partial}{\partial X^i}
    =
    \frac{\partial}{\partial x^i}.
\end{equation}
Derivatives with respect to the complex coordinates are understood as
complex-linear combinations of derivatives along the underlying real
spacetime. They therefore act on fields defined on the physical Lorentzian
slice and do not require those fields to admit a holomorphic extension
to the surrounding complex coordinate space. These relations will become
important when the construction is extended to spacetime-dependent
metrics and covariant derivatives.

Representing spacetime events by complex vectors also requires specifying
how the spacetime metric is represented on the resulting complex vector
space. As shown in the following subsection, both complex-bilinear and
Hermitian structures arise naturally. Although they reproduce the same
spacetime interval for the complex vectors corresponding to real spacetime
displacements, they define distinct extensions of that geometry within
the surrounding complex vector space.

\subsection{Bilinear and Hermitian structures}
\label{sec:structures}

Once spacetime events are represented by complex vectors, the spacetime
interval must be expressed consistently in the resulting complex vector
space. For two infinitesimally separated events, the complex displacement is
\begin{equation}
    dX^\mu = (ic\,dt,d\mathbf{x}).
\end{equation}
The conventional spacetime interval may be written
\begin{equation}
    ds^2 = dx^T g\,dx,
\end{equation}
where \(g=\eta=\operatorname{diag}(-1,1,1,1)\) in Minkowski spacetime.
Using
\begin{equation}
    dX=S\,dx,
    \qquad
    dx=S^{-1}dX,
\end{equation}
gives
\begin{align}
    ds^2
    &=(S^{-1}dX)^T g(S^{-1}dX) \nonumber\\
    &=dX^T(S^{-1})^T gS^{-1}dX.
\end{align}
This motivates the complex-bilinear form
\begin{equation}
    \boxed{
    h=(S^{-1})^T gS^{-1}
    },
\end{equation}
for which
\begin{equation}
    ds^2=dX^T h\,dX.
\end{equation}

For the Minkowski metric and
\begin{equation}
    S=\operatorname{diag}(i,1,1,1),
\end{equation}
the bilinear form reduces to
\begin{equation}
    h=I.
\end{equation}
The interval therefore takes the particularly simple form
\begin{equation}
    ds^2=dX^T dX
        =(ic\,dt)^2+d\mathbf{x}^2
        =-c^2dt^2+d\mathbf{x}^2.
\end{equation}
Thus the Lorentzian sign is carried by the imaginary temporal component
rather than appearing explicitly in the matrix \(h\).

The identity matrix in this expression should not, however, be interpreted
as a positive-definite Euclidean metric on physical spacetime. The
coordinates \(X^\mu\) are complex representations of the original real
Lorentzian coordinates and therefore do not range independently over
arbitrary complex values. The restrictions they satisfy, and their
geometrical significance, will be examined in the following subsection.

A distinct extension of the spacetime metric to the complex vector space is
obtained by using a Hermitian rather than bilinear contraction. The ordinary
Euclidean Hermitian product gives
\begin{equation}
    dX^\dagger dX
    =c^2dt^2+d\mathbf{x}^2,
\end{equation}
and therefore does not reproduce the Lorentzian interval. We instead define
a Hermitian form \(G\) by
\begin{equation}
    ds^2=dX^\dagger G\,dX.
\end{equation}
Requiring this expression to reproduce the conventional interval gives
\begin{align}
    ds^2
    &=(S^{-1}dX)^\dagger g(S^{-1}dX) \nonumber\\
    &=dX^\dagger(S^\dagger)^{-1}gS^{-1}dX,
\end{align}
and hence
\begin{equation}
    \boxed{
    G=(S^\dagger)^{-1}gS^{-1}
    }.
\end{equation}
For the transformation used here,
\begin{equation}
    S^\dagger=S^{-1},
\end{equation}
so that
\begin{equation}
    G=SgS^{-1}.
\end{equation}
In Minkowski spacetime, \(S\) and \(\eta\) are diagonal and commute, yielding
\begin{equation}
    G=\eta.
\end{equation}
Although \(G\) is Hermitian, it is not positive definite. Its indefinite
signature is essential to reproducing the Lorentzian interval and
distinguishes it from the positive-definite inner product ordinarily
used in quantum-mechanical Hilbert spaces.

Consequently,
\begin{equation}
    dX^\dagger G\,dX
    =(-ic\,dt)(-1)(ic\,dt)+d\mathbf{x}^2
    =-c^2dt^2+d\mathbf{x}^2.
\end{equation}

The bilinear and Hermitian forms therefore define distinct structures on the
complex vector space, although they reproduce the same interval for vectors
corresponding to real spacetime displacements:
\begin{equation}
    \boxed{
    dX^\dagger G\,dX
    =
    dX^T h\,dX
    =
    ds^2
    }.
    \label{eq:equivalent_interval_forms}
\end{equation}
The bilinear form \(h\) extends the metric without complex conjugation,
whereas \(G\) provides a sesquilinear extension through conjugation of its
first argument. Their equivalence for physical spacetime vectors does not
imply that the two forms are identical on the surrounding complex vector
space. The origin and significance of this distinction depend on the
restrictions satisfied by the complex vectors representing physical
spacetime events, to which we now turn.

\subsection{The physical slice}
\label{sec:physical_slice}

Although the complex representation places spacetime vectors in
$\mathbb{C}^4$, physical spacetime events do not correspond to arbitrary
vectors in this space. By construction,
\begin{equation}
    X^\mu=S^\mu{}_\nu x^\nu,
    \qquad
    x^\mu\in\mathbb{R}^4,
\end{equation}
so the complex vectors representing physical events satisfy
\begin{equation}
    X^0\in i\mathbb{R},
    \qquad
    X^i\in\mathbb{R}.
    \label{eq:physical_slice}
\end{equation}
We refer to this four-real-dimensional subset of $\mathbb{C}^4$ as the
physical slice. The map $S$ establishes a one-to-one correspondence
between this slice and the conventional real representation of
Minkowski spacetime.

The role of the physical slice is particularly important in interpreting
the bilinear form introduced in the previous subsection. In flat
spacetime, $h=I$, so that
\begin{equation}
    ds^2=dX^T dX.
\end{equation}
Despite its Euclidean appearance, this expression is not positive
definite on physical spacetime displacements. For
\begin{equation}
    dX^\mu=(ic\,dt,d\mathbf{x}),
\end{equation}
one obtains
\begin{equation}
    dX^T dX
    =
    -c^2dt^2+d\mathbf{x}^2.
\end{equation}
On the physical slice, this contraction is real, and the usual
timelike, null, and spacelike classification is therefore preserved:
\begin{equation}
\begin{aligned}
    dX^T dX &<0 &&\text{timelike},\\
    dX^T dX &=0 &&\text{null},\\
    dX^T dX &>0 &&\text{spacelike}.
\end{aligned}
\end{equation}
These inequalities apply to displacements on the physical slice;
for arbitrary vectors in $\mathbb{C}^4$, the bilinear contraction
need not be real. For example, the nonzero physical displacement
\begin{equation}
    dX^\mu=(ic\,dt,c\,dt,0,0)
\end{equation}
satisfies
\begin{equation}
    dX^T dX=0,
\end{equation}
demonstrating directly that the identity matrix appearing in the
bilinear form does not define a positive-definite Euclidean geometry
on the physical slice.

The same causal structure is encoded by the Hermitian metric. Since
$G=\eta$ in flat spacetime,
\begin{equation}
    dX^\dagger G\,dX
    =
    -c^2dt^2+d\mathbf{x}^2
    =
    dX^T h\,dX.
\end{equation}
Thus neither the introduction of complex spacetime vectors nor the
appearance $h=I$ alters the Lorentzian causal structure. The factors
of $i$ carried by the temporal components of physical vectors
account for the different algebraic forms of the two contractions.

It is therefore useful to distinguish the ambient complex vector
space from the physical spacetime it represents. The space
$\mathbb{C}^4$ contains vectors that do not correspond to physical
spacetime events under the reality conditions of
Eq.~\eqref{eq:physical_slice}. Physical Minkowski spacetime is
represented by the image $S(\mathbb{R}^4)$ within this larger space.
On this physical slice, the bilinear form $h$ and Hermitian form $G$
provide equivalent expressions for the Lorentzian interval, despite
defining distinct structures on the surrounding complex vector space.
This distinction will remain important when the construction is
extended to relativistic transformations and, later, to curved
spacetime.

% ============================================================
\section{Lorentz Transformations and Relativistic Dynamics}
\label{sec:dynamics}
% ============================================================
\subsection{Lorentz boosts in the complex representation}
\label{sec:boosts}

We now consider Lorentz transformations acting on the complex spacetime
vectors introduced in the previous section.  For a boost with velocity $v$
along the $x$ direction, the conventional Lorentz transformation may be
written
\begin{equation}
    x'=\Lambda x,
\end{equation}
where
\begin{equation}
    \Lambda=
    \begin{pmatrix}
        \gamma & -\gamma\beta & 0 & 0\\
        -\gamma\beta & \gamma & 0 & 0\\
        0&0&1&0\\
        0&0&0&1
    \end{pmatrix},
    \qquad
    \beta=\frac{v}{c},
    \qquad
    \gamma=\frac{1}{\sqrt{1-\beta^2}}.
    \label{eq:real_lorentz_boost}
\end{equation}
Since $X=Sx$ and $X'=Sx'$, the corresponding transformation in the
complex representation is
\begin{equation}
    X'=UX,
    \qquad
    U=S\Lambda S^{-1}.
    \label{eq:complex_lorentz_transformation}
\end{equation}
For the boost of Eq.~\eqref{eq:real_lorentz_boost}, this gives
\begin{equation}
    U=
    \begin{pmatrix}
        \gamma & -i\gamma\beta & 0 & 0\\
        i\gamma\beta & \gamma & 0 & 0\\
        0&0&1&0\\
        0&0&0&1
    \end{pmatrix}.
    \label{eq:complex_lorentz_boost}
\end{equation}
Thus the usual hyperbolic mixing of the temporal and spatial coordinates
appears as a complex rotation in the $(X^0,X^1)$ plane.

The transformation preserves the physical slice introduced in
Sec.~\ref{sec:physical_slice}.  If
\begin{equation}
    X^\mu=(ict,x,y,z),
\end{equation}
then Eq.~\eqref{eq:complex_lorentz_boost} gives
\begin{equation}
    X'^0
    =
    i\gamma(ct-\beta x),
    \qquad
    X'^1
    =
    \gamma(x-\beta ct),
\end{equation}
with $X'^2=y$ and $X'^3=z$.  The transformed temporal component therefore
remains purely imaginary while the spatial components remain real.  Hence
\begin{equation}
    U\mathcal{P}=\mathcal{P},
\end{equation}
where $\mathcal{P}=S(\mathbb{R}^4)$ denotes the physical slice.

The metric-preserving properties of $U$ may be expressed using either of
the two structures introduced in Sec.~\ref{sec:structures}.  In the conventional real
representation, Lorentz invariance is expressed by
\begin{equation}
    \Lambda^T g\Lambda=g.
    \label{eq:real_lorentz_invariance}
\end{equation}
For the complex-bilinear form,
\begin{equation}
    h=(S^{-1})^TgS^{-1},
\end{equation}
Eq.~\eqref{eq:real_lorentz_invariance} becomes
\begin{equation}
    \boxed{
    U^T hU=h.
    }
    \label{eq:bilinear_lorentz_invariance}
\end{equation}
In Minkowski spacetime $h=I$, so this reduces to
\begin{equation}
    \boxed{
    U^TU=I.
    }
    \label{eq:complex_orthogonality}
\end{equation}
The complex Lorentz boost is therefore a complex-orthogonal
transformation.  Its inverse may be written
\begin{equation}
    U^{-1}=h^{-1}U^Th,
    \label{eq:bilinear_inverse}
\end{equation}
which in flat spacetime reduces simply to
\begin{equation}
    \boxed{
    U^{-1}=U^T.
    }
    \label{eq:transpose_inverse}
\end{equation}
For Eq.~\eqref{eq:complex_lorentz_boost}, transposition changes the signs
of the imaginary off-diagonal elements and therefore produces the inverse
boost, corresponding to $\beta\rightarrow-\beta$.

The same transformation preserves the Hermitian metric $G$.  From
\begin{equation}
    G=(S^\dagger)^{-1}gS^{-1},
\end{equation}
the Lorentz condition may equivalently be written
\begin{equation}
    \boxed{
    U^\dagger G U=G.
    }
    \label{eq:hermitian_lorentz_invariance}
\end{equation}
Consequently,
\begin{equation}
    \boxed{
    U^{-1}=G^{-1}U^\dagger G.
    }
    \label{eq:hermitian_inverse}
\end{equation}
This is the metric-dependent analogue of the familiar relation between the
inverse and Hermitian conjugate of a unitary transformation.

It is important that Eq.~\eqref{eq:hermitian_lorentz_invariance} does not
make the Lorentz boost unitary in the ordinary sense.  Indeed, the boost
matrix in Eq.~\eqref{eq:complex_lorentz_boost} is Hermitian,
\begin{equation}
    U^\dagger=U,
\end{equation}
but for $\beta\neq0$,
\begin{equation}
    U^\dagger U
    =
    U^2
    \neq I.
\end{equation}
Rather, the transformation preserves the indefinite Hermitian form
defined by $G$, which retains the Lorentzian signature of the original
spacetime metric:
\begin{equation}
    (UX)^\dagger G(UY)=X^\dagger GY.
\end{equation}
When the Hermitian metric is the identity, the more general preservation
condition
\begin{equation}
    U^\dagger G U=G
\end{equation}
reduces to
\begin{equation}
    \boxed{
    U^\dagger I U
    =
    U^\dagger U
    =
    I.
    }
\end{equation}
Ordinary unitarity may therefore be viewed mathematically as the special
case of Hermitian metric preservation in which the metric itself is the
identity.

The transformation of the metric under a general change of complex basis
makes this distinction explicit.  If
\begin{equation}
    X'=UX,
\end{equation}
then $X=U^{-1}X'$, and invariance of the Hermitian interval gives
\begin{equation}
    X^\dagger GX
    =
    X'^\dagger
    (U^{-1})^\dagger G U^{-1}
    X'.
\end{equation}
Thus the components of the Hermitian metric in the transformed basis are
\begin{equation}
    G'
    =
    (U^{-1})^\dagger G U^{-1}.
    \label{eq:metric_basis_transformation}
\end{equation}
For a Lorentz transformation, Eq.~\eqref{eq:hermitian_lorentz_invariance}
implies
\begin{equation}
    G'=G.
\end{equation}
The same statement holds for the bilinear form,
\begin{equation}
    h'
    =
    (U^{-1})^T h U^{-1}
    =
    h.
\end{equation}
Lorentz transformations therefore preserve both representations of the
spacetime interval while mapping the physical slice onto itself.

More generally, an operator $A$ acting on the complex spacetime vectors
transforms under a change of basis by the usual similarity transformation,
\begin{equation}
    A'=UAU^{-1}.
    \label{eq:operator_similarity}
\end{equation}
The inverse in this expression may be written using either preserved
metric structure:
\begin{equation}
    \boxed{
    U^{-1}
    =
    h^{-1}U^Th
    =
    G^{-1}U^\dagger G.
    }
    \label{eq:equivalent_inverse_relations}
\end{equation}
In flat spacetime, where $h=I$ and $G=\eta$, this becomes
\begin{equation}
    \boxed{
    U^{-1}
    =
    U^T
    =
    \eta U^\dagger\eta.
    }
    \label{eq:flat_inverse_relations}
\end{equation}
The transpose, Hermitian conjugate, and inverse therefore remain distinct
operations, but their relationship is fixed by the bilinear and Hermitian
metric structures.

\subsection{Imaginary proper-time parameter}
\label{sec:imaginary_tau}

For a timelike worldline, the conventional proper time $\tau$ is defined by
\begin{equation}
    ds^2=-c^2d\tau^2.
    \label{eq:proper_time_interval}
\end{equation}
The quantity $\tau$ has its usual operational meaning as the real-valued
proper-time magnitude registered by an ideal clock.  In the complex
spacetime representation, however, it is natural to associate with this
measurement the imaginary proper-time parameter
\begin{equation}
    \boxed{
    \Sigma=ic\tau.
    }
    \label{eq:complex_proper_time}
\end{equation}
Its differential is
\begin{equation}
    d\Sigma=ic\,d\tau,
\end{equation}
and therefore
\begin{equation}
    \boxed{
    d\Sigma^2=-c^2d\tau^2=ds^2.
    }
    \label{eq:complex_proper_interval}
\end{equation}
Thus, for a timelike worldline, the square of the imaginary proper-time
differential is equal to the invariant spacetime interval.

This relation parallels the representation of the temporal coordinate
itself.  Just as a real coordinate-time interval $dt$ is represented by
the imaginary temporal displacement
\begin{equation}
    dX^0=ic\,dt,
\end{equation}
the real proper-time magnitude $d\tau$ is represented along the worldline
by
\begin{equation}
    d\Sigma=ic\,d\tau.
\end{equation}
The reality of a clock reading therefore does not require the corresponding
temporal displacement in the complex spacetime representation to be real.

Since $\tau$ is Lorentz invariant, $\Sigma$ is likewise invariant under
Lorentz transformations. It provides a natural parameter along any
timelike worldline and remains an affine parameter when the worldline
is a timelike geodesic. Differentiation with respect to $\Sigma$ is
related to differentiation with respect to conventional proper time by

\begin{equation}
    \boxed{
    \frac{d}{d\Sigma}
    =
    \frac{1}{ic}\frac{d}{d\tau}
    =
    -\frac{i}{c}\frac{d}{d\tau}.
    }
    \label{eq:complex_proper_derivative}
\end{equation}
The factor $-i/c$ in Eq.~\eqref{eq:complex_proper_derivative} will play an
important role in the transformation of four-velocity and four-momentum.
In particular, transforming the imaginary proper-time parameter together with the
spacetime coordinates produces a momentum representation different from
that obtained by transforming the spacetime components alone.

Equation~\eqref{eq:complex_proper_time} may be regarded purely as part of
an equivalent complex representation of Lorentzian relativity.  It also
leaves open a stronger interpretation: the imaginary character assigned
to temporal components may reflect a physically meaningful distinction
between time and space rather than only a choice of mathematical
representation.  Nothing in the construction developed here requires
this interpretation, however, and the mathematical results that follow
are independent of it.

\subsection{Four-velocity and complementary momentum}
\label{sec:four-velocity}

Using $\Sigma$ to parameterize a timelike worldline leads naturally
to a tangent vector in the complex spacetime representation. We define

\begin{equation}
    W^\mu \equiv \frac{dX^\mu}{d\Sigma}.
    \label{eq:complex_tangent}
\end{equation}
Using
\begin{equation}
    X^\mu=(ict,\mathbf{x}),
    \qquad
    d\Sigma=ic\,d\tau,
\end{equation}
together with
\begin{equation}
    \frac{dt}{d\tau}=\gamma,
    \qquad
    \frac{d\mathbf{x}}{d\tau}=\gamma\mathbf{v},
\end{equation}
gives
\begin{equation}
    W^\mu
    =
    \left(
    \frac{ic\,dt}{ic\,d\tau},
    \frac{d\mathbf{x}}{ic\,d\tau}
    \right)
    =
    \boxed{
    \left(
    \gamma,
    -i\gamma\frac{\mathbf{v}}{c}
    \right).
    }
    \label{eq:complex_four_velocity}
\end{equation}
The temporal component of $W^\mu$ is therefore real, while its spatial
components are imaginary. This is complementary to the spacetime vector
$X^\mu=(ict,\mathbf{x})$, for which the temporal component is imaginary
and the spatial components are real.

This structure can also be related directly to the conventional
four-velocity
\begin{equation}
    u^\mu
    =
    \frac{dx^\mu}{d\tau}
    =
    (c\gamma,\gamma\mathbf{v}).
\end{equation}
Since $X=Sx$ and
\begin{equation}
    \frac{d}{d\Sigma}
    =
    \frac{1}{ic}\frac{d}{d\tau},
\end{equation}
we obtain
\begin{equation}
    W
    =
    \frac{1}{ic}S u
    =
    -\frac{i}{c}S u.
    \label{eq:complex_velocity_transform}
\end{equation}
Thus the transformation of the tangent vector contains not only the
factor $S$ associated with the complex spacetime representation, but
also the additional factor $1/i$ arising from the use of the complex
proper-time parameter.

The normalization of $W^\mu$ may be expressed using either of the two
metric structures introduced in Sec.~\ref{sec:structures}. For the Hermitian form,
\begin{equation}
    dX^\dagger G\,dX
    =
    ds^2
    =
    -c^2d\tau^2.
\end{equation}
Since
\begin{equation}
    d\Sigma^\ast d\Sigma
    =
    c^2d\tau^2,
\end{equation}
division by $d\Sigma^\ast d\Sigma$ gives
\begin{equation}
    \boxed{
    W^\dagger G W=-1.
    }
    \label{eq:complex_velocity_hermitian_norm}
\end{equation}
In flat spacetime, where $G=\eta$, this may be verified explicitly:
\begin{equation}
    W^\dagger G W
    =
    -\gamma^2
    +
    \gamma^2\frac{v^2}{c^2}
    =
    -\gamma^2(1-\beta^2)
    =
    -1.
\end{equation}

The corresponding bilinear normalization takes a different algebraic
form. Since
\begin{equation}
    dX^T h\,dX
    =
    ds^2
    =
    d\Sigma^2,
\end{equation}
division by $d\Sigma^2$ gives
\begin{equation}
    \boxed{
    W^T h W=1.
    }
    \label{eq:complex_velocity_bilinear_norm}
\end{equation}
For Minkowski spacetime, $h=I$, and therefore
\begin{equation}
    W^T W
    =
    \gamma^2
    +
    \left(-i\gamma\frac{\mathbf{v}}{c}\right)^2
    =
    \gamma^2
    -
    \gamma^2\frac{v^2}{c^2}
    =
    1.
\end{equation}
Thus the same timelike tangent vector has Hermitian norm $-1$ and
bilinear norm $+1$. The difference reflects the complex conjugation
appearing in the Hermitian contraction and is consistent with the two
equivalent representations of the invariant interval developed in
Sec.~\ref{sec:structures}.

Multiplying the tangent vector by $mc$ defines the corresponding
momentum vector
\begin{equation}
    \Pi^\mu
    \equiv
    mc\,W^\mu
    =
    mc\frac{dX^\mu}{d\Sigma}.
    \label{eq:complementary_momentum_definition}
\end{equation}
Using
\begin{equation}
    E=\gamma mc^2,
    \qquad
    \mathbf{p}=\gamma m\mathbf{v},
\end{equation}
gives
\begin{equation}
    \boxed{
    \Pi^\mu
    =
    \left(
    \frac{E}{c},
    -i\mathbf{p}
    \right).
    }
    \label{eq:complementary_momentum}
\end{equation}
We will refer to $\Pi^\mu$ as the \emph{complementary momentum}. Its
complex structure is complementary to that of the spacetime vector:
\begin{equation}
    X^\mu=(ict,\mathbf{x}),
    \qquad
    \Pi^\mu=
    \left(
    \frac{E}{c},
    -i\mathbf{p}
    \right).
    \label{eq:position_momentum_complementarity}
\end{equation}
The imaginary factor that appears in the temporal component of position
therefore appears instead in the spatial components of the corresponding
momentum vector.

The origin of this structure can be seen particularly clearly by
comparison with the conventional four-momentum
\begin{equation}
    p^\mu
    =
    m u^\mu
    =
    \left(
    \frac{E}{c},
    \mathbf{p}
    \right).
\end{equation}
From Eq.~\eqref{eq:complex_velocity_transform},
\begin{equation}
    \boxed{
    \Pi=-iS p.
    }
    \label{eq:complementary_momentum_transform}
\end{equation}
Thus $\Pi$ is not obtained simply by applying the spacetime
transformation $S$ to the conventional four-momentum. The additional
factor $-i$ arises because the tangent vector is defined with respect
to the imaginary proper-time parameter $\Sigma=ic\tau$. Since this
relation is invertible, the complementary momentum contains exactly
the same physical information as the conventional four-momentum; it
is an alternative representation rather than an independent dynamical
quantity. Applying only $S$ would instead produce
\begin{equation}
    Sp=
    \left(
    i\frac{E}{c},
    \mathbf{p}
    \right),
\end{equation}
whereas transforming both the coordinates and the imaginary proper-time parameter
produces
\begin{equation}
    \Pi=
    \left(
    \frac{E}{c},
    -i\mathbf{p}
    \right).
\end{equation}

The complementary momentum inherits both normalization relations from
$W^\mu$:
\begin{equation}
    \boxed{
    \Pi^\dagger G\Pi=-m^2c^2,
    \qquad
    \Pi^T h\Pi=m^2c^2.
    }
    \label{eq:complementary_momentum_norms}
\end{equation}
In flat spacetime, the Hermitian contraction gives
\begin{equation}
    -\frac{E^2}{c^2}
    +
    \mathbf{p}^2
    =
    -m^2c^2,
\end{equation}
while the bilinear contraction gives
\begin{equation}
    \left(\frac{E}{c}\right)^2
    +
    (-i\mathbf{p})^2
    =
    \frac{E^2}{c^2}
    -
    \mathbf{p}^2
    =
    m^2c^2.
\end{equation}
Both are equivalent to the usual relativistic energy--momentum
relation
\begin{equation}
    E^2
    =
    \mathbf{p}^2c^2
    +
    m^2c^4.
\end{equation}
The complex redistribution of factors of $i$ therefore changes the
algebraic representation of the momentum vector without changing its
physical invariants.

\subsection{Hamilton--Jacobi structure and invariant action}
\label{sec:hamilton_jacobi}

The complementary momentum introduced above can also be related directly
to the relativistic action through the Hamilton--Jacobi formulation.
Let $\mathcal{S}$ denote Hamilton's principal function, in order to
distinguish it from the transformation matrix $S$.  In the conventional
real-coordinate representation, its differential is
\begin{equation}
    d\mathcal{S}
    =
    p_\mu\,dx^\mu,
\end{equation}
where
\begin{equation}
    p_\mu
    =
    \frac{\partial\mathcal{S}}{\partial x^\mu}
\end{equation}
is the canonical momentum covector.  In flat spacetime, with signature
$(-,+,+,+)$,
\begin{equation}
    p_\mu
    =
    \left(
    -\frac{E}{c},
    \mathbf{p}
    \right).
\end{equation}

Because the complex spacetime coordinates are related to the conventional
coordinates by
\begin{equation}
    X=Sx,
\end{equation}
covectors transform according to the inverse transpose.  Defining the
canonical momentum covector in the complex representation by
\begin{equation}
    P_\mu
    \equiv
    \frac{\partial\mathcal{S}}{\partial X^\mu},
\end{equation}
one therefore has
\begin{equation}
    P
    =
    (S^{-1})^T p.
\end{equation}
For $S=\operatorname{diag}(i,1,1,1)$, this gives in flat spacetime
\begin{equation}
    P_\mu
    =
    \left(
    i\frac{E}{c},
    \mathbf{p}
    \right).
\end{equation}
The differential of the action consequently retains its invariant
canonical form,
\begin{equation}
    d\mathcal{S}
    =
    P_\mu\,dX^\mu
    =
    P^T dX.
\end{equation}

The canonical covector $P_\mu$ should be distinguished from the
complementary momentum vector $\Pi^\mu$ introduced in the preceding
subsection.  The two are related through the complex-bilinear form $h$.
Using the relation for the conventional contravariant four-momentum
established above,
\begin{equation}
    \Pi^\mu=-iS^\mu{}_\nu p^\nu,
\end{equation}
together with
\begin{equation}
    h=(S^{-1})^TgS^{-1},
\end{equation}
gives
\begin{equation}
    P_\mu
    =
    i h_{\mu\nu}\Pi^\nu,
\end{equation}
or equivalently,
\begin{equation}
    \Pi^\mu
    =
    -i h^{\mu\nu}P_\nu.
\end{equation}
The invariant differential of the action may therefore be written as
\begin{equation}
    \boxed{
    \Pi^T h\,dX
    =
    -i\,d\mathcal{S}
    }.
\end{equation}

A corresponding expression follows from the Hermitian structure.  On the
physical complex representation of spacetime,
\begin{equation}
    \boxed{
    \Pi^\dagger G\,dX
    =
    i\,d\mathcal{S}
    }.
\end{equation}
Thus the bilinear and Hermitian pairings encode the same real action
differential with opposite factors of $i$.  For real
$d\mathcal{S}$, they are related by complex conjugation,
\begin{equation}
    \left(\Pi^T h\,dX\right)^*
    =
    \Pi^\dagger G\,dX.
\end{equation}

The same relationship appears naturally in the relativistic
Hamilton--Jacobi equation for a freely propagating massive particle.
In the conventional representation,

\begin{equation}
    g^{\mu\nu}
    \frac{\partial\mathcal{S}}{\partial x^\mu}
    \frac{\partial\mathcal{S}}{\partial x^\nu}
    +
    m^2c^2
    =
    0.
\end{equation}
Transforming to the complex spacetime representation gives
\begin{equation}
    \boxed{
    h^{\mu\nu}
    \frac{\partial\mathcal{S}}{\partial X^\mu}
    \frac{\partial\mathcal{S}}{\partial X^\nu}
    +
    m^2c^2
    =
    0
    }.
\end{equation}
In flat spacetime, where $h_{\mu\nu}=\delta_{\mu\nu}$, this becomes
\begin{equation}
    \sum_{\mu=0}^{3}
    \left(
    \frac{\partial\mathcal{S}}{\partial X^\mu}
    \right)^2
    +
    m^2c^2
    =
    0.
\end{equation}
The Lorentzian sign difference has not disappeared; rather, it is carried
by the imaginary temporal derivative,
\begin{equation}
    \frac{\partial\mathcal{S}}{\partial X^0}
    =
    i\frac{E}{c}.
\end{equation}

Finally, substituting
\begin{equation}
    P_\mu
    =
    i h_{\mu\nu}\Pi^\nu
\end{equation}
into the Hamilton--Jacobi equation yields
\begin{equation}
    \boxed{
    h_{\mu\nu}\Pi^\mu\Pi^\nu
    =
    m^2c^2
    },
\end{equation}
which is precisely the complex-bilinear mass-shell relation obtained
previously from the normalization of the tangent vector.  The
complementary momentum therefore admits two mutually consistent
interpretations: it arises kinematically from differentiation with
respect to the imaginary proper-time displacement $\Sigma=ic\tau$, and
dynamically from the gradient of Hamilton's principal function according
to
\begin{equation}
    \boxed{
    \Pi^\mu
    =
    -i h^{\mu\nu}
    \frac{\partial\mathcal{S}}{\partial X^\nu}
    }.
\end{equation}

The action, canonical momentum, and mass-shell condition thus retain
their standard invariant content in the complex representation, while
the factors that distinguish temporal and spatial components are carried
by the complex structure of the coordinates and complementary momentum.
This action formulation will also provide a useful point of comparison
with relativistic wave equations considered later.

% ============================================================
\section{Uniform Acceleration and Rindler Coordinates}
\label{sec:rindler}

\subsection{Uniformly accelerated worldlines}
\label{sec:accel_worldlines}

Uniform acceleration provides a useful setting in which to examine the
geometric meaning of Lorentz transformations in the complex spacetime
representation. Consider motion in the $x$ direction with constant
positive proper acceleration $\alpha>0$. Introducing the rapidity
parameter
\begin{equation}
    \eta = \frac{\alpha\tau}{c},
\end{equation}
the conventional Minkowski coordinates of a uniformly accelerated
worldline may be written as
\begin{equation}
    ct = \rho \sinh\eta,
    \qquad
    x = \rho \cosh\eta,
    \qquad
    \rho = \frac{c^2}{\alpha}.
\end{equation}
The trajectory therefore satisfies
\begin{equation}
    x^2-c^2t^2=\rho^2,
\end{equation}
and is the familiar hyperbola associated with constant proper
acceleration. We have chosen the right-hand branch, for which the
observer passes through $t=0$ at $x=\rho$.

In the complex spacetime representation, the same trajectory becomes
\begin{equation}
    X(\eta)
    =
    \rho
    \begin{pmatrix}
        i\sinh\eta\\
        \cosh\eta
    \end{pmatrix},
\end{equation}
where the transverse coordinates have been suppressed. Since the
complex-bilinear form is $h=I$ in Minkowski spacetime,
\begin{equation}
    X^T X
    =
    \rho^2
    \left[
        (i\sinh\eta)^2+\cosh^2\eta
    \right]
    =
    \rho^2.
\end{equation}
Thus the hyperbolic worldline takes the formally circular form
\begin{equation}
    (X^0)^2+(X^1)^2=\rho^2.
\end{equation}
This does not make the trajectory a circle in a real Euclidean plane.
For real $\eta$, the temporal component $X^0=i\rho\sinh\eta$ remains
purely imaginary, while $X^1=\rho\cosh\eta$ remains real. The trajectory
therefore lies on the physical complex representation of Lorentzian
spacetime introduced in Sec.~\ref{sec:physical_slice}.

The relation between this form and an ordinary rotation can be made
explicit. Consider the standard Euclidean rotation matrix
\begin{equation}
    R(\theta)
    =
    \begin{pmatrix}
        \cos\theta & -\sin\theta\\
        \sin\theta & \cos\theta
    \end{pmatrix}.
\end{equation}
For an imaginary angle $\theta=i\eta$,
\begin{equation}
    \cos(i\eta)=\cosh\eta,
    \qquad
    \sin(i\eta)=i\sinh\eta,
    \label{eq:imaginary_angle_identities}
\end{equation}
so that
\begin{equation}
    R(i\eta)
    =
    \begin{pmatrix}
        \cosh\eta & -i\sinh\eta\\
        i\sinh\eta & \cosh\eta
    \end{pmatrix}.
\end{equation}
This is precisely the complex Lorentz boost matrix obtained in
Sec.~\ref{sec:boosts}:
\begin{equation}
    U(\eta)=R(i\eta).
\end{equation}
A Lorentz boost may therefore be represented algebraically by the
analytic continuation of a Euclidean rotation matrix to an imaginary
angle. The parameter $\eta$ remains real, however, and the resulting
matrix is complex orthogonal rather than an ordinary real Euclidean
rotation. The transformation continues to preserve the physical
Lorentzian slice. This analytic continuation of the rotation matrix
should therefore not be confused with a continuation of physical
spacetime to a Euclidean geometry.

There is also an important distinction between the passive coordinate
transformation described by $U(\eta)$ and the active evolution of the
uniformly accelerated worldline. At $\eta=0$, the accelerated observer
is located at
\begin{equation}
    X(0)
    =
    \begin{pmatrix}
        0\\
        \rho
    \end{pmatrix}.
\end{equation}
The worldline at later rapidity is obtained from
\begin{equation}
    X(\eta)
    =
    U(-\eta)X(0)
    =
    R(-i\eta)X(0),
\end{equation}
since
\begin{equation}
    U(-\eta)
    =
    \begin{pmatrix}
        \cosh\eta & i\sinh\eta\\
        -i\sinh\eta & \cosh\eta
    \end{pmatrix}.
\end{equation}
Acting on $X(0)$ therefore gives
\begin{equation}
    X(\eta)
    =
    \rho
    \begin{pmatrix}
        i\sinh\eta\\
        \cosh\eta
    \end{pmatrix}.
\end{equation}
The opposite signs appearing in the passive coordinate boost and the
active evolution reflect the usual distinction between transforming
the coordinate frame and transforming the physical vector.

The tangent vector introduced in Sec.~\ref{sec:four-velocity}
follows directly from this trajectory. Since
\begin{equation}
    \Sigma=ic\tau
    =
    i\frac{c^2}{\alpha}\eta
    =
    i\rho\eta,
\end{equation}
we have
\begin{equation}
    W
    =
    \frac{dX}{d\Sigma}
    =
    \begin{pmatrix}
        \cosh\eta\\
        -i\sinh\eta
    \end{pmatrix}.
\end{equation}
The corresponding complementary momentum is therefore
\begin{equation}
    \Pi(\eta)
    =
    mc
    \begin{pmatrix}
        \cosh\eta\\
        -i\sinh\eta
    \end{pmatrix}.
\end{equation}
Its bilinear normalization is
\begin{equation}
    \Pi^T\Pi
    =
    m^2c^2
    \left(
        \cosh^2\eta-\sinh^2\eta
    \right)
    =
    m^2c^2,
\end{equation}
in agreement with the mass-shell relation derived in
Sec.~\ref{sec:four-velocity}.

The evolution may also be expressed in terms of a constant generator.
Define
\begin{equation}
    K_x
    =
    \begin{pmatrix}
        0 & i\\
        -i & 0
    \end{pmatrix},
    \qquad
    K_x^2=I.
\end{equation}
It follows that
\begin{equation}
    e^{\eta K_x}
    =
    I\cosh\eta+K_x\sinh\eta
    =
    U(-\eta).
\end{equation}
The active accelerated trajectory may therefore be written
\begin{equation}
    X(\eta)=e^{\eta K_x}X(0).
\end{equation}
Since $\eta=\alpha\tau/c$, differentiation gives
\begin{equation}
    \frac{dX}{d\tau}
    =
    \frac{\alpha}{c}K_x X.
\end{equation}
The complementary momentum evolves under the same generator:
\begin{equation}
    \Pi(\eta)=e^{\eta K_x}\Pi(0),
\end{equation}
and hence
\begin{equation}
    \frac{d\Pi}{d\tau}
    =
    \frac{\alpha}{c}K_x\Pi.
\end{equation}

Uniform acceleration is therefore represented by a continuous
family of complex-orthogonal transformations generated by $K_x$.
The physical rapidity $\eta$ is real, but in the analytically
continued rotation matrix it corresponds to the imaginary angle
$\theta=-i\eta$ for the active transformation. The trigonometric
functions of this complex angle become hyperbolic functions of
the real rapidity according to
Eq.~\eqref{eq:imaginary_angle_identities}, with the signs appropriate
to the active transformation. The resulting trajectory nevertheless
remains entirely within the physical Lorentzian representation.
In the following subsection, this single accelerated worldline
will be generalized to the family of worldlines defining Rindler
coordinates.

\subsection{Rindler coordinates in the complex representation}

The uniformly accelerated trajectory considered above is one member of
a family of hyperbolic worldlines covering the right Rindler wedge~\cite{Rindler1966}.
To construct the corresponding coordinate system, we now promote
$\rho$ from the constant labeling a single trajectory to a spatial
coordinate, while retaining $\eta$ as the dimensionless Rindler time
coordinate. In conventional Minkowski coordinates, the transformation
is
\begin{equation}
    ct=\rho\sinh\eta,
    \qquad
    x=\rho\cosh\eta,
    \qquad
    \rho>0.
    \label{eq:rindler_real}
\end{equation}
Each curve of constant $\rho$ is therefore a uniformly accelerated
worldline satisfying
\begin{equation}
    x^2-c^2t^2=\rho^2.
    \label{eq:rindler_hyperbolae}
\end{equation}
The transformation covers the right Rindler wedge $x>|ct|$, with the
null boundaries $x=\pm ct$ corresponding to the Rindler horizons.

In the complex spacetime representation, Eq.~\eqref{eq:rindler_real}
becomes
\begin{equation}
    X^0=i\rho\sinh\eta,
    \qquad
    X^1=\rho\cosh\eta.
    \label{eq:rindler_complex}
\end{equation}
For real $\rho$ and $\eta$, $X^0$ remains purely imaginary and $X^1$
remains real. The Rindler transformation therefore maps the real
coordinate pair $(\eta,\rho)$ into the same physical complex slice
used throughout the preceding sections; no Euclidean continuation has
been performed.

The corresponding differentials are
\begin{align}
    dX^0
    &=
    i\sinh\eta\,d\rho
    +i\rho\cosh\eta\,d\eta,
    \\
    dX^1
    &=
    \cosh\eta\,d\rho
    +\rho\sinh\eta\,d\eta.
\end{align}
Since the complex-bilinear form is $h=I$ in Minkowski spacetime, the
line element in the $(X^0,X^1)$ plane is
\begin{equation}
    ds^2=(dX^0)^2+(dX^1)^2.
\end{equation}
Substituting the differentials above gives
\begin{align}
    ds^2
    &=
    -\left(
        \sinh\eta\,d\rho
        +\rho\cosh\eta\,d\eta
    \right)^2
    +
    \left(
        \cosh\eta\,d\rho
        +\rho\sinh\eta\,d\eta
    \right)^2
    \nonumber\\
    &=
    d\rho^2-\rho^2d\eta^2.
    \label{eq:rindler_metric}
\end{align}
The mixed terms cancel, while
$\cosh^2\eta-\sinh^2\eta=1$. Thus the formally Euclidean bilinear
expression $(dX^0)^2+(dX^1)^2$ produces the standard Lorentzian
Rindler metric when restricted to the physical complex slice.

Including the transverse coordinates gives
\begin{equation}
    ds^2
    =
    -\rho^2d\eta^2
    +d\rho^2+dy^2+dz^2.
    \label{eq:rindler_metric_full}
\end{equation}
The Lorentzian signature has therefore not been introduced by an
explicit negative entry in the bilinear form. Instead, it arises from
the imaginary temporal differential $dX^0$ carried by the physical
complex representation.

The physical meaning of the coordinate $\rho$ follows by considering a
worldline at fixed $\rho$, $y$, and $z$. Along such a trajectory,
Eq.~\eqref{eq:rindler_metric_full} reduces to
\begin{equation}
    ds^2=-\rho^2d\eta^2.
\end{equation}
For a timelike worldline, $ds^2=-c^2d\tau^2$, and hence
\begin{equation}
    d\tau=\frac{\rho}{c}\,d\eta.
    \label{eq:rindler_proper_time}
\end{equation}
Comparison with the relation $\eta=\alpha\tau/c$ used for the trajectory in
Sec.~\ref{sec:accel_worldlines} then gives
\begin{equation}
    \alpha(\rho)=\frac{c^2}{\rho}.
    \label{eq:rindler_acceleration}
\end{equation}
Observers at different values of $\rho$ therefore undergo different
constant proper accelerations, with the acceleration increasing as
$\rho$ decreases toward the coordinate boundary associated with the
Rindler horizons. As discussed below, approaching a nonzero point on
either horizon requires a correlated limit involving both $\rho$
and $\eta$.

The horizon itself is most transparent in the conventional coordinates.
From Eq.~\eqref{eq:rindler_real},
\begin{equation}
    x+ct=\rho e^\eta,
    \qquad
    x-ct=\rho e^{-\eta}.
    \label{eq:rindler_null_coordinates}
\end{equation}
For finite $\eta$, the limit $\rho\rightarrow0$ approaches only the
origin. To approach a nonzero point on the future Rindler horizon,
$x=ct>0$, one instead requires the correlated limit
\begin{equation}
    \rho\rightarrow0,
    \qquad
    \eta\rightarrow+\infty,
    \qquad
    \rho e^\eta\rightarrow L>0,
\end{equation}
where $L$ is a finite positive constant. In this limit,
$x-ct\rightarrow0$ and $x+ct\rightarrow L$, so the limiting point
satisfies $x=ct=L/2$.

Similarly, a nonzero point on the past Rindler horizon, $x=-ct>0$,
is approached when
\begin{equation}
    \rho\rightarrow0,
    \qquad
    \eta\rightarrow-\infty,
    \qquad
    \rho e^{-\eta}\rightarrow L>0.
\end{equation}
Here $x+ct\rightarrow0$ and $x-ct\rightarrow L$, giving the limiting
point $x=-ct=L/2$. If the corresponding limiting value of $L$ is
zero, the trajectory instead approaches the origin.

The degeneration of Eq.~\eqref{eq:rindler_metric_full} at
$\rho=0$ therefore represents a coordinate horizon rather than a
singularity of Minkowski spacetime.

The same geometry can be expressed directly in terms of the complex
coordinates. Equation~\eqref{eq:rindler_complex} gives
\begin{equation}
    (X^0)^2+(X^1)^2=\rho^2,
\end{equation}
so each constant-$\rho$ Rindler trajectory again has the algebraic form
of a circle under the complex-bilinear contraction. Nevertheless,
$X^0\in i\mathbb{R}$ and $X^1\in\mathbb{R}$ for real $\eta$, and
Eq.~\eqref{eq:rindler_metric} retains Lorentzian signature. The
circle-like algebraic form therefore does not by itself imply a
Euclidean geometry.

Rindler coordinates provide a particularly direct demonstration of
the distinction between the complex representation developed here and
a Euclidean continuation. The bilinear form has the Euclidean
appearance $h=I$, and constant-$\rho$ trajectories satisfy a
formally circular relation, yet the restriction to the physical
complex slice produces the ordinary Lorentzian Rindler wedge, its
hyperbolic accelerated trajectories, and its null horizons. In the
following subsection, we will examine what changes when this physical
reality condition is replaced by one in which the corresponding
temporal coordinate is real.

\subsection{Lorentzian and Euclidean slices}
\label{sec:rindler_slices}

The Rindler construction provides a natural setting in which to
distinguish the complex representation of Lorentzian spacetime from
a genuinely Euclidean section. For real $\rho$ and $\eta$, the
complex Rindler coordinates are
\begin{equation}
    X_L(\rho,\eta)
    =
    \begin{pmatrix}
        i\rho\sinh\eta\\
        \rho\cosh\eta
    \end{pmatrix},
\end{equation}
where the subscript $L$ denotes the Lorentzian spacetime considered
thus far. Although the complex-bilinear form is $h=I$, the temporal
coordinate satisfies $X_L^0\in i\mathbb{R}$, and the induced line
element is
\begin{equation}
    dX_L^T h\,dX_L
    =
    d\rho^2-\rho^2d\eta^2.
\end{equation}
Thus the formally Euclidean bilinear expression retains Lorentzian
signature when restricted to the physical slice.

A genuinely Euclidean section can be obtained in this example by
analytically continuing the Rindler parametrization. As shown in
Sec.~\ref{sec:accel_worldlines}, the accelerated trajectory may be
expressed using the complex-rotation matrix,
\begin{equation}
    X_L(\rho,\eta)
    =
    R(-i\eta)
    \begin{pmatrix}
        0\\
        \rho
    \end{pmatrix}.
\end{equation}
To obtain an ordinary rotation through a real angle, introduce
\begin{equation}
    \eta=i\theta,
    \qquad
    \theta\in\mathbb{R},
\end{equation}
so that $-i\eta=\theta$. Unlike the original change of representation
$X^0=ict$, this step is an analytic continuation of the Rindler
parametrization beyond the Lorentzian physical slice. It is well
defined here because the coordinate transformation is analytic in
$\eta$.

The continued parametrization becomes
\begin{equation}
    X_E(\rho,\theta)
    =
    R(\theta)
    \begin{pmatrix}
        0\\
        \rho
    \end{pmatrix}
    =
    \begin{pmatrix}
        -\rho\sin\theta\\
        \rho\cos\theta
    \end{pmatrix}.
\end{equation}
Both components of $X_E$ are now real. The same complex-bilinear
form $h=I$ therefore gives
\begin{equation}
    dX_E^T h\,dX_E
    =
    d\rho^2+\rho^2d\theta^2,
    \label{eq:euclidean_rindler_line_element}
\end{equation}
which is the ordinary Euclidean metric in polar coordinates.
The constant-$\rho$ trajectories now describe actual circles
in a real Euclidean plane, rather than hyperbolae represented
in complex coordinates.

The Lorentzian and Euclidean geometries in this example are
distinguished by their reality conditions. On the Lorentzian slice,
\begin{equation}
    X^0\in i\mathbb{R},
    \qquad
    X^1\in\mathbb{R},
\end{equation}
whereas on the Euclidean slice,
\begin{equation}
    X^0\in\mathbb{R},
    \qquad
    X^1\in\mathbb{R}.
\end{equation}
The signature of the induced real geometry is therefore not
determined by the matrix $h=I$ alone. For this flat-spacetime
construction, the same complex-bilinear form is compatible with
both reality conditions, and the distinct signatures arise from
the real subspaces on which that form is evaluated.

The passage between these slices is consequently not an invertible
coordinate transformation within the original real Lorentzian
spacetime. Rather, analytic continuation permits the selection
of a different real subspace of the ambient complex vector space,
on which the bilinear form induces a different real metric.

This distinction requires additional care in more general
geometries. In particular, when a Lorentzian metric contains
off-diagonal temporal--spatial components, as in Kerr spacetime,
the corresponding complex-bilinear metric generally contains
imaginary cross terms. Simply changing the temporal reality
condition while holding that bilinear metric fixed need not
produce a real Euclidean metric. A compatible analytic
continuation of the metric data may also be necessary, and
a suitable real Euclidean section need not always exist.

The associated Hermitian structures provide a complementary
description of the distinction between the two slices. On the
Lorentzian slice, the Hermitian form introduced in
Sec.~\ref{sec:structures} is
\begin{equation}
    G_L
    =
    \begin{pmatrix}
        -1&0\\
        0&1
    \end{pmatrix},
\end{equation}
where only the $(X^0,X^1)$ sector is displayed. The complex
Lorentz transformation preserves this form according to
\begin{equation}
    U^\dagger G_L U=G_L.
\end{equation}
Thus the complex boost preserves the indefinite Hermitian
representation of the Lorentzian metric.

On the Lorentzian slice, $dX_L^0\in i\mathbb{R}$, so that
\begin{equation}
    (dX_L^0)^*=-dX_L^0.
\end{equation}
The negative temporal entry in $G_L$ compensates for the sign
introduced by complex conjugation. On the Euclidean slice,
however, $dX_E^0\in\mathbb{R}$, and therefore
\begin{equation}
    (dX_E^0)^*=dX_E^0.
\end{equation}
Since $dX_E$ is real, $dX_E^\dagger=dX_E^T$. Requiring the
Hermitian contraction to reproduce the Euclidean line element,
\begin{equation}
    ds_E^2=dX_E^\dagger G_E\,dX_E,
\end{equation}
therefore gives
\begin{equation}
    G_E
    =
    \begin{pmatrix}
        1&0\\
        0&1
    \end{pmatrix}.
\end{equation}
The real rotation $R(\theta)$ then satisfies
\begin{equation}
    R^\dagger(\theta)G_E R(\theta)=G_E,
\end{equation}
which reduces to the ordinary Euclidean relation
\begin{equation}
    R^T(\theta)R(\theta)=I.
\end{equation}

For the present flat-spacetime construction, the bilinear
form remains
\begin{equation}
    h_L=h_E=I,
\end{equation}
while the associated Hermitian forms $G_L$ and $G_E$ have
different signatures. Thus the change of real geometry is
encoded in the reality condition in the bilinear formulation
and is also reflected explicitly in the corresponding
Hermitian metric.

This relationship can be expressed more precisely by
introducing the conjugation operations whose fixed points
define the two real slices. For the Lorentzian representation,
define
\begin{equation}
    \sigma_L(X^0,X^1)
    =
    \left(-X^{0*},X^{1*}\right).
\end{equation}
The condition $\sigma_L(X)=X$ selects the Lorentzian slice,
with $X^0\in i\mathbb{R}$ and $X^1\in\mathbb{R}$. The
Euclidean slice is instead selected by ordinary complex
conjugation,
\begin{equation}
    \sigma_E(X^0,X^1)
    =
    \left(X^{0*},X^{1*}\right),
\end{equation}
whose fixed points have both coordinates real.

Writing these conjugation operations in the form
\begin{equation}
    \sigma(X)=CX^*,
\end{equation}
the corresponding matrices are
\begin{equation}
    C_L
    =
    \begin{pmatrix}
        -1&0\\
        0&1
    \end{pmatrix},
    \qquad
    C_E=I.
\end{equation}
For the present example, where $h=I$, the Hermitian forms
are related to the bilinear form by
\begin{equation}
    G_L=C_L^T h,
    \qquad
    G_E=C_E^T h.
\end{equation}

The distinction between Lorentzian and Euclidean sections
can therefore be expressed as a distinction between the
antilinear reality structures selecting their respective
real subspaces. In this flat example, the bilinear metric
remains unchanged, while the associated Hermitian metric
changes from an indefinite Lorentzian form to a
positive-definite Euclidean form.

The relationship between the bilinear metric, Hermitian
metric, and physical reality structure will be generalized
to curved spacetime in Sec.~\ref{sec:gr}. The additional
difficulties associated with Euclidean continuation in
curved spacetime will be considered in
Sec.~\ref{sec:wick_rotation}.

% ============================================================
\section{Extension to Curved Spacetime}
\label{sec:gr}
% ============================================================

\subsection{Complexified spacetime geometry}
\label{sec:geometry}

The flat-spacetime construction developed above extends naturally to a general
Lorentzian spacetime. Let $M$ be a four-dimensional real manifold equipped
with a Lorentzian metric $g_{\mu\nu}(x)$ of signature $(-,+,+,+)$. At each
point $p\in M$, the real tangent space $T_pM$ may be embedded in its
complexification,
\begin{equation}
T_pM \longrightarrow T_pM\otimes\mathbb{C},
\end{equation}
and the complex spacetime representation is defined locally by the same
linear map used in flat spacetime,
\begin{equation}
X^\mu=S^\mu{}_\nu x^\nu,
\qquad
S^\mu{}_\nu=\operatorname{diag}(i,1,1,1).
\end{equation}

In a general curved spacetime, this construction is understood relative
to a chosen real coordinate chart. The quantities $X^\mu$ provide complex
representations of the corresponding real coordinate values, while the
same matrix $S$ defines the associated transformation of tangent and
cotangent components. The construction does not require a single global
coordinate chart, a preferred global time coordinate, or a preferred
foliation of spacetime. It is not, in general, a real coordinate
transformation on the underlying Lorentzian manifold.

At the tangent-space level, the construction may be regarded as a
complex-linear change of component representation within the complexified
tangent and cotangent spaces. For a real tangent vector $V^\mu$, define
its complex representation by
\begin{equation}
V_X^\mu=S^\mu{}_\nu V^\nu.
\end{equation}
The corresponding complex vectors do not range independently over the
complexified tangent space. Just as in Minkowski spacetime, they form the
image of the original real tangent space under $S$. This physical real
subspace may be characterized by an antilinear involution. Since
\begin{equation}
V=S^{-1}V_X
\end{equation}
must have real components, complex conjugation gives
\begin{equation}
(S^{-1}V_X)^*=S^{-1}V_X.
\end{equation}
Multiplying by $S$ therefore gives
\begin{equation}
V_X=S(S^{-1})^*V_X^*.
\end{equation}
Defining
\begin{equation}
C\equiv S(S^{-1})^*,
\end{equation}
the physical reality condition may be written
\begin{equation}
V_X=C V_X^*.
\end{equation}
For the imaginary-time representation used here,
\begin{equation}
C=\operatorname{diag}(-1,1,1,1),
\end{equation}
so this condition requires the temporal component to be purely imaginary and
the spatial components to be real. The corresponding antilinear map
\begin{equation}
\sigma(V_X)=C V_X^*
\end{equation}
satisfies $\sigma^2=I$, and the physical Lorentzian tangent space is represented
by its fixed-point set. The reality structure $C$ is determined by the
chosen transformation $S$; it is not an additional independent physical
field. Different local representations may have different component
expressions for this structure while describing the same underlying
real tangent space.

The Lorentzian metric likewise admits two natural representations on the
complexified tangent space. Substituting
\begin{equation}
V=S^{-1}V_X,
\qquad
W=S^{-1}W_X
\end{equation}
into the real bilinear contraction $g(V,W)$ gives
\begin{equation}
g(V,W)
=
V_X^T (S^{-1})^T g S^{-1} W_X.
\end{equation}
The complex-bilinear representation of the metric is therefore
\begin{equation}
h=(S^{-1})^T g S^{-1},
\end{equation}
or, in components,
\begin{equation}
h_{\mu\nu}
:=
(S^{-1})^\alpha{}_\mu
(S^{-1})^\beta{}_\nu
g_{\alpha\beta}.
\end{equation}
Because $g_{\alpha\beta}=g_{\beta\alpha}$, the complex-bilinear metric remains
symmetric,
\begin{equation}
h_{\mu\nu}=h_{\nu\mu}.
\end{equation}

The corresponding Hermitian representation is obtained by requiring the same
real metric contraction to be expressed sesquilinearly on physical complex
vectors. As in Sec.~\ref{sec:structures},
\begin{equation}
G=(S^\dagger)^{-1}gS^{-1},
\end{equation}
or
\begin{equation}
G_{\bar\mu\nu}
:=
\left[(S^{-1})^\alpha{}_\mu\right]^*
(S^{-1})^\beta{}_\nu
g_{\alpha\beta}.
\end{equation}
A barred index has been introduced here to indicate that the first slot of
$G_{\bar\mu\nu}$ belongs to the conjugate cotangent space. The Hermitian
symmetry of this form is therefore
\begin{equation}
G_{\bar\mu\nu}=G_{\bar\nu\mu}^*,
\end{equation}
rather than the ordinary symmetry relation obeyed by $h_{\mu\nu}$.

The relationship observed in the Rindler example now follows in general.
Using
\begin{equation}
C=S(S^{-1})^*,
\end{equation}
one finds
\begin{equation}
G_{\bar\alpha\nu}
=
C^\mu{}_{\bar\alpha}h_{\mu\nu}.
\end{equation}
In matrix notation,
\begin{equation}
\boxed{G=C^T h.}
\end{equation}
Equivalently,
\begin{equation}
h=(C^{-1})^T G.
\end{equation}
Thus the complex-bilinear and Hermitian forms do not contain different
physical metric information. Once the physical reality structure is
specified, either determines the other. In particular, the pair
$(h,C)$ determines $G$, while the pair $(G,C)$ determines $h$.
Since $C$ itself is fixed by the chosen complex representation,
these relations introduce no additional gravitational degrees of
freedom.

The correspondence between Hermitian forms, symmetric
complex-bilinear forms, and antilinear operators also appears in
the general mathematical study of such structures, including
their canonical forms~\cite{SykesZelenko2020}. In the present
construction, the antilinear operator is specifically the
involution defining the physical Lorentzian real subspace.

The relationship between the bilinear and Hermitian forms also
explains why they behave differently under a change of real section.
The bilinear form $h$ extends the metric
without complex conjugation, whereas the Hermitian form combines the
bilinear metric structure with the conjugation operation that identifies
the physical real subspace. In the flat Rindler example of
Sec.~\ref{sec:rindler_slices}, $h=I$ was compatible with both the
Lorentzian and Euclidean reality conditions, while the corresponding
Hermitian form changed from $G_L=\eta$ to $G_E=I$. For a general
spacetime, however, the bilinear metric itself need not remain unchanged
when the real section is analytically continued. As discussed in
Sec.~\ref{sec:rindler_slices}, such a continuation is distinct from
the equivalent complex representation of the original Lorentzian
geometry developed here.

For example, with $S=\operatorname{diag}(i,1,1,1)$, the components of the
complex-bilinear metric are
\begin{equation}
h_{00}=-g_{00},
\qquad
h_{0i}=h_{i0}=-i g_{0i},
\qquad
h_{ij}=g_{ij},
\end{equation}
whereas the Hermitian representation has
\begin{equation}
G_{\bar 0 0}=g_{00},
\qquad
G_{\bar 0 i}=i g_{0i},
\qquad
G_{\bar i 0}=-i g_{i0},
\qquad
G_{\bar i j}=g_{ij}.
\end{equation}
The temporal--spatial cross terms are therefore purely imaginary in both
representations but appear with the symmetry appropriate to each structure:
$h_{\mu\nu}$ is complex symmetric, whereas $G_{\bar\mu\nu}$ is Hermitian.
This distinction will be particularly important for rotating spacetimes such
as Kerr.

Both representations nevertheless recover exactly the same Lorentzian metric
on the physical real subspace:
\begin{equation}
V_X^T h W_X
=
V_X^\dagger G W_X
=
V^T g W,
\end{equation}
for physical vectors $V_X=SV$ and $W_X=SW$. These expressions are
evaluated using the metric components at the corresponding physical
spacetime point. No holomorphic extension of the metric away from
the physical Lorentzian slice is required.

The choice between $h$ and $G$ is therefore not a choice between
physically complete and incomplete metric descriptions. Rather,
the two forms encode the same Lorentzian geometry with different
complex index structures. The following subsections will show how
the inverse metric and the differential-geometric structures of general
relativity are represented in each formulation.

\subsection{Metric and inverse metric}
\label{sec:metric}

The distinction between the bilinear and Hermitian metric representations
becomes particularly important when indices are raised and lowered. For the
complex-bilinear metric, the inverse is defined in the usual way by
\begin{equation}
h^{\mu\alpha}h_{\alpha\nu}
=
\delta^\mu{}_\nu.
\end{equation}
Since
\begin{equation}
h=(S^{-1})^T g S^{-1},
\end{equation}
its inverse is
\begin{equation}
h^{-1}=Sg^{-1}S^T,
\end{equation}
or, in components,
\begin{equation}
h^{\mu\nu}
=
S^\mu{}_\alpha
S^\nu{}_\beta
g^{\alpha\beta}.
\end{equation}
The inverse bilinear metric is therefore symmetric,
\begin{equation}
h^{\mu\nu}=h^{\nu\mu},
\end{equation}
and raises and lowers ordinary complex tensor indices in the same algebraic
manner as the real Lorentzian metric.

The inverse Hermitian metric has a different index structure. Since
$G_{\bar\mu\nu}$ pairs a conjugate covector index with an ordinary covector
index, its inverse is written
\begin{equation}
G^{\mu\bar\nu},
\end{equation}
and is defined by
\begin{equation}
G^{\mu\bar\alpha}G_{\bar\alpha\nu}
=
\delta^\mu{}_\nu,
\qquad
G_{\bar\mu\alpha}G^{\alpha\bar\nu}
=
\delta_{\bar\mu}{}^{\bar\nu}.
\end{equation}
From
\begin{equation}
G=(S^\dagger)^{-1}gS^{-1},
\end{equation}
one obtains
\begin{equation}
G^{-1}=Sg^{-1}S^\dagger.
\end{equation}
Thus the inverse Hermitian metric satisfies
\begin{equation}
G^{\mu\bar\nu}
=
\left(G^{\nu\bar\mu}\right)^*.
\end{equation}

The relation between the two inverse metrics follows from the corresponding
relation between the covariant forms. From Sec.~\ref{sec:geometry},
\begin{equation}
G_{\bar\alpha\nu}
=
C^\mu{}_{\bar\alpha}h_{\mu\nu}.
\end{equation}
Multiplying by $G^{\rho\bar\alpha}$ gives
\begin{equation}
G^{\rho\bar\alpha}
C^\mu{}_{\bar\alpha}
h_{\mu\nu}
=
\delta^\rho{}_\nu.
\end{equation}
Comparison with
\begin{equation}
h^{\rho\mu}h_{\mu\nu}
=
\delta^\rho{}_\nu
\end{equation}
therefore yields
\begin{equation}
\boxed{
h^{\rho\mu}
=
G^{\rho\bar\alpha}
C^\mu{}_{\bar\alpha}.
}
\end{equation}
In matrix notation, this relation is
\begin{equation}
h^{-1}=G^{-1}C^T.
\end{equation}
The physical reality structure therefore provides the conversion between the
Hermitian inverse metric and the ordinary complex-bilinear inverse metric,
just as it relates their covariant counterparts.

For the particular representation
\begin{equation}
S=\operatorname{diag}(i,1,1,1),
\qquad
C=\operatorname{diag}(-1,1,1,1),
\end{equation}
the components of the inverse bilinear metric are
\begin{equation}
h^{00}=-g^{00},
\qquad
h^{0i}=h^{i0}=i g^{0i},
\qquad
h^{ij}=g^{ij}.
\end{equation}
The corresponding inverse Hermitian metric has
\begin{equation}
G^{0\bar 0}=g^{00},
\qquad
G^{0\bar i}=i g^{0i},
\qquad
G^{i\bar 0}=-i g^{i0},
\qquad
G^{i\bar j}=g^{ij}.
\end{equation}
As for the covariant metrics, temporal--spatial cross terms therefore
distinguish particularly clearly between the symmetric bilinear and
Hermitian representations.

For an ordinary complex covector $A_\mu$, index raising with the bilinear
metric is defined directly by
\begin{equation}
A^\mu=h^{\mu\nu}A_\nu.
\end{equation}
The same operation may be expressed using the Hermitian metric only after the
reality structure is inserted:
\begin{equation}
A^\mu
=
G^{\mu\bar\alpha}
C^\nu{}_{\bar\alpha}A_\nu.
\end{equation}
Likewise, lowering an ordinary complex vector index may be written
\begin{equation}
A_\mu=h_{\mu\nu}A^\nu
=
(C^{-1})^{\bar\alpha}{}_\mu
G_{\bar\alpha\nu}A^\nu.
\end{equation}
Thus $G_{\bar\mu\nu}$ does not act as an ordinary symmetric metric tensor
with two interchangeable covariant indices. Its barred and unbarred slots
must remain distinct. In particular, raising or lowering an ordinary
tensor index using the Hermitian representation requires the appropriate
conversion supplied by the physical reality structure $C$; a barred index
cannot simply be treated as interchangeable with an ordinary tensor index.
The resulting contractions are equivalent to those obtained directly
from the complex-bilinear metric $h$.

This observation explains why the bilinear representation is particularly
convenient for the differential geometry developed below. The metric
$h_{\mu\nu}$ and its inverse $h^{\mu\nu}$ have precisely the index structure
required by the conventional tensor calculus of general relativity, whereas
the Hermitian representation retains explicitly the distinction between the
complex vector space and its conjugate. This difference is one of algebraic
organization rather than physical content: together with $C$, the Hermitian
metric contains exactly the information required to reconstruct $h$ and its
inverse. In the following subsection, this equivalence will be extended to
the Levi-Civita connection.

\subsection{Levi-Civita connection}
\label{sec:levi-civita}

The metric relations established above may now be extended from algebraic
contractions to covariant differentiation. In the conventional real
representation, the Levi-Civita connection associated with
$g_{\mu\nu}$ is
\begin{equation}
\Gamma^\rho{}_{\mu\nu}
=
\frac{1}{2}g^{\rho\sigma}
\left(
\partial_\mu g_{\sigma\nu}
+
\partial_\nu g_{\sigma\mu}
-
\partial_\sigma g_{\mu\nu}
\right).
\end{equation}
The connection is uniquely characterized by vanishing torsion and metric
compatibility.

In the complex representation, the symmetric bilinear metric
$h_{\mu\nu}$ has the same tensorial index structure as the conventional
metric. Its Levi-Civita connection may therefore be written directly as
\begin{equation}
\boxed{
\widetilde{\Gamma}^\rho{}_{\mu\nu}
=
\frac{1}{2}h^{\rho\lambda}
\left(
\partial_\mu h_{\lambda\nu}
+
\partial_\nu h_{\lambda\mu}
-
\partial_\lambda h_{\mu\nu}
\right)
}
\end{equation}
Here and below, derivatives are taken with respect to the complex
coordinates $X^\mu$,
\begin{equation}
\partial_\mu
\equiv
\frac{\partial}{\partial X^\mu}
=
(S^{-1})^\alpha{}_\mu
\frac{\partial}{\partial x^\alpha}.
\end{equation}
As in Sec.~\ref{sec:complex_coordinates}, these derivatives are
understood as complex-linear combinations of derivatives along the
underlying real spacetime. They act on the metric components and other
fields defined on the physical Lorentzian slice, without requiring a
holomorphic extension into the surrounding complex coordinate space.

Because $S$ is constant, substitution of
\begin{equation}
h_{\mu\nu}
=
(S^{-1})^\alpha{}_\mu
(S^{-1})^\beta{}_\nu
g_{\alpha\beta}
\end{equation}
into the preceding expression gives
\begin{equation}
\boxed{
\widetilde{\Gamma}^\rho{}_{\mu\nu}
=
S^\rho{}_\alpha
(S^{-1})^\beta{}_\mu
(S^{-1})^\gamma{}_\nu
\Gamma^\alpha{}_{\beta\gamma}
}
\end{equation}
The connection coefficients in the complex representation are therefore
precisely those obtained by transforming the conventional Levi-Civita
connection under the constant complex change of representation. No new
connection or gravitational dynamics has been introduced.

For the imaginary-time transformation
\begin{equation}
S=\operatorname{diag}(i,1,1,1),
\end{equation}
this relation makes the redistribution of factors of $i$ explicit. A
temporal upper index contributes a factor of $i$, while each temporal lower
index contributes a factor of $-i$:
\begin{equation}
\widetilde{\Gamma}^\rho{}_{\mu\nu}
=
i^{\delta_{\rho 0}}
(-i)^{\delta_{\mu 0}+\delta_{\nu 0}}
\Gamma^\rho{}_{\mu\nu},
\end{equation}
with no summation over the displayed indices in this expression. Thus the
complex connection contains the same information as the conventional
connection, expressed in the transformed basis.

The same connection can also be reconstructed entirely from the Hermitian
metric and the physical reality structure. From Secs.~\ref{sec:geometry} and \ref{sec:metric},
\begin{equation}
h_{\mu\nu}
=
(C^{-1})^{\bar\alpha}{}_\mu
G_{\bar\alpha\nu},
\end{equation}
and
\begin{equation}
h^{\rho\lambda}
=
G^{\rho\bar\alpha}
C^\lambda{}_{\bar\alpha}.
\end{equation}
Substitution into the bilinear Levi-Civita expression gives
\begin{align}
\widetilde{\Gamma}^\rho{}_{\mu\nu}
={}&
\frac{1}{2}
G^{\rho\bar\alpha}
C^\lambda{}_{\bar\alpha}
\Bigg[
\partial_\mu
\left(
(C^{-1})^{\bar\beta}{}_\lambda
G_{\bar\beta\nu}
\right)
\nonumber\\
&\qquad+
\partial_\nu
\left(
(C^{-1})^{\bar\beta}{}_\lambda
G_{\bar\beta\mu}
\right)
-
\partial_\lambda
\left(
(C^{-1})^{\bar\beta}{}_\mu
G_{\bar\beta\nu}
\right)
\Bigg].
\end{align}
This expression demonstrates explicitly that the Hermitian representation,
together with the physical conjugation structure, contains all of the
information required to reconstruct the Levi-Civita connection.

For the representation used throughout this work, $C$ is constant.
The preceding expression then reduces to
\begin{align}
\widetilde{\Gamma}^\rho{}_{\mu\nu}
={}&
\frac{1}{2}
G^{\rho\bar\alpha}
\left(
\partial_\mu G_{\bar\alpha\nu}
+
\partial_\nu G_{\bar\alpha\mu}
\right)
\nonumber\\
&-
\frac{1}{2}
G^{\rho\bar\alpha}
C^\lambda{}_{\bar\alpha}
(C^{-1})^{\bar\beta}{}_\mu
\partial_\lambda G_{\bar\beta\nu}.
\end{align}
Although less compact than the corresponding expression in terms of $h$,
this is the same Levi-Civita connection, not a distinct Hermitian
connection. Because the connection is induced from the real Lorentzian
Levi-Civita connection, it preserves the physical reality structure.
Metric compatibility may consequently be expressed in terms of $G$ as
well as $h$, provided the covariant derivative acts on the conjugate
index of $G$ through the corresponding induced conjugate connection.

It is important in particular that the conventional Levi-Civita formula
cannot in general be obtained by simply replacing $h_{\mu\nu}$ with
$G_{\bar\mu\nu}$. An expression such as
\begin{equation}
\frac{1}{2}G^{\rho\bar\sigma}
\left(
\partial_\mu G_{\bar\sigma\nu}
+
\partial_\nu G_{\bar\sigma\mu}
-
\partial_\sigma G_{\bar\mu\nu}
\right)
\end{equation}
does not respect the barred and unbarred index structure of the Hermitian
metric and therefore does not represent the Levi-Civita connection of the
physical Lorentzian geometry in general. The conjugation map $C$ is required
to convert the Hermitian metric into the symmetric bilinear structure on
which the ordinary Levi-Civita construction acts.

The bilinear representation is therefore algebraically adapted to the
standard differential geometry of general relativity. It absorbs the
physical conjugation map into the symmetric tensor $h_{\mu\nu}$, allowing
the familiar Levi-Civita formula to retain its ordinary form. The Hermitian
representation instead retains explicitly the distinction between the
complex vector space and its conjugate. This distinction changes the
bookkeeping but not the physical geometry: the pair $(G,C)$ reconstructs
the same connection as $h$.

Metric compatibility is consequently preserved in the complex
representation:
\begin{equation}
\widetilde{\nabla}_\lambda h_{\mu\nu}=0,
\end{equation}
and the connection remains torsion-free,
\begin{equation}
\widetilde{\Gamma}^\rho{}_{\mu\nu}
=
\widetilde{\Gamma}^\rho{}_{\nu\mu}.
\end{equation}
These properties establish the complex representation of the same
Levi-Civita geometry carried by the original Lorentzian metric. The
corresponding geodesic equation will be considered in the following
subsection.

\subsection{Geodesics and conserved quantities}
\label{sec:geodesics}

Having established the Levi-Civita connection in the complex
representation, the corresponding geodesic equation follows directly.
From this point onward, unless otherwise specified, $V^\mu$ will
denote the components of a tangent vector in the complex representation,
rather than the conventional real-coordinate components distinguished
in Sec.~\ref{sec:geometry}. Let $\lambda$ be an affine parameter along
a geodesic and define
\begin{equation}
V^\mu=\frac{dX^\mu}{d\lambda}.
\end{equation}
The geodesic equation is then
\begin{equation}
\boxed{
\frac{d^2X^\rho}{d\lambda^2}
+
\widetilde{\Gamma}^\rho{}_{\mu\nu}
\frac{dX^\mu}{d\lambda}
\frac{dX^\nu}{d\lambda}
=0
}
\end{equation}
or, equivalently,
\begin{equation}
V^\nu\widetilde{\nabla}_\nu V^\mu=0.
\end{equation}

The equivalence of this equation with the conventional real-coordinate
geodesic equation can be shown directly. Let
\begin{equation}
X^\mu=S^\mu{}_\alpha x^\alpha.
\end{equation}
Since $S$ is constant,
\begin{equation}
\frac{dX^\mu}{d\lambda}
=
S^\mu{}_\alpha\frac{dx^\alpha}{d\lambda},
\qquad
\frac{d^2X^\mu}{d\lambda^2}
=
S^\mu{}_\alpha\frac{d^2x^\alpha}{d\lambda^2}.
\end{equation}
Using the transformed connection derived in Sec.~\ref{sec:levi-civita},
\begin{equation}
\widetilde{\Gamma}^\rho{}_{\mu\nu}
=
S^\rho{}_\alpha
(S^{-1})^\beta{}_\mu
(S^{-1})^\gamma{}_\nu
\Gamma^\alpha{}_{\beta\gamma},
\end{equation}
the complex geodesic equation becomes
\begin{equation}
S^\rho{}_\alpha
\left[
\frac{d^2x^\alpha}{d\lambda^2}
+
\Gamma^\alpha{}_{\beta\gamma}
\frac{dx^\beta}{d\lambda}
\frac{dx^\gamma}{d\lambda}
\right]
=0.
\end{equation}
Because $S$ is invertible, this is equivalent to
\begin{equation}
\frac{d^2x^\alpha}{d\lambda^2}
+
\Gamma^\alpha{}_{\beta\gamma}
\frac{dx^\beta}{d\lambda}
\frac{dx^\gamma}{d\lambda}
=0.
\end{equation}
Thus the complex representation describes the same geodesics as
the conventional Lorentzian formulation. Only their representation in the
complexified tangent space has changed.

For a timelike geodesic, the imaginary proper-time parameter introduced in
Sec.~\ref{sec:imaginary_tau} provides a natural affine parameter,
\begin{equation}
\Sigma=ic\tau.
\end{equation}
The corresponding tangent vector is
\begin{equation}
W^\mu=\frac{dX^\mu}{d\Sigma},
\end{equation}
and the geodesic equation takes the form
\begin{equation}
\boxed{
\frac{dW^\rho}{d\Sigma}
+
\widetilde{\Gamma}^\rho{}_{\mu\nu}
W^\mu W^\nu
=0.
}
\end{equation}
Since $\Sigma$ differs from the conventional proper time $\tau$ only by the
constant factor $ic$, it remains an affine parameter whenever $\tau$ is
affine.

As in Sec.~\ref{sec:four-velocity}, the tangent vector satisfies the bilinear and Hermitian
normalization conditions
\begin{equation}
W^\mu h_{\mu\nu}W^\nu=1,
\end{equation}
and
\begin{equation}
W^\dagger G W=-1.
\end{equation}
These relations remain valid in curved spacetime because they follow from
the invariant proper-time relation
\begin{equation}
ds^2=d\Sigma^2=-c^2d\tau^2
\end{equation}
together with the two equivalent representations of the metric on the
physical Lorentzian subspace.

Multiplying $W^\mu$ by $mc$ gives the complementary momentum
\begin{equation}
\Pi^\mu=mcW^\mu.
\end{equation}
For a freely falling massive particle, the geodesic equation may therefore
be written
\begin{equation}
\boxed{
\frac{d\Pi^\rho}{d\Sigma}
+
\widetilde{\Gamma}^\rho{}_{\mu\nu}
\Pi^\mu W^\nu
=0,
}
\end{equation}
or equivalently,
\begin{equation}
W^\nu\widetilde{\nabla}_\nu\Pi^\mu=0.
\end{equation}
The complementary momentum is therefore parallel transported along a
timelike geodesic, just as the conventional four-momentum is parallel
transported in the real Lorentzian representation.

The bilinear mass-shell condition likewise generalizes directly:
\begin{equation}
\boxed{
h_{\mu\nu}\Pi^\mu\Pi^\nu=m^2c^2.
}
\end{equation}
Its Hermitian counterpart is
\begin{equation}
\Pi^\dagger G\Pi=-m^2c^2.
\end{equation}
Metric compatibility ensures that these quantities remain constant under
parallel transport.

Spacetime symmetries and their associated conserved quantities may also be
expressed directly in the complex representation. Let $\xi^\mu$ be a Killing
vector in the conventional Lorentzian representation, and define its complex
representation by
\begin{equation}
\Xi^\mu
=
S^\mu{}_\nu\xi^\nu.
\end{equation}
The transformed Killing vector satisfies
\begin{equation}
\widetilde{\nabla}_{\mu}\Xi_{\nu}
+
\widetilde{\nabla}_{\nu}\Xi_{\mu}
=
0,
\end{equation}
where
\begin{equation}
\Xi_{\mu}=h_{\mu\nu}\Xi^{\nu}.
\end{equation}
Along an affinely parameterized geodesic, consider the contraction
\begin{equation}
\mathcal{K}
=
\Xi_{\mu}V^{\mu}.
\end{equation}
Differentiating along the geodesic gives
\begin{align}
\frac{d\mathcal{K}}{d\lambda}
&=
V^{\nu}\widetilde{\nabla}_{\nu}
\left(\Xi_{\mu}V^{\mu}\right)
\nonumber\\
&=
V^{\mu}V^{\nu}\widetilde{\nabla}_{\nu}\Xi_{\mu}
+
\Xi_{\mu}V^{\nu}\widetilde{\nabla}_{\nu}V^{\mu}.
\end{align}
The second term vanishes by the geodesic equation. Since
$V^{\mu}V^{\nu}$ is symmetric in its indices while
\begin{equation}
\widetilde{\nabla}_{(\mu}\Xi_{\nu)}=0,
\end{equation}
the first term also vanishes. Therefore
\begin{equation}
\frac{d\mathcal{K}}{d\lambda}=0.
\end{equation}
Every Killing symmetry of the original Lorentzian spacetime therefore
produces the corresponding conserved quantity in the complex
representation.

For a massive particle parameterized by $\Sigma$, it is convenient to write
the corresponding conserved quantity in terms of the complementary momentum:
\begin{equation}
\mathcal{K}_{\Pi}
=
\Xi_{\mu}\Pi^{\mu}
=
h_{\mu\nu}\Xi^{\mu}\Pi^{\nu},
\end{equation}
so that
\begin{equation}
\frac{d\mathcal{K}_{\Pi}}{d\Sigma}=0.
\end{equation}
Using
\begin{equation}
\Pi^\mu=-iS^\mu{}_\nu p^\nu,
\end{equation}
together with the transformation of the Killing covector, gives
\begin{equation}
\mathcal{K}_{\Pi}
=
\Xi_\mu\Pi^\mu
=
-i\,\xi_\mu p^\mu.
\end{equation}
Thus the complementary-momentum conserved quantity differs from the
conventional Killing constant by the universal phase $-i$.  The two
expressions encode the same physical constant of motion, with the factor of
$i$ redistributed into its complex representation.

The same conserved quantity may also be reconstructed from the Hermitian
metric and the physical reality structure. Using
\begin{equation}
h_{\mu\nu}
=
(C^{-1})^{\bar\alpha}{}_\mu
G_{\bar\alpha\nu},
\end{equation}
one obtains
\begin{equation}
\mathcal{K}_\Pi
=
(C^{-1})^{\bar\alpha}{}_\mu
G_{\bar\alpha\nu}
\Xi^\mu\Pi^\nu.
\end{equation}
Thus the same geodesic dynamics and conserved quantities may be
expressed using either metric representation, with the physical reality
structure providing the necessary conversions between barred and
unbarred indices.

For stationary and axisymmetric spacetimes, the temporal and rotational
Killing vectors generate the conserved quantities conventionally
identified with energy and angular momentum. Their physical content is
unchanged in the complex representation, although factors of $i$ are
redistributed among the corresponding vectors, covectors, and
complementary momentum. The Kerr spacetime considered in
Sec.~\ref{sec:kerr} will provide an explicit example.

\subsection{Curvature}
\label{sec:curvature}

Once the Levi-Civita connection has been constructed, the curvature tensors
in the complex representation follow in the usual manner. The Riemann
curvature tensor is defined by
\begin{equation}
\boxed{
\widetilde{R}^\rho{}_{\sigma\mu\nu}
=
\partial_\mu
\widetilde{\Gamma}^\rho{}_{\nu\sigma}
-
\partial_\nu
\widetilde{\Gamma}^\rho{}_{\mu\sigma}
+
\widetilde{\Gamma}^\rho{}_{\mu\lambda}
\widetilde{\Gamma}^\lambda{}_{\nu\sigma}
-
\widetilde{\Gamma}^\rho{}_{\nu\lambda}
\widetilde{\Gamma}^\lambda{}_{\mu\sigma}
}
\end{equation}
where the derivatives are taken with respect to the complex coordinates
$X^\mu$.

Because the transformation
\begin{equation}
X^\mu=S^\mu{}_\alpha x^\alpha
\end{equation}
is linear and $S$ is constant, the Riemann tensor transforms directly as
\begin{equation}
\boxed{
\widetilde{R}^\rho{}_{\sigma\mu\nu}
=
S^\rho{}_\alpha
(S^{-1})^\beta{}_\sigma
(S^{-1})^\gamma{}_\mu
(S^{-1})^\delta{}_\nu
R^\alpha{}_{\beta\gamma\delta}.
}
\end{equation}
Thus the curvature represented by
$\widetilde{R}^\rho{}_{\sigma\mu\nu}$ is precisely the curvature of the
original Lorentzian spacetime expressed in the complex basis.

For the imaginary-time transformation
\begin{equation}
S=\operatorname{diag}(i,1,1,1),
\end{equation}
the redistribution of factors of $i$ may again be displayed explicitly:
\begin{equation}
\widetilde{R}^\rho{}_{\sigma\mu\nu}
=
i^{\delta_{\rho0}}
(-i)^{\delta_{\sigma0}+\delta_{\mu0}+\delta_{\nu0}}
R^\rho{}_{\sigma\mu\nu},
\end{equation}
with no summation over the displayed indices in this expression. The
complex components may therefore differ from their conventional
counterparts by factors of $i$ or by signs, while encoding the same
curvature tensor.

The Ricci tensor is obtained by contraction of the first and third indices,
\begin{equation}
\boxed{
\widetilde{R}_{\sigma\nu}
=
\widetilde{R}^\rho{}_{\sigma\rho\nu}.
}
\end{equation}
Using the transformation law above gives
\begin{equation}
\widetilde{R}_{\mu\nu}
=
(S^{-1})^\alpha{}_\mu
(S^{-1})^\beta{}_\nu
R_{\alpha\beta}.
\end{equation}
The Ricci tensor therefore transforms as an ordinary covariant rank-two
tensor under the constant complex change of representation. Since the
original Levi-Civita Ricci tensor is symmetric,
\begin{equation}
R_{\alpha\beta}=R_{\beta\alpha},
\end{equation}
its complex representation remains symmetric:
\begin{equation}
\widetilde{R}_{\mu\nu}
=
\widetilde{R}_{\nu\mu}.
\end{equation}

It is worth emphasizing that neither the construction of the Riemann
tensor from the connection nor the contraction producing the Ricci tensor
requires a further choice between the bilinear and Hermitian metric
representations. Once the Levi-Civita connection has been reconstructed,
these curvature tensors follow directly from it. The metric re-enters when
the Ricci tensor is contracted to form the curvature scalar.

Using the complex-bilinear inverse metric, the Ricci scalar is
\begin{equation}
\boxed{
\widetilde{R}
=
h^{\mu\nu}\widetilde{R}_{\mu\nu}.
}
\end{equation}
Substituting
\begin{equation}
h^{\mu\nu}
=
S^\mu{}_\alpha
S^\nu{}_\beta
g^{\alpha\beta}
\end{equation}
and the transformed Ricci tensor gives
\begin{align}
h^{\mu\nu}\widetilde{R}_{\mu\nu}
&=
S^\mu{}_\alpha
S^\nu{}_\beta
g^{\alpha\beta}
(S^{-1})^\gamma{}_\mu
(S^{-1})^\delta{}_\nu
R_{\gamma\delta}
\nonumber\\
&=
g^{\alpha\beta}R_{\alpha\beta}.
\end{align}
Hence
\begin{equation}
\boxed{
\widetilde{R}=R,
}
\end{equation}
where the tilde on the left denotes the scalar calculated in the complex
representation. As expected for a scalar, its value is unchanged.

The same contraction can be written entirely in terms of the Hermitian
metric and the physical reality structure. From Sec.~\ref{sec:metric},
\begin{equation}
h^{\mu\nu}
=
G^{\mu\bar\alpha}
C^\nu{}_{\bar\alpha},
\end{equation}
and therefore
\begin{equation}
\boxed{
\widetilde{R}
=
G^{\mu\bar\alpha}
C^\nu{}_{\bar\alpha}
\widetilde{R}_{\mu\nu}.
}
\end{equation}
The matrix $C$ again supplies the conversion required between the barred
index of the Hermitian inverse metric and the ordinary covariant index of
the Ricci tensor. The bilinear and Hermitian formulations consequently
produce the same curvature scalar.

The preservation of curvature under the complex representation may also
be understood geometrically through parallel transport. For a vector
$V^\rho$, the commutator of covariant derivatives is
\begin{equation}
\left[
\widetilde{\nabla}_\mu,
\widetilde{\nabla}_\nu
\right]
V^\rho
=
\widetilde{R}^\rho{}_{\sigma\mu\nu}V^\sigma.
\end{equation}
Since both the vector and the connection are related invertibly to their
conventional Lorentzian counterparts, the resulting change under parallel
transport around an infinitesimal closed loop is likewise the complex
representation of the same physical curvature. Complexification changes
the components used to describe this holonomy, but not the underlying
geometric effect.

In particular, a spacetime that is flat in the conventional
representation remains flat in the complex representation:
\begin{equation}
R^\rho{}_{\sigma\mu\nu}=0
\quad\Longleftrightarrow\quad
\widetilde{R}^\rho{}_{\sigma\mu\nu}=0.
\end{equation}
Likewise, a Ricci-flat spacetime satisfies
\begin{equation}
R_{\mu\nu}=0
\quad\Longleftrightarrow\quad
\widetilde{R}_{\mu\nu}=0.
\end{equation}
The latter result will be particularly useful for the Kerr spacetime
examined in Sec.~\ref{sec:kerr}.

The complex representation therefore preserves the complete local
curvature structure of the original Lorentzian geometry. The Riemann and
Ricci tensors transform covariantly, while scalar contractions remain
invariant and may be expressed using either metric representation.
These results provide the final geometric ingredients needed to express
the Einstein field equations in the complex representation.

\subsection{Einstein equations and conservation}
\label{sec:einstein}

The curvature construction of the preceding subsection provides the
remaining ingredients needed to express the Einstein field equations in
the complex spacetime representation. In the conventional Lorentzian
formulation, the Einstein tensor is
\begin{equation}
\mathcal{G}_{\mu\nu}
=
R_{\mu\nu}
-
\frac{1}{2}g_{\mu\nu}R.
\end{equation}
To distinguish the Einstein tensor from the Hermitian metric $G$, we use
the calligraphic symbol $\mathcal{G}_{\mu\nu}$ throughout this section.

In the complex representation, the corresponding tensor is obtained by
replacing the real metric and Ricci tensor by their complex-bilinear
representations:
\begin{equation}
\boxed{
\widetilde{\mathcal{G}}_{\mu\nu}
:=
\widetilde{R}_{\mu\nu}
-
\frac{1}{2}h_{\mu\nu}\widetilde{R}.
}
\end{equation}
Since
\begin{equation}
\widetilde{R}_{\mu\nu}
=
(S^{-1})^\alpha{}_\mu
(S^{-1})^\beta{}_\nu
R_{\alpha\beta},
\end{equation}
and
\begin{equation}
h_{\mu\nu}
=
(S^{-1})^\alpha{}_\mu
(S^{-1})^\beta{}_\nu
g_{\alpha\beta},
\end{equation}
while the Ricci scalar is invariant, it follows immediately that
\begin{equation}
\boxed{
\widetilde{\mathcal{G}}_{\mu\nu}
=
(S^{-1})^\alpha{}_\mu
(S^{-1})^\beta{}_\nu
\mathcal{G}_{\alpha\beta}.
}
\end{equation}
The Einstein tensor therefore transforms as an ordinary covariant
rank-two tensor under the constant complex change of representation.

The same tensor may be written entirely in terms of the Hermitian metric
and the physical reality structure. Using
\begin{equation}
h_{\mu\nu}
=
(C^{-1})^{\bar\alpha}{}_\mu
G_{\bar\alpha\nu}
\end{equation}
and
\begin{equation}
\widetilde{R}
=
G^{\rho\bar\beta}
C^\sigma{}_{\bar\beta}
\widetilde{R}_{\rho\sigma},
\end{equation}
one obtains
\begin{equation}
\boxed{
\widetilde{\mathcal{G}}_{\mu\nu}
=
\widetilde{R}_{\mu\nu}
-
\frac{1}{2}
(C^{-1})^{\bar\alpha}{}_\mu
G_{\bar\alpha\nu}
G^{\rho\bar\beta}
C^\sigma{}_{\bar\beta}
\widetilde{R}_{\rho\sigma}.
}
\end{equation}
Thus the Hermitian representation, together with the physical conjugation
structure, reconstructs the same Einstein tensor as the
complex-bilinear representation. As in the preceding sections, the
difference between the two formulations is one of index structure rather
than physical content.

The matter sector transforms in the same manner. Let
$T_{\mu\nu}$ denote the conventional stress-energy tensor. Its
complex-bilinear representation is
\begin{equation}
\boxed{
\widetilde{T}_{\mu\nu}
=
(S^{-1})^\alpha{}_\mu
(S^{-1})^\beta{}_\nu
T_{\alpha\beta}.
}
\end{equation}
For the imaginary-time transformation
\begin{equation}
S=\operatorname{diag}(i,1,1,1),
\end{equation}
the components therefore satisfy
\begin{equation}
\widetilde{T}_{00}=-T_{00},
\qquad
\widetilde{T}_{0i}
=
\widetilde{T}_{i0}
=
-iT_{0i},
\qquad
\widetilde{T}_{ij}=T_{ij}.
\end{equation}
The factors of $i$ appearing in mixed temporal--spatial components are
the same factors required by the transformed metric and curvature
tensors. They reflect the complex representation of the tensor indices
rather than any change in the physical matter distribution.

The conventional Einstein field equations,
\begin{equation}
\mathcal{G}_{\mu\nu}
=
\frac{8\pi G_N}{c^4}T_{\mu\nu},
\end{equation}
therefore become
\begin{equation}
\boxed{
\widetilde{\mathcal{G}}_{\mu\nu}
=
\frac{8\pi G_N}{c^4}
\widetilde{T}_{\mu\nu}.
}
\end{equation}
Because both sides transform according to the same covariant rank-two
rule, this equation is exactly equivalent to the conventional Einstein
equations. The complex representation introduces no additional field
equations and modifies neither the gravitational coupling nor the
physical stress-energy content.

If a cosmological constant is included, the same conclusion follows from
\begin{equation}
\mathcal{G}_{\mu\nu}
+
\Lambda g_{\mu\nu}
=
\frac{8\pi G_N}{c^4}T_{\mu\nu}.
\end{equation}
Its complex representation is
\begin{equation}
\boxed{
\widetilde{\mathcal{G}}_{\mu\nu}
+
\Lambda h_{\mu\nu}
=
\frac{8\pi G_N}{c^4}
\widetilde{T}_{\mu\nu}.
}
\end{equation}
The cosmological constant is a scalar and is therefore unchanged, while
the metric multiplying it is represented by $h_{\mu\nu}$.

Covariant conservation is preserved as well. The contracted Bianchi
identity in the complex representation gives
\begin{equation}
\widetilde{\nabla}_\mu
\widetilde{\mathcal{G}}^\mu{}_\nu
=0.
\end{equation}
The mixed Einstein tensor is obtained using the bilinear inverse metric,
\begin{equation}
\widetilde{\mathcal{G}}^\mu{}_\nu
=
h^{\mu\rho}
\widetilde{\mathcal{G}}_{\rho\nu}.
\end{equation}
The field equations therefore imply
\begin{equation}
\boxed{
\widetilde{\nabla}_\mu
\widetilde{T}^\mu{}_\nu
=0,
}
\end{equation}
where
\begin{equation}
\widetilde{T}^\mu{}_\nu
=
h^{\mu\rho}\widetilde{T}_{\rho\nu}.
\end{equation}
This is precisely the complex representation of the conventional
local conservation law
\begin{equation}
\nabla_\mu T^\mu{}_\nu=0.
\end{equation}

The same mixed tensors may again be constructed using the Hermitian
metric together with the physical reality structure. In particular,
\begin{equation}
\widetilde{T}^\mu{}_\nu
=
G^{\mu\bar\alpha}
C^\rho{}_{\bar\alpha}
\widetilde{T}_{\rho\nu},
\end{equation}
and similarly
\begin{equation}
\widetilde{\mathcal{G}}^\mu{}_\nu
=
G^{\mu\bar\alpha}
C^\rho{}_{\bar\alpha}
\widetilde{\mathcal{G}}_{\rho\nu}.
\end{equation}
Thus the conservation equations can be expressed using either metric
representation, provided the barred and unbarred index spaces are
converted consistently.

The transformation of the stress-energy tensor is especially useful for
interpreting the complex components physically. Although quantities such
as $\widetilde{T}_{0i}$ may be imaginary, the corresponding conventional
energy and momentum fluxes remain real. The physical tensor is recovered
invertibly through
\begin{equation}
T_{\alpha\beta}
=
S^\mu{}_\alpha
S^\nu{}_\beta
\widetilde{T}_{\mu\nu}.
\end{equation}
Complex components in the transformed representation therefore do not
indicate complex-valued physical energy density, pressure, or momentum
flux. They arise from expressing a real stress-energy tensor in the
complex spacetime basis.

The results of this section complete the extension of the imaginary-time
representation to the local geometry and dynamics of general relativity.
The metric, Levi-Civita connection, geodesics, curvature tensors,
Einstein equations, and covariant conservation laws retain their
conventional physical content under the invertible complex change of
representation. The complex-bilinear metric $h_{\mu\nu}$ is
algebraically adapted to ordinary tensor calculus, while the Hermitian
metric $G_{\bar\mu\nu}$, together with the physical reality structure
$C$, provides an equivalent description with a distinct complex index
structure.

The construction therefore represents Lorentzian general relativity
rather than a modified theory of gravity. We now test its consistency
in two geometrically distinct exact spacetimes: the stationary,
rotating Kerr solution and the time-dependent
Friedmann--Lema\^{i}tre--Robertson--Walker cosmology.

% ============================================================
\section{Exact Spacetime Tests}
\label{sec:tests}
% ============================================================

\subsection{Kerr spacetime}
\label{sec:kerr}

The Kerr spacetime~\cite{Kerr1963} provides a stringent test of the complex representation 
developed in the preceding section. Unlike a static diagonal geometry, the
Kerr metric contains an off-diagonal temporal--azimuthal component associated
with frame dragging.  The transformation of this component therefore tests
directly the complex symmetric and Hermitian metric structures introduced in
Sec.~\ref{sec:geometry}, while the vacuum character of the solution provides a corresponding
test of the transformed curvature tensors and Einstein equations.

We use Boyer--Lindquist coordinates~\cite{BoyerLindquist1967},
\begin{equation}
x^\mu=(x^0,r,\theta,\phi)=(ct,r,\theta,\phi),
\end{equation}
and define
\begin{equation}
r_g=\frac{G_N M}{c^2},
\qquad
a=\frac{J}{Mc},
\end{equation}
together with
\begin{equation}
\Sigma_{\rm K}=r^2+a^2\cos^2\theta,
\qquad
\Delta=r^2-2 r_g r+a^2.
\end{equation}
The subscript on $\Sigma_{\rm K}$ distinguishes the Kerr metric function
from the imaginary proper-time parameter $\Sigma=ic\tau$ introduced in
Sec.~\ref{sec:imaginary_tau}. Here $M$ and $J$ are the mass and angular momentum of the source,
respectively, and $r_g$ is the corresponding gravitational length parameter.  With signature $(-,+,+,+)$, the Kerr line element is
\begin{align}
ds^2={}&
-\left(1-\frac{2 r_g r}{\Sigma_{\rm K}}\right)(dx^0)^2
-\frac{4 r_g ar\sin^2\theta}{\Sigma_{\rm K}}\,dx^0d\phi
\nonumber\\
&+\frac{\Sigma_{\rm K}}{\Delta}\,dr^2
+\Sigma_{\rm K}\,d\theta^2
\nonumber\\
&+\left(
r^2+a^2+
\frac{2 r_g a^2r\sin^2\theta}{\Sigma_{\rm K}}
\right)
\sin^2\theta\,d\phi^2.
\end{align}
In particular, the off-diagonal component is
\begin{equation}
g_{0\phi}=g_{\phi 0}
=
-\frac{2 r_g ar\sin^2\theta}{\Sigma_{\rm K}}.
\end{equation}

Applying the complex spacetime representation
\begin{equation}
X^0=ix^0=ict,
\qquad
X^1=r,\qquad
X^2=\theta,\qquad
X^3=\phi,
\end{equation}
the complex-bilinear metric is
\begin{equation}
h=(S^{-1})^TgS^{-1}.
\end{equation}
As established in Sec.~\ref{sec:geometry}, its components satisfy
\begin{equation}
h_{00}=-g_{00},
\qquad
h_{0i}=h_{i0}=-ig_{0i},
\qquad
h_{ij}=g_{ij}.
\end{equation}
For Kerr this gives
\begin{align}
ds^2={}&
\left(1-\frac{2 r_g r}{\Sigma_{\rm K}}\right)(dX^0)^2
+\frac{4i r_g ar\sin^2\theta}{\Sigma_{\rm K}}\,dX^0d\phi
\nonumber\\
&+\frac{\Sigma_{\rm K}}{\Delta}\,dr^2
+\Sigma_{\rm K}\,d\theta^2
\nonumber\\
&+\left(
r^2+a^2+
\frac{2 r_g a^2r\sin^2\theta}{\Sigma_{\rm K}}
\right)
\sin^2\theta\,d\phi^2.
\end{align}
The temporal--azimuthal component has therefore become
\begin{equation}
h_{0\phi}=h_{\phi0}
=
\frac{2i r_g ar\sin^2\theta}{\Sigma_{\rm K}},
\end{equation}
which is purely imaginary while preserving the complex symmetry
$h_{\mu\nu}=h_{\nu\mu}$.

The physical Lorentzian metric is recovered immediately by restricting this
expression to the physical slice.  Since
\begin{equation}
dX^0=i\,dx^0,
\end{equation}
the temporal term becomes
\begin{equation}
\left(1-\frac{2 r_g r}{\Sigma_{\rm K}}\right)(dX^0)^2
=
-\left(1-\frac{2 r_g r}{\Sigma_{\rm K}}\right)(dx^0)^2,
\end{equation}
while the mixed term gives
\begin{equation}
\frac{4i r_g ar\sin^2\theta}{\Sigma_{\rm K}}\,dX^0d\phi
=
-\frac{4 r_g ar\sin^2\theta}{\Sigma_{\rm K}}\,dx^0d\phi.
\end{equation}
Thus both the temporal sign and the frame-dragging cross term of the
conventional Kerr metric are recovered exactly.  The imaginary
temporal--azimuthal component of $h_{\mu\nu}$ therefore reflects the complex
representation of the Lorentzian geometry rather than a complex-valued
physical metric.

The corresponding Hermitian representation is
\begin{equation}
G=(S^\dagger)^{-1}gS^{-1}.
\end{equation}
Its nontrivial temporal components are
\begin{equation}
G_{\bar 0 0}
=
-\left(1-\frac{2 r_g r}{\Sigma_{\rm K}}\right),
\end{equation}
and
\begin{equation}
G_{\bar 0\phi}
=
-\frac{2i r_g ar\sin^2\theta}{\Sigma_{\rm K}},
\qquad
G_{\bar\phi 0}
=
\frac{2i r_g ar\sin^2\theta}{\Sigma_{\rm K}},
\end{equation}
while the purely spatial components agree with those of the conventional Kerr
metric.  Hence
\begin{equation}
G_{\bar\mu\nu}=G_{\bar\nu\mu}^{\,*},
\end{equation}
as required for the Hermitian metric.

The Kerr geometry therefore displays explicitly the distinction between the
two metric structures developed in Sec.~\ref{sec:gr}.  The bilinear metric is complex
symmetric, with
\begin{equation}
h_{0\phi}=h_{\phi0},
\end{equation}
whereas the Hermitian metric satisfies
\begin{equation}
G_{\bar 0\phi}=G_{\bar\phi0}^{\,*}.
\end{equation}
Together with the Lorentzian reality structure
\begin{equation}
C=\operatorname{diag}(-1,1,1,1),
\end{equation}
they obey
\begin{equation}
G=C^Th,
\end{equation}
and therefore encode the same physical Kerr metric.

The nontrivial off-diagonal structure of the Kerr metric also provides a
useful test of the transformed Levi-Civita connection.  As shown in
Sec.~\ref{sec:levi-civita}, the connection coefficients in the complex representation satisfy
\begin{equation}
\widetilde{\Gamma}^{\rho}{}_{\mu\nu}
=
i^{\delta_{\rho 0}}
(-i)^{\delta_{\mu 0}+\delta_{\nu 0}}
\Gamma^{\rho}{}_{\mu\nu}.
\end{equation}
For the Kerr metric, the nonvanishing connection coefficients realize all
four phase factors permitted by this relation.  Representative components
are
\begin{align}
\widetilde{\Gamma}^{r}{}_{rr}
&=
\Gamma^{r}{}_{rr},
\\
\widetilde{\Gamma}^{r}{}_{00}
&=
-\Gamma^{r}{}_{00},
\\
\widetilde{\Gamma}^{r}{}_{0\phi}
&=
-i\Gamma^{r}{}_{0\phi},
\\
\widetilde{\Gamma}^{0}{}_{r\phi}
&=
+i\Gamma^{0}{}_{r\phi}.
\end{align}
Thus the factors of $i$ appearing in the complex Kerr metric propagate
consistently into the connection according to the tensor-index structure
derived in Sec.~\ref{sec:levi-civita}.  A direct calculation from $h_{\mu\nu}$ reproduces
these transformed coefficients, while reconstruction from the Hermitian
metric $G_{\bar\mu\nu}$ together with the reality structure $C$ gives the
same result.  The complete set of nonvanishing Kerr connection coefficients
and their transformation factors is summarized in Appendix~\ref{app:kerr}.

The curvature provides a still stronger test because it depends both on
derivatives of the connection and on products of connection coefficients.
For the fully covariant Riemann tensor, the general transformation law may
be written
\begin{equation}
\widetilde{R}_{\rho\sigma\mu\nu}
=
(S^{-1})^{\alpha}{}_{\rho}
(S^{-1})^{\beta}{}_{\sigma}
(S^{-1})^{\gamma}{}_{\mu}
(S^{-1})^{\delta}{}_{\nu}
R_{\alpha\beta\gamma\delta}.
\end{equation}
For $S=\operatorname{diag}(i,1,1,1)$, each temporal covariant index
therefore contributes a factor of $-i$.  The antisymmetry of the Riemann
tensor within each index pair implies that a nonvanishing component can
contain at most two temporal indices.  The curvature components consequently
fall into the three classes
\begin{align}
\widetilde{R}_{0i0j}
&=
-R_{0i0j},
\\
\widetilde{R}_{0ijk}
&=
-iR_{0ijk},
\\
\widetilde{R}_{ijkl}
&=
R_{ijkl},
\end{align}
where $i,j,k,l$ denote spatial indices.  These relations are satisfied by
the Kerr curvature obtained directly from the transformed connection.  The
individual component calculations are given in Appendix~\ref{app:kerr}.

Although the transformed Riemann tensor contains both real and imaginary
components, its contractions retain the invariant content of the
conventional Kerr geometry.  In particular, the Ricci tensor transforms as
\begin{equation}
\widetilde{R}_{\mu\nu}
=
(S^{-1})^{\alpha}{}_{\mu}
(S^{-1})^{\beta}{}_{\nu}
R_{\alpha\beta}.
\end{equation}
Since the Kerr spacetime is a vacuum solution,
\begin{equation}
R_{\mu\nu}=0,
\end{equation}
and hence
\begin{equation}
\widetilde{R}_{\mu\nu}=0.
\end{equation}
The Ricci scalar therefore also vanishes,
\begin{equation}
\widetilde{R}=R=0,
\end{equation}
so that the transformed Kerr metric satisfies the vacuum Einstein equations
in the complex representation.

Ricci flatness does not, of course, imply that the spacetime is flat.  A
useful invariant test is provided by the Kretschmann scalar,
\begin{equation}
K
=
R_{\alpha\beta\gamma\delta}
R^{\alpha\beta\gamma\delta}.
\end{equation}
In the complex representation we similarly define
\begin{equation}
\widetilde{K}
=
\widetilde{R}_{\alpha\beta\gamma\delta}
\widetilde{R}^{\alpha\beta\gamma\delta}.
\end{equation}
Each factor of $S^{-1}$ carried by a covariant Riemann index is canceled by
the corresponding factor of $S$ carried by the raised tensor.  Consequently,
\begin{equation}
\widetilde{K}=K.
\end{equation}
For Kerr,
\begin{equation}
K
=
\frac{
48 r_g^2
\left(
r^6
-15a^2r^4\cos^2\theta
+15a^4r^2\cos^4\theta
-a^6\cos^6\theta
\right)
}{
\Sigma_{\rm K}^6
}.
\end{equation}
The complex representation therefore preserves not only the vacuum
condition but also the generally nonzero Kretschmann scalar of the Kerr
geometry.  In summary,
\begin{equation}
\boxed{
\widetilde{R}_{\mu\nu}=0,
\qquad
\widetilde{R}=0,
\qquad
\widetilde{K}=K.
}
\end{equation}

The stationary and axisymmetric symmetries of the Kerr spacetime provide a
corresponding test of the conserved quantities developed in
Sec.~\ref{sec:geodesics}. Consider a freely falling test particle of rest
mass $m$, with conventional four-momentum $p^\alpha$. Here $m$ denotes the
particle rest mass used in Sec.~\ref{sec:four-velocity}, while $r_g$ is the
gravitational length parameter of the black hole.
In the conventional Boyer--Lindquist representation, the stationary
and axial Killing vectors are
\begin{equation}
\xi_{(t)}^\mu=(1,0,0,0),
\qquad
\xi_{(\phi)}^\mu=(0,0,0,1),
\end{equation}
respectively.  With $x^0=ct$, their contractions with the canonical
four-momentum give
\begin{equation}
p_\mu\xi_{(t)}^\mu=p_0=-\frac{E}{c},
\qquad
p_\mu\xi_{(\phi)}^\mu=p_\phi=L_z.
\end{equation}

Under the complex transformation, the Killing vectors become
\begin{equation}
\Xi_{(t)}^\mu
=
S^\mu{}_\nu\xi_{(t)}^\nu
=
(i,0,0,0),
\qquad
\Xi_{(\phi)}^\mu
=
S^\mu{}_\nu\xi_{(\phi)}^\nu
=
(0,0,0,1).
\end{equation}
The canonical momentum covector introduced in Sec.~\ref{sec:hamilton_jacobi} transforms according
to
\begin{equation}
P_\mu
=
(S^{-1})^\nu{}_\mu p_\nu.
\end{equation}
Consequently, its contraction with a transformed Killing vector is exactly
invariant:
\begin{align}
P_\mu\Xi^\mu
&=
(S^{-1})^\alpha{}_\mu p_\alpha
S^\mu{}_\beta\xi^\beta
\nonumber\\
&=
p_\alpha\xi^\alpha.
\end{align}
The physical Kerr constants of motion therefore retain their conventional
values directly in the canonical formulation:
\begin{equation}
\boxed{
E=-cP_\mu\Xi_{(t)}^\mu,
\qquad
L_z=P_\mu\Xi_{(\phi)}^\mu.
}
\end{equation}
Here $E$ denotes the conserved Killing energy associated with the
stationary symmetry of Kerr spacetime, rather than the energy measured
locally by an arbitrary observer. In particular, the stationary
Killing vector becomes spacelike within the ergoregion, although its
associated conserved quantity remains well defined.

This exact preservation should be distinguished from the corresponding
contraction involving the complementary momentum. As derived in
Sec.~\ref{sec:geodesics},
\begin{equation}
\Pi^\mu=-iS^\mu{}_\nu p^\nu,
\end{equation}
and hence
\begin{equation}
\Xi_\mu\Pi^\mu
=
-i\,\xi_\mu p^\mu.
\end{equation}
For the stationary and axial Killing symmetries this gives
\begin{equation}
\Xi^{(t)}_\mu\Pi^\mu
=
i\frac{E}{c},
\qquad
\Xi^{(\phi)}_\mu\Pi^\mu
=
-iL_z.
\end{equation}
Equivalently,
\begin{equation}
\boxed{
E=-ic\,\Xi^{(t)}_\mu\Pi^\mu,
\qquad
L_z=i\,\Xi^{(\phi)}_\mu\Pi^\mu.
}
\end{equation}
Thus the canonical momentum covector preserves the numerical values of the
usual Killing contractions, whereas the complementary momentum carries the
additional universal phase associated with differentiation with respect to
the imaginary proper-time parameter $\Sigma=ic\tau$.  Both representations
encode the same conserved energy and axial angular momentum.

Kerr spacetime possesses an additional conserved quantity that does
not arise from a Killing vector~\cite{Carter1968}. Its complete
integrability is associated with a symmetric rank-two Killing tensor
$K_{\mu\nu}$ satisfying
\begin{equation}
\nabla_{(\rho}K_{\mu\nu)}=0.
\end{equation}
The corresponding quadratic constant of motion may be written
\begin{equation}
\mathcal{Q}
=
K^{\mu\nu}p_\mu p_\nu
=
K_{\mu\nu}p^\mu p^\nu.
\end{equation}
Here $\mathcal{Q}$ denotes the invariant associated with the Kerr Killing
tensor; conventions for the quantity commonly called the Carter constant
may differ by combinations of the other conserved quantities.

In the complex representation, define the transformed Killing tensor by
\begin{equation}
\widetilde{K}^{\mu\nu}
=
S^\mu{}_\alpha
S^\nu{}_\beta
K^{\alpha\beta},
\end{equation}
or, equivalently, in covariant form,
\begin{equation}
\widetilde{K}_{\mu\nu}
=
(S^{-1})^\alpha{}_\mu
(S^{-1})^\beta{}_\nu
K_{\alpha\beta}.
\end{equation}
Because $S$ is constant and the Levi-Civita connection transforms as derived
in Sec.~\ref{sec:levi-civita}, the Killing-tensor equation is preserved:
\begin{equation}
\widetilde{\nabla}_{(\rho}
\widetilde{K}_{\mu\nu)}
=
0.
\end{equation}

The quadratic invariant formed with the canonical momentum covector is then
\begin{align}
\widetilde{K}^{\mu\nu}P_\mu P_\nu
&=
S^\mu{}_\alpha S^\nu{}_\beta K^{\alpha\beta}
(S^{-1})^\gamma{}_\mu p_\gamma
(S^{-1})^\delta{}_\nu p_\delta
\nonumber\\
&=
K^{\alpha\beta}p_\alpha p_\beta
=
\mathcal{Q}.
\end{align}
Thus, just as for the energy and axial angular momentum,
\begin{equation}
\boxed{
\widetilde{K}^{\mu\nu}P_\mu P_\nu=\mathcal{Q},
}
\end{equation}
so the canonical complex representation preserves the quadratic Killing
invariant exactly.

The complementary-momentum representation again differs by a phase.  Using
\begin{equation}
\Pi^\mu=-iS^\mu{}_\alpha p^\alpha,
\end{equation}
one obtains
\begin{align}
\widetilde{K}_{\mu\nu}\Pi^\mu\Pi^\nu
&=
(S^{-1})^\alpha{}_\mu
(S^{-1})^\beta{}_\nu
K_{\alpha\beta}
(-iS^\mu{}_\gamma p^\gamma)
(-iS^\nu{}_\delta p^\delta)
\nonumber\\
&=
-\,
K_{\alpha\beta}p^\alpha p^\beta
=
-\mathcal{Q}.
\end{align}
Therefore
\begin{equation}
\boxed{
\widetilde{K}_{\mu\nu}\Pi^\mu\Pi^\nu=-\mathcal{Q}.
}
\end{equation}
The additional minus sign is the quadratic counterpart of the factor $-i$
appearing for a Killing-vector contraction: each complementary momentum
contributes one factor of $-i$.

The preservation of the Killing tensor also implies that the separability
structure of Kerr geodesic motion is unchanged. The complex
Hamilton--Jacobi equation for a test particle of rest mass $m$,
\begin{equation}
h^{\mu\nu}
\frac{\partial \mathcal{S}}{\partial X^\mu}
\frac{\partial \mathcal{S}}{\partial X^\nu}
+
m^2c^2
=
0,
\end{equation}
is the transformed form of the conventional Kerr Hamilton--Jacobi equation.
Its constants associated with stationarity, axial symmetry, and the
rank-two Killing tensor are therefore the same $E$, $L_z$, and
$\mathcal{Q}$ that characterize the conventional geodesic problem.  The
complex representation consequently preserves not only the metric and
curvature of Kerr spacetime, but also the hidden symmetry responsible for
the integrability of its geodesic motion.

\subsection{FLRW Spacetime}
\label{sec:flrw}

The Friedmann--Lema\^{i}tre--Robertson--Walker (FLRW) spacetime
provides a complementary test of the complex representation.
Whereas the Kerr metric provides a stationary vacuum solution with an off-diagonal
temporal--azimuthal component, the FLRW metric is diagonal but explicitly
time dependent, with nonzero Ricci curvature in the presence of matter.
It therefore provides a useful test of the transformation of temporal
derivatives, the Einstein equations, and covariant conservation.

Using coordinates
\begin{equation}
x^\mu=(x^0,r,\theta,\phi)=(ct,r,\theta,\phi),
\end{equation}
the FLRW line element may be written as
\begin{equation}
ds^2
=
-(dx^0)^2
+
a^2(t)
\left[
\frac{dr^2}{1-kr^2}
+r^2d\theta^2
+r^2\sin^2\theta\,d\phi^2
\right],
\end{equation}
where $a(t)$ is the cosmological scale factor and $k$ is the spatial
curvature parameter.  Introducing the complex temporal coordinate
\begin{equation}
X^0=ix^0=ict,
\qquad
X^i=x^i,
\end{equation}
gives
\begin{equation}
t=-\frac{iX^0}{c}.
\end{equation}
Thus, in the complex representation, the scale factor is understood as
\begin{equation}
a(t)=a\left(-\frac{iX^0}{c}\right),
\end{equation}
where this expression is evaluated on the physical Lorentzian slice
$X^0=ict$. As in Sec.~\ref{sec:complex_coordinates}, it represents
the original real time dependence in complex coordinates and does not
require a holomorphic extension of $a(t)$ away from the physical slice.
A dot will continue to denote differentiation with respect to the
physical cosmic time $t$. In particular,
\begin{equation}
\frac{\partial}{\partial X^0}
=
-\frac{i}{c}\frac{\partial}{\partial t}.
\end{equation}

Applying the transformation
\begin{equation}
h=(S^{-1})^TgS^{-1},
\qquad
S=\operatorname{diag}(i,1,1,1),
\end{equation}
the complex-bilinear metric takes the form
\begin{equation}
h_{\mu\nu}
=
\operatorname{diag}
\left(
1,\,
\frac{a^2}{1-kr^2},\,
a^2r^2,\,
a^2r^2\sin^2\theta
\right).
\end{equation}
The corresponding line element is therefore
\begin{equation}
ds^2
=
(dX^0)^2
+
a^2(t)
\left[
\frac{dr^2}{1-kr^2}
+r^2d\theta^2
+r^2\sin^2\theta\,d\phi^2
\right].
\end{equation}
Although this expression is diagonal and contains no explicitly
imaginary metric components, it does not define a Euclidean spacetime
on the physical slice.  Since
\begin{equation}
dX^0=ic\,dt,
\end{equation}
the temporal contribution satisfies
\begin{equation}
(dX^0)^2=-c^2dt^2,
\end{equation}
and restriction to the physical slice immediately recovers the
Lorentzian FLRW line element.

The associated Hermitian metric,
\begin{equation}
G=(S^\dagger)^{-1}gS^{-1},
\end{equation}
is
\begin{equation}
G_{\bar\mu\nu}
=
\operatorname{diag}
\left(
-1,\,
\frac{a^2}{1-kr^2},\,
a^2r^2,\,
a^2r^2\sin^2\theta
\right).
\end{equation}
Thus the component matrix of $G$ has the same form as the conventional
Lorentzian FLRW metric, while the bilinear form $h$ has a positive
temporal component.  As in the general construction of
Sec.~\ref{sec:gr}, these two forms encode the same physical
geometry on the physical slice.  FLRW therefore illustrates particularly
clearly that the Lorentzian character of the complex representation need
not appear as a negative or explicitly imaginary component of
$h_{\mu\nu}$; it is encoded jointly in the bilinear form and in the
reality condition $X^0\in i\mathbb{R}$.

The explicit dependence of the scale factor on the complex temporal
coordinate provides a direct test of the transformation of the
connection.  Since
\begin{equation}
\partial_0 a
\equiv
\frac{\partial a}{\partial X^0}
=
-\frac{i}{c}\dot a,
\end{equation}
the time-dependent components of the Levi-Civita connection are
\begin{equation}
\widetilde{\Gamma}^{0}{}_{ij}
=
\frac{i a\dot a}{c}\gamma_{ij},
\qquad
\widetilde{\Gamma}^{i}{}_{0j}
=
-\frac{i}{c}\frac{\dot a}{a}\delta^i{}_j,
\end{equation}
where
\begin{equation}
\gamma_{ij}
=
\operatorname{diag}
\left(
\frac{1}{1-kr^2},
r^2,
r^2\sin^2\theta
\right)
\end{equation}
denotes the metric of the constant-curvature spatial slices, so that
$h_{ij}=a^2\gamma_{ij}$.  In terms of the corresponding connection
coefficients in the conventional representation,
\begin{equation}
\widetilde{\Gamma}^{0}{}_{ij}
=
i\Gamma^{0}{}_{ij},
\qquad
\widetilde{\Gamma}^{i}{}_{0j}
=
-i\Gamma^{i}{}_{0j}.
\end{equation}
The purely spatial connection coefficients are unchanged,
\begin{equation}
\widetilde{\Gamma}^{i}{}_{jk}
=
\Gamma^{i}{}_{jk},
\end{equation}
since the spatial derivatives do not act on $a(t)$.  These relations
agree directly with the general transformation law
\begin{equation}
\widetilde{\Gamma}^{\rho}{}_{\mu\nu}
=
S^\rho{}_\alpha
(S^{-1})^\beta{}_\mu
(S^{-1})^\gamma{}_\nu
\Gamma^\alpha{}_{\beta\gamma}
\end{equation}
derived in Sec.~\ref{sec:gr}.

The resulting Ricci tensor is
\begin{equation}
\widetilde R_{00}
=
\frac{3\ddot a}{ac^2},
\qquad
\widetilde R_{0i}=0,
\end{equation}
and
\begin{equation}
\widetilde R_{ij}
=
\left[
\frac{a\ddot a+2\dot a^2}{c^2}
+2k
\right]\gamma_{ij}.
\end{equation}
These components obey
\begin{equation}
\widetilde R_{00}=-R_{00},
\qquad
\widetilde R_{0i}=-iR_{0i}=0,
\qquad
\widetilde R_{ij}=R_{ij},
\end{equation}
as required by the covariant tensor transformation law.  In particular,
the factors of $i$ introduced by temporal differentiation and by the
connection coefficients combine so that the transformed Ricci tensor is
again real.

Contracting with the inverse bilinear metric gives
\begin{equation}
\widetilde R
=
h^{\mu\nu}\widetilde R_{\mu\nu}
=
\frac{6}{c^2}
\left(
\frac{\ddot a}{a}
+\frac{\dot a^2}{a^2}
+\frac{kc^2}{a^2}
\right).
\end{equation}
Thus
\begin{equation}
\widetilde R=R,
\end{equation}
in agreement with the scalar invariance established in
Sec.~\ref{sec:gr}.  Unlike the Kerr vacuum solution, for which the Ricci
scalar vanishes, the FLRW spacetime therefore provides a nontrivial
check that the scalar curvature is preserved by the complex
representation.

We next consider the Einstein equations in the presence of a homogeneous
and isotropic perfect fluid, taking the cosmological constant to vanish
($\Lambda=0$). The extension to nonzero $\Lambda$ follows directly from
the general transformation established in Sec.~\ref{sec:einstein}.
Let $\varepsilon$ denote the energy density measured by a comoving
observer and $p$ the isotropic pressure. In the conventional FLRW
representation, the nonzero covariant components of the stress-energy
tensor are
\begin{equation}
T_{00}=\varepsilon,
\qquad
T_{ij}=p\,g_{ij}.
\end{equation}
Under the transformation to the complex representation,
\begin{equation}
\widetilde T_{\mu\nu}
=
(S^{-1})^\alpha{}_\mu
(S^{-1})^\beta{}_\nu
T_{\alpha\beta},
\end{equation}
these become
\begin{equation}
\widetilde T_{00}=-\varepsilon,
\qquad
\widetilde T_{0i}=0,
\qquad
\widetilde T_{ij}=p\,h_{ij}.
\end{equation}
The negative sign in the temporal component is a consequence of the
complex basis transformation and does not represent a negative physical
energy density.

Using
\begin{equation}
\widetilde{\mathcal G}_{\mu\nu}
=
\widetilde R_{\mu\nu}
-\frac{1}{2}h_{\mu\nu}\widetilde R,
\end{equation}
the temporal component of the Einstein tensor is
\begin{equation}
\widetilde{\mathcal G}_{00}
=
-\frac{3}{c^2}
\left(
H^2+\frac{kc^2}{a^2}
\right),
\qquad
H\equiv\frac{\dot a}{a},
\end{equation}
while the spatial components are
\begin{equation}
\widetilde{\mathcal G}_{ij}
=
-\frac{1}{c^2}
\left(
2\frac{\ddot a}{a}
+H^2
+\frac{kc^2}{a^2}
\right)h_{ij}.
\end{equation}
Thus
\begin{equation}
\widetilde{\mathcal G}_{00}
=
-\mathcal G_{00},
\qquad
\widetilde{\mathcal G}_{ij}
=
\mathcal G_{ij},
\end{equation}
so that the Einstein tensor and stress-energy tensor acquire the same
transformation factors.

The transformed Einstein equations,
\begin{equation}
\widetilde{\mathcal G}_{\mu\nu}
=
\frac{8\pi G_N}{c^4}\widetilde T_{\mu\nu},
\end{equation}
therefore give, from the temporal component,
\begin{equation}
-\frac{3}{c^2}
\left(
H^2+\frac{kc^2}{a^2}
\right)
=
-\frac{8\pi G_N}{c^4}\varepsilon.
\end{equation}
The common minus sign cancels, yielding
\begin{equation}
\boxed{
H^2+\frac{kc^2}{a^2}
=
\frac{8\pi G_N}{3c^2}\varepsilon
},
\end{equation}
which is the standard first Friedmann equation.

The spatial components similarly give
\begin{equation}
2\frac{\ddot a}{a}
+H^2
+\frac{kc^2}{a^2}
=
-\frac{8\pi G_N}{c^2}p.
\end{equation}
Combining this result with the first Friedmann equation gives
\begin{equation}
\boxed{
\frac{\ddot a}{a}
=
-\frac{4\pi G_N}{3c^2}
\left(
\varepsilon+3p
\right)
},
\end{equation}
the standard acceleration equation.  If the energy density is written
as $\varepsilon=\rho c^2$, these equations take the equivalent forms
\begin{equation}
H^2+\frac{kc^2}{a^2}
=
\frac{8\pi G_N}{3}\rho,
\end{equation}
and
\begin{equation}
\frac{\ddot a}{a}
=
-\frac{4\pi G_N}{3}
\left(
\rho+\frac{3p}{c^2}
\right).
\end{equation}

The transformed field equations therefore reproduce the conventional
FLRW dynamics without modification.  In particular, the sign change in
the temporal component of the Einstein tensor is accompanied by the
same sign change in the temporal component of the stress-energy tensor,
so that the relation between spacetime curvature and physical matter
content is preserved exactly.

Finally, the covariant conservation law provides an independent test of
the transformed matter sector.  Raising the first index with the inverse
bilinear metric gives
\begin{equation}
\widetilde T^\mu{}_\nu
=
\operatorname{diag}
\left(
-\varepsilon,p,p,p
\right),
\end{equation}
which has the same component form as the conventional mixed
stress-energy tensor.  For the temporal component, the transformed
conservation equation is
\begin{equation}
\widetilde\nabla_\mu \widetilde T^\mu{}_0
=
\partial_\mu\widetilde T^\mu{}_0
+
\widetilde\Gamma^\mu{}_{\mu\lambda}
\widetilde T^\lambda{}_0
-
\widetilde\Gamma^\lambda{}_{\mu0}
\widetilde T^\mu{}_\lambda.
\end{equation}
Using
\begin{equation}
\partial_0=-\frac{i}{c}\frac{\partial}{\partial t},
\qquad
\widetilde\Gamma^i{}_{i0}=-\frac{iH}{c},
\end{equation}
gives
\begin{equation}
\widetilde\nabla_\mu \widetilde T^\mu{}_0
=
\frac{i}{c}
\left[
\dot\varepsilon
+
3H(\varepsilon+p)
\right].
\end{equation}
Thus the condition
\begin{equation}
\widetilde\nabla_\mu \widetilde T^\mu{}_0=0
\end{equation}
is equivalent to
\begin{equation}
\boxed{
\dot\varepsilon
+
3H(\varepsilon+p)
=
0
},
\end{equation}
which is the standard FLRW continuity equation.  In terms of the mass
density $\rho=\varepsilon/c^2$, this becomes
\begin{equation}
\dot\rho
+
3H
\left(
\rho+\frac{p}{c^2}
\right)
=
0.
\end{equation}
The overall factor of $i/c$ in the transformed temporal conservation
equation reflects the transformation of its remaining covariant temporal
index and has no effect on the physical conservation law.

The Kerr and FLRW solutions therefore provide complementary tests of the
complex representation in curved spacetime.  Kerr tests a stationary,
rotating vacuum geometry with an off-diagonal temporal--azimuthal metric
component, for which the bilinear and Hermitian representations acquire
explicitly imaginary components.  FLRW instead tests a diagonal but
explicitly time-dependent geometry with nonzero Ricci curvature, a
perfect-fluid stress-energy tensor, and nontrivial covariant
conservation.  Although its bilinear metric is real and superficially
positive in all diagonal components, the Lorentzian character of the
physical spacetime is retained through the reality condition
$X^0\in i\mathbb{R}$.

Together, these examples demonstrate that the complex representation
preserves the metric geometry, connection, curvature, Einstein
equations, and conservation laws across two qualitatively different
exact solutions of general relativity.  The factors of $i$ introduced
by temporal components and derivatives reorganize the component
expressions but cancel or combine according to the appropriate tensor
transformation laws, leaving the underlying physical predictions
unchanged.

\section{Discussion}

\subsection{Relation to Wick Rotation and Euclidean Continuation}
\label{sec:wick_rotation}

The appearance of an imaginary temporal coordinate naturally invites
comparison with Wick rotation.  The two constructions are related, but
they should not be identified.  Throughout the preceding sections, the
replacement
\begin{equation}
x^0=ct
\qquad\longrightarrow\qquad
X^0=ict
\end{equation}
has represented a change in the complex representation of the same
Lorentzian spacetime.  The physical reality condition remains
\begin{equation}
X^0\in i\mathbb{R},
\qquad
X^i\in\mathbb{R},
\end{equation}
and the invariant interval, causal structure, equations of motion, and
curved-spacetime geometry remain those of the original Lorentzian
theory.  Multiplication of the temporal coordinate by $i$ therefore
does not, by itself, constitute a Wick rotation.

This distinction is especially transparent in flat spacetime.  The
complex-bilinear metric is
\begin{equation}
h_{\mu\nu}=\delta_{\mu\nu},
\end{equation}
so that
\begin{equation}
ds^2
=
(dX^0)^2+d\mathbf X^2.
\end{equation}
Despite its Euclidean-looking component matrix, restriction to the
physical slice gives
\begin{equation}
dX^0=ic\,dt
\end{equation}
and hence
\begin{equation}
ds^2=-c^2dt^2+d\mathbf x^2.
\end{equation}
The Lorentzian signature is therefore not determined by the signs of
the component matrix of $h$ alone.  It is encoded jointly in the
complex-bilinear form and in the real structure that selects the
physical slice.

The Rindler construction of Sec.~\ref{sec:rindler} makes the relation
to Euclidean continuation more explicit.  In the Lorentzian
representation, the Rindler coordinates satisfy
\begin{equation}
X_L^0=i\rho\sinh\eta,
\qquad
X_L^1=\rho\cosh\eta,
\end{equation}
with
\begin{equation}
ds_L^2=d\rho^2-\rho^2d\eta^2.
\end{equation}
Under the continuation
\begin{equation}
\eta=i\theta,
\end{equation}
these become
\begin{equation}
X_E^0=-\rho\sin\theta,
\qquad
X_E^1=\rho\cos\theta,
\end{equation}
and therefore
\begin{equation}
ds_E^2=d\rho^2+\rho^2d\theta^2.
\end{equation}
In this case the ambient bilinear form remains
\begin{equation}
h=I
\end{equation}
on both sides of the continuation. To extend the Euclidean metric
smoothly through $\rho=0$, the angular coordinate must be identified
with period $2\pi$. What changes is the corresponding reality
condition: the Lorentzian slice has $X^0\in i\mathbb{R}$,
whereas the Euclidean slice has $X^0\in\mathbb{R}$. Equivalently, the
two geometries correspond to different real structures within the same
complexified coordinate space.

This provides a useful way to separate several operations that are
often grouped together under the language of imaginary time.  The map
\begin{equation}
x^\mu\longmapsto X^\mu=S^\mu{}_\nu x^\nu
\end{equation}
is an invertible change of representation of a Lorentzian geometry.
Complexification enlarges the space in which this representation is
defined.  A reality condition then selects a real slice of that complex
space.  Euclidean continuation, when it exists, may relate the
Lorentzian real structure to a different real structure associated with
a Euclidean geometry.  These operations are conceptually distinct even
when, in simple examples, they can all be expressed through factors of
$i$.

The distinction becomes important in general curved spacetimes.
Rindler spacetime provides an especially simple example because the
same bilinear form $h=I$ is compatible with both the Lorentzian and
Euclidean reality conditions.  This need not remain true when the
Lorentzian metric contains temporal--spatial cross terms.  For example,
the Kerr metric examined in Sec.~\ref{sec:kerr} contains a nonzero
$g_{0\phi}$ component, which becomes an explicitly imaginary
$h_{0\phi}$ component in the complex representation.  Simply declaring
$X^0$ to be real while holding this bilinear form fixed would therefore
leave an imaginary temporal--azimuthal cross term and would not, by
itself, define a real Euclidean metric.  In such cases a Euclidean
continuation must include a compatible continuation of the metric data,
rather than only a change in the reality condition.

The familiar substitution associated with Wick rotation may therefore
be viewed as a shorthand for a more general relationship between
Lorentzian and Euclidean geometries. In the simplest cases, that
relationship can be expressed entirely as a change of real slice within
a fixed complex-bilinear geometry. More generally, both the reality
condition and the geometric data may require compatible continuation,
and a globally regular real Euclidean section need not exist.
The complex representation developed here does not remove these
well-known subtleties of curved-spacetime Wick rotation, but it provides
a framework in which the distinct roles of complex representation,
real structure, and analytic continuation can be stated separately.

A complementary approach to imaginary-time structures in black hole
spacetimes avoids the construction of Euclidean sections altogether.
In particular, Araneda and Whiting have used complex coordinates
associated with null geodesics and $\alpha$-surfaces to derive
imaginary-time periodicity in Rindler, Schwarzschild, and Kerr
geometries~\cite{AranedaWhiting2024}. Their construction concerns
the global analytic structure of these spacetimes, rather than
an invertible complex representation of local Lorentzian geometry.

Against this broader background, the effectiveness of imaginary-time
methods need not be regarded solely as the consequence of a formal
substitution. Within the present construction, the factor of $i$ associated with the
temporal direction is already present in the Lorentzian representation
before any Euclidean continuation is performed.  The same structure
appears consistently in coordinate displacements, proper temporal
displacement, derivatives, momentum variables, and the transformation
of tensorial quantities.  Euclidean continuation then acts upon an
already complex representation rather than introducing complex time
for the first time.

This observation admits a possible structural interpretation.  The
success of Wick rotation may reflect an underlying relationship between
Lorentzian and Euclidean geometries that is naturally expressed through
a common complex framework, with their distinction encoded in the
choice of compatible real structure and geometric data.  The results
of the present work do not establish that this complex structure is
ontologically fundamental, nor do they require physical time itself to
be intrinsically imaginary.  They do, however, show that imaginary
temporal variables can arise as part of a systematic and exactly
equivalent representation of Lorentzian relativity, rather than only
as an auxiliary device introduced for Euclidean continuation.
Whether this structure should be regarded purely as a useful representation
 or as evidence of a deeper geometric role for complex time remains an 
open interpretive question. 

\subsection{Relativistic Wave Equations and the Nonrelativistic Limit}
\label{sec:waves}

The complex representation also provides a useful perspective on wave
dynamics.  Up to this point, the consequences of the transformation
$X^0=ict$ have been examined primarily through spacetime geometry,
relativistic dynamics, and the Hamilton--Jacobi formulation.  We now
ask how the same transformation acts on wave equations and on the
quantities that characterize their solutions.

This question may be approached without introducing any additional
complexification of the temporal coordinate.  Time has already been
represented by $X^0=ict$, while the physical Lorentzian spacetime is
selected by the corresponding reality condition.  We will first
examine how a relativistic wave equation and its phase structure are
represented in these coordinates, then consider the associated wave
covector and its relation to the Hamilton--Jacobi quantities developed
in Sec.~\ref{sec:hamilton_jacobi}.  The nonrelativistic limit will then
provide a particularly useful test of the construction, leading
naturally to the relation between oscillatory and diffusion-like
evolution.  As in the preceding discussion of Euclidean continuation,
the distinction between the ambient complex coordinate space and the
real section on which an equation is evaluated will play a central
role.

Consider first the Klein--Gordon equation for a free scalar field in
flat spacetime.  In the conventional Minkowski representation, with
signature $(-,+,+,+)$, it may be written
\begin{equation}
\left(
g^{\mu\nu}\partial_\mu\partial_\nu
-\frac{m^2c^2}{\hbar^2}
\right)\Psi=0.
\end{equation}
Under the constant complex transformation used throughout this work,
the inverse metric is represented by $h^{\mu\nu}$ and derivatives are
taken with respect to $X^\mu$.  The equation therefore becomes
\begin{equation}
\boxed{
\left(
h^{\mu\nu}
\frac{\partial^2}{\partial X^\mu\partial X^\nu}
-\frac{m^2c^2}{\hbar^2}
\right)\Psi=0.
}
\end{equation}
In flat spacetime, $h^{\mu\nu}=\delta^{\mu\nu}$, so that
\begin{equation}
\left(
\sum_{\mu=0}^{3}
\frac{\partial^2}{\partial (X^\mu)^2}
-\frac{m^2c^2}{\hbar^2}
\right)\Psi=0.
\end{equation}
The differential operator therefore has a formally Euclidean
appearance.  As with the bilinear line element, however, this
appearance does not imply that the physical equation has become
Euclidean.  On the Lorentzian physical slice,
\begin{equation}
X^0=ict,
\qquad
\frac{\partial}{\partial X^0}
=
-\frac{i}{c}\frac{\partial}{\partial t},
\end{equation}
and hence
\begin{equation}
\frac{\partial^2}{\partial (X^0)^2}
=
-\frac{1}{c^2}\frac{\partial^2}{\partial t^2}.
\end{equation}
The preceding equation therefore reproduces the conventional
Lorentzian Klein--Gordon equation exactly.  The distinction between
temporal and spatial derivatives is carried by the reality condition
on $X^0$ rather than by an explicit negative entry in
$h^{\mu\nu}$.

The relation to the Hamilton--Jacobi structure developed in
Sec.~\ref{sec:hamilton_jacobi} may be seen by writing the scalar field
in amplitude--phase form,
\begin{equation}
\Psi=Ae^{i\mathcal{S}/\hbar},
\end{equation}
where $A$ and $\mathcal{S}$ are taken to be real on the physical
Lorentzian spacetime.  Substitution into the Klein--Gordon equation and
separation into real and imaginary parts gives
\begin{equation}
h^{\mu\nu}
\frac{\partial\mathcal{S}}{\partial X^\mu}
\frac{\partial\mathcal{S}}{\partial X^\nu}
+m^2c^2
=
\hbar^2
\frac{
h^{\mu\nu}\partial_\mu\partial_\nu A
}{A},
\end{equation}
together with
\begin{equation}
\partial_\mu
\left(
A^2 h^{\mu\nu}
\frac{\partial\mathcal{S}}{\partial X^\nu}
\right)=0.
\end{equation}
In the semiclassical (WKB or eikonal) approximation, the amplitude
varies sufficiently slowly that the term proportional to $\hbar^2$
may be neglected relative to the leading phase-gradient terms.
The first equation then reduces to
\begin{equation}
h^{\mu\nu}
\frac{\partial\mathcal{S}}{\partial X^\mu}
\frac{\partial\mathcal{S}}{\partial X^\nu}
+m^2c^2=0,
\end{equation}
which is precisely the relativistic Hamilton--Jacobi equation obtained
in Sec.~\ref{sec:hamilton_jacobi}.  Using
\begin{equation}
\Pi^\mu
=
-i h^{\mu\nu}
\frac{\partial\mathcal{S}}{\partial X^\nu},
\end{equation}
this becomes
\begin{equation}
h_{\mu\nu}\Pi^\mu\Pi^\nu=m^2c^2,
\end{equation}
the complementary-momentum mass-shell relation.

This correspondence is suggestive, but its limitation is equally
important.  The complex spacetime representation determines neither
the amplitude $A$ nor the scale $\hbar$, and the replacement of a
classical Hamilton principal function by the phase of a wavefunction
introduces quantum structure not contained in classical relativity.
The amplitude equation above likewise carries information absent from
the classical Hamilton--Jacobi theory.  The appearance of the same
action function and complex phase structure should therefore be
understood as a structural correspondence rather than as a derivation
of quantum mechanics.

The same correspondence may be expressed in terms of the wave
covector.  In conventional coordinates, a plane-wave phase may be
written
\begin{equation}
\Phi
=
k_\mu x^\mu
=
-\omega t+\mathbf{k}\cdot\mathbf{x},
\end{equation}
where
\begin{equation}
k_\mu
=
\left(
-\frac{\omega}{c},
\mathbf{k}
\right).
\end{equation}
All four components therefore have dimensions of inverse length.  The
wave covector transforms in the same manner as the canonical
Hamilton--Jacobi covector,
\begin{equation}
K_\mu
=
(S^{-1})^\nu{}_\mu k_\nu,
\end{equation}
giving
\begin{equation}
\boxed{
K_\mu
=
\left(
\frac{i\omega}{c},
\mathbf{k}
\right).
}
\end{equation}
The phase itself is unchanged:
\begin{equation}
K_\mu X^\mu
=
k_\mu x^\mu
=
-\omega t+\mathbf{k}\cdot\mathbf{x}.
\end{equation}
Consequently,
\begin{equation}
e^{i k_\mu x^\mu}
=
e^{i K_\mu X^\mu}
=
\exp\left(
-\frac{\omega X^0}{c}
+i\mathbf{k}\cdot\mathbf{x}
\right).
\end{equation}
The explicit factor of $i$ multiplying the temporal Fourier frequency
has therefore been absorbed into the complex temporal coordinate and
the transformed temporal component of the wave covector.

For a quantum plane wave, the de Broglie relations imply
\begin{equation}
p_\mu=\hbar k_\mu,
\end{equation}
and hence the transformed canonical quantities satisfy
\begin{equation}
\boxed{
P_\mu=\hbar K_\mu.
}
\end{equation}
A complementary wave vector may likewise be defined by
\begin{equation}
\mathcal{K}^\mu
=
-i h^{\mu\nu}K_\nu,
\end{equation}
so that
\begin{equation}
\boxed{
\mathcal{K}^\mu
=
\left(
\frac{\omega}{c},
-i\mathbf{k}
\right),
\qquad
\Pi^\mu=\hbar\mathcal{K}^\mu.
}
\end{equation}
The wave-vector structure therefore mirrors exactly the distinction
between the canonical momentum covector and complementary momentum
introduced in Sec.~\ref{sec:hamilton_jacobi}.  In particular, the
usual oscillatory plane wave may equivalently be written
\begin{equation}
\boxed{
\Psi
\propto
e^{-\mathcal{K}^T h X}.
}
\end{equation}
Although this expression has the appearance of a real exponential,
its exponent is purely imaginary on the physical Lorentzian slice
for real $\omega$ and $\mathbf{k}$. It therefore describes the same
oscillatory plane wave as the conventional representation.

A further connection appears in the nonrelativistic limit of the
Klein--Gordon equation.  For a positive-energy solution, separate the
rapid phase associated with the rest energy by writing
\begin{equation}
\Psi
=
e^{-imc^2t/\hbar}\psi
=
e^{-mcX^0/\hbar}\psi.
\end{equation}
The second form follows directly from $X^0=ict$ and
rewrites the usual oscillatory rest-energy phase as an
exponential whose exponent is linear in the complex
temporal coordinate with a real coefficient. On the
physical Lorentzian slice, $X^0$ is purely imaginary,
so the exponent remains purely imaginary and the
factor retains unit modulus. 

Substitution into the flat-spacetime Klein--Gordon equation gives
\begin{equation}
\frac{\partial^2\psi}{\partial(X^0)^2}
-\frac{2mc}{\hbar}
\frac{\partial\psi}{\partial X^0}
+\nabla^2\psi=0.
\end{equation}
In the nonrelativistic limit, the envelope $\psi$ varies slowly
compared with the rest-energy scale, so that the second derivative
with respect to $X^0$ may be neglected relative to the first.  The
remaining equation is
\begin{equation}
\boxed{
\hbar c\frac{\partial\psi}{\partial X^0}
=
\frac{\hbar^2}{2m}\nabla^2\psi.
}
\end{equation}
Restricting this equation to the physical Lorentzian slice gives
\begin{equation}
\hbar c
\left(
-\frac{i}{c}\frac{\partial\psi}{\partial t}
\right)
=
\frac{\hbar^2}{2m}\nabla^2\psi,
\end{equation}
or
\begin{equation}
\boxed{
i\hbar\frac{\partial\psi}{\partial t}
=
-\frac{\hbar^2}{2m}\nabla^2\psi,
}
\end{equation}
which is the free-particle Schr\"odinger equation.

The role of the complex temporal coordinate becomes especially
transparent if one defines
\begin{equation}
\chi\equiv\frac{X^0}{c}.
\end{equation}
The nonrelativistic evolution equation then takes the form
\begin{equation}
\hbar\frac{\partial\psi}{\partial\chi}
=
\frac{\hbar^2}{2m}\nabla^2\psi,
\end{equation}
or, with the free-particle Hamiltonian
\begin{equation}
H=-\frac{\hbar^2}{2m}\nabla^2,
\end{equation}
\begin{equation}
\boxed{
\hbar\frac{\partial\psi}{\partial\chi}
=
-H\psi.
}
\end{equation}
No explicit factor of $i$ appears in this equation.  Its physical
interpretation nevertheless depends both on the reality condition
imposed on $\chi$ and on the spectral character of the solution.

On the Lorentzian physical slice,
\begin{equation}
\chi=it,
\end{equation}
so that
\begin{equation}
\frac{\partial}{\partial\chi}
=
-i\frac{\partial}{\partial t},
\end{equation}
and the equation becomes
\begin{equation}
i\hbar\frac{\partial\psi}{\partial t}=H\psi.
\end{equation}
Evolution along the physical temporal direction is therefore the
ordinary unitary Schr\"odinger evolution. If instead $\chi$ is
restricted to a real Euclidean temporal variable $\tau_E$, the same
complex evolution equation formally becomes
\begin{equation}
\hbar\frac{\partial\psi}{\partial\tau_E}
=
-H\psi,
\end{equation}
which has the form of a heat or diffusion equation in Euclidean time.
Unlike the original transformation $X^0=ict$, this change of real
section requires a suitable continuation of the wave dynamics beyond
the Lorentzian physical slice. Such a continuation is not guaranteed
for arbitrary solutions, although the free-particle Hamiltonian
considered here admits the standard forward Euclidean-time
evolution. Under these conditions, unitary and diffusion-like
evolution arise from different real restrictions of the same complex
evolution equation. This provides the direct analogue, within the
present representation, of the conventional relation between
Schr\"odinger evolution and Wick rotation.

It is important, however, to distinguish this analytic continuation
from the description of physical diffusion itself.  A diffusive
process observed in ordinary physical time does not require the
temporal coordinate to be moved from the Lorentzian physical slice to
a real Euclidean section.  Consider the ordinary diffusion equation
\begin{equation}
\frac{\partial u}{\partial t}
=
D\nabla^2 u.
\end{equation}
Since $\chi=it$ on the physical slice, this equation may be written
entirely within the same complex temporal representation as
\begin{equation}
\boxed{
i\frac{\partial u}{\partial\chi}
=
D\nabla^2u.
}
\end{equation}
No change of temporal reality condition has been made.

The distinction becomes especially clear in Fourier space.  For a
mode written in the conventional form
\begin{equation}
u
\propto
e^{i\mathbf{k}\cdot\mathbf{x}-i\omega t},
\end{equation}
the diffusion equation gives the dispersion relation
\begin{equation}
\boxed{
\omega=-iDk^2.
}
\end{equation}
Thus the exponential decay of a diffusive mode may be understood as
arising from an imaginary frequency:
\begin{equation}
e^{-i\omega t}
=
e^{-Dk^2t}.
\end{equation}
The physical temporal coordinate remains represented by $X^0=ict$;
the dissipative character of the solution is instead carried by the
complex frequency.

This result has a particularly simple expression in terms of the
transformed wave covector.  Since
\begin{equation}
K_0=\frac{i\omega}{c},
\end{equation}
an oscillatory mode with real $\omega$ has an imaginary transformed
temporal wave number,
\begin{equation}
\omega\in\mathbb{R}
\qquad\Longrightarrow\qquad
K_0\in i\mathbb{R}.
\end{equation}
For a diffusive mode, by contrast,
\begin{equation}
\omega=-iDk^2
\qquad\Longrightarrow\qquad
K_0=\frac{Dk^2}{c}\in\mathbb{R}.
\end{equation}
Oscillatory and exponentially decaying behavior may therefore coexist
within the same physical temporal reality condition.  The distinction
between them is not determined by the temporal coordinate alone, but
also by the spectral character of the dynamics.

The two routes to diffusion-like behavior should consequently be
distinguished.  Analytic continuation of the Schr\"odinger equation
from the Lorentzian physical slice to a real Euclidean temporal
section converts unitary evolution into exponential evolution while
retaining the underlying spectral relation.  Physical diffusion, on
the other hand, may remain entirely on the Lorentzian physical slice,
with exponential decay encoded through the imaginary-frequency
dispersion relation.  The mathematical connection between
Schr\"odinger evolution and diffusion therefore does not imply that
physical diffusion requires a fundamentally real temporal coordinate.

This distinction provides a dynamical counterpart to the geometric
discussion in Sec.~\ref{sec:rindler_slices} and
Sec.~\ref{sec:wick_rotation}.  A change of real section provides the
analogue of Euclidean continuation, while a complex frequency permits
dissipative evolution without any change in the physical reality
condition.  In particular, the representation $X^0=ict$ does not
require physical evolution to be oscillatory: oscillatory or
exponential behavior depends jointly on the temporal coordinate and
the spectral properties of the mode.

This result should again be interpreted with appropriate restraint.
The construction does not explain the origin of the quantum state,
the Born rule, the value of $\hbar$, or the Hilbert-space structure of
quantum theory, and it does not derive quantum mechanics from general
relativity. The Klein--Gordon equation already contains the quantum
structure required for the nonrelativistic limit. Within its complex
temporal representation, however, the factor of $i$ governing ordinary
Schr\"odinger time evolution arises from restricting the transformed
equation to the physical Lorentzian slice,
\begin{equation}
X^0=ict
\end{equation}
and applying the corresponding derivative relation. Thus the factor
is not introduced as an additional prescription within the transformed
equation, although its appearance does not constitute an independent
derivation of quantum dynamics. The diffusion-like form conventionally
associated with imaginary-time evolution appears under Euclidean
continuation, whereas physical diffusion may be represented on the
same Lorentzian physical slice through a complex dispersion relation.

The recurrence of this structure in the spacetime interval,
proper-time displacement, complementary momentum, Hamilton--Jacobi
theory, the transformed wave covector, Euclidean continuation, and now
relativistic and nonrelativistic wave equations does not by itself
determine its physical interpretation.  It does, however, motivate
asking whether the complex temporal structure should be regarded
solely as an equivalent mathematical representation or whether it
reflects a deeper relationship among relativistic geometry, Euclidean
continuation, wave propagation, and quantum dynamics.  We consider
this interpretive question, together with the limitations of the
construction, in the following subsection.

\subsection{Physical Interpretation and Limitations}

The construction developed in this work is, at its most conservative
level, an exactly equivalent representation of Lorentzian relativity.
Because the transformation $X=Sx$ is invertible, the physical content
of the theory is preserved when its associated geometric and
dynamical quantities are transformed consistently. The representation
therefore introduces no new empirical predictions by itself; any
observable result can be mapped back to its conventional Lorentzian
form.

Exact equivalence, however, does not imply that the alternative
representation is structurally uninformative.  In the conventional
formulation, the distinction between time and space is displayed
explicitly through the Lorentzian signature of the metric.  In the
complex representation, that distinction is redistributed among the
complex temporal coordinate, the physical real structure, the
bilinear and Hermitian metric forms, and the transformation properties
of quantities carrying temporal indices.  In flat spacetime the
bilinear metric may even take the formally Euclidean form $h=I$, while
the physical spacetime remains Lorentzian because its temporal
coordinate satisfies $X^0\in i\mathbb{R}$.  
In the complex-bilinear representation, the Lorentzian distinction
is thus encoded through the physical reality condition rather than
an explicit negative temporal entry in $h$. The associated Hermitian
metric $G$, by contrast, retains the indefinite Lorentzian signature
explicitly.

This redistribution persists beyond the spacetime interval itself.
Proper temporal displacement is represented by
$\Sigma=ic\tau$, the tangent to a timelike worldline acquires the
corresponding complex structure, and the complementary momentum takes
the form
\begin{equation}
\Pi^\mu
=
\left(
\frac{E}{c},
-i\mathbf{p}
\right).
\end{equation}
The Hamilton--Jacobi formulation then relates this quantity to the
gradient of the action through
\begin{equation}
\Pi^\mu
=
-i h^{\mu\nu}
\frac{\partial\mathcal{S}}{\partial X^\nu},
\end{equation}
while the wave covector and complementary wave vector reproduce the
same pattern.  These relations show that the complex temporal
representation is not confined to the coordinates themselves.  Once
the transformation is applied consistently, factors of $i$ recur in
the associated kinematic, canonical, and wave quantities according to
their tensorial and differential roles.

The corresponding structures in relativistic wave dynamics are
suggestive, particularly the relationships among the
Hamilton--Jacobi action, the wave covector, and the nonrelativistic
Schr\"odinger equation. Their common complex temporal representation
does not, however, establish a physical connection between spacetime
geometry and the foundations of quantum mechanics. In particular, the
construction does not determine $\hbar$, supply a probability
interpretation, or account for the Born rule or quantum measurement.

Nor does the Hermitian metric on complexified spacetime define a
quantum Hilbert-space inner product. Its role is to represent the
Lorentzian spacetime metric together with the appropriate physical
conjugation structure. Any deeper relationship with quantum theory
would require additional physical principles beyond the change of
representation developed here.

A corresponding distinction is required in interpreting the temporal
coordinate itself.  An ideal clock continues to register a real
proper-time magnitude $\tau$.  In the present representation, the
associated proper temporal displacement is represented by
$\Sigma=ic\tau$, just as coordinate time is represented by
$X^0=ict$.  Nothing in the mathematical equivalence requires the
measured clock reading $\tau$ or $t$ to become a complex-valued
observable.  The statement that the temporal direction has an
imaginary algebraic representation is therefore distinct from the
stronger ontological claim that physical time is intrinsically
imaginary.

A further limitation concerns the nonuniqueness of the complex
representation itself. The transformation
$S=\operatorname{diag}(i,1,1,1)$ is defined relative to a chosen real
coordinate chart, and its particular component form does not select
a preferred temporal coordinate, observer, or spacetime foliation.
As discussed in Sec.~\ref{sec:geometry}, the
corresponding physical reality structure is determined by the
representation rather than by an additional physical field. The
equivalence established here therefore does not, by itself, identify
a coordinate-independent complex temporal structure as physically
preferred. Such an interpretation would require additional geometric
or physical justification.

At the same time, mathematical equivalence alone does not establish
that the complex temporal structure is physically meaningless.
Equivalent representations necessarily agree on observables, but
this agreement does not determine which mathematical structures,
if any, should be regarded as fundamental. The recurring role of
imaginary temporal quantities in relativistic geometry, momentum,
action, and wave dynamics may therefore motivate further inquiry:
whether these relationships reflect only the mathematical freedom
of an equivalent representation or suggest a deeper geometric
significance of complex time remains an open question.

\section{Conclusion}
\label{sec:conclusion}

We have developed a systematic complex representation of Lorentzian
relativity based on the temporal coordinate transformation
$X^0=ict$. Rather than interpreting this transformation as a
Euclideanization of spacetime, we have treated it as an invertible
change of representation in which physical Lorentzian spacetime is
identified locally with an appropriate real slice of a complex
coordinate space. Within this framework, the spacetime interval may
be expressed using either a complex-bilinear metric $h$ or an
associated Hermitian metric $G$, with the Lorentzian distinction
between time and space encoded through the combined structure of
the metric, complex coordinates, and physical reality condition.

The construction extends consistently to relativistic dynamics,
including Lorentz transformations, proper time, momentum, and the
Hamilton--Jacobi formulation. In curved spacetime, the metric,
Levi-Civita connection, curvature tensors, geodesics, stress-energy
tensor, and Einstein equations retain their conventional content
under the complex transformation. The Kerr and FLRW spacetimes
provide complementary tests of this framework, demonstrating its
consistency for rotating geometries with temporal--spatial cross
terms and time-dependent cosmological geometries with matter.

The same framework reveals structural connections among action,
momentum, and wave propagation. The complementary 
momentum and wave vector satisfy analogous relations to the
action and wave phase, respectively. The Klein--Gordon equation
retains its Lorentzian content in the complex representation, and
its nonrelativistic limit recovers the Schr\"odinger equation on
the physical Lorentzian slice. Expressing these equations in the
complex temporal coordinate also clarifies the distinction between
Euclidean-time evolution, obtained through suitable analytic
continuation, and physical diffusion, which may occur without
leaving the Lorentzian slice.

These results introduce no new empirical predictions, nor do they
establish that physical time is intrinsically imaginary. Proper time
remains the real quantity $\tau$ measured by an ideal clock, while
$\Sigma=ic\tau$ provides its representation as an imaginary
proper-time parameter. Likewise, the mathematical connections to
quantum wave dynamics do not constitute a derivation of quantum
mechanics or establish that the Hermitian spacetime metric defines
a quantum Hilbert space.

The significance of this representation therefore lies not in new
physical predictions, but in the relationships that become apparent
when the transformation is carried through consistently. The
distinction between temporal and spatial directions is preserved,
even when the bilinear metric takes a formally Euclidean appearance,
through the complex coordinates and the reality structure selecting
physical spacetime. The recurring connections among proper time,
momentum, action, and wave propagation demonstrate that the historical
substitution $ct\rightarrow ict$ can be developed into more than a
formal device for producing Euclidean-looking expressions. Applied
consistently, it provides a coherent complex representation of
Lorentzian relativity that makes explicit the interplay among
spacetime geometry, relativistic dynamics, and the physical reality
conditions distinguishing time from space.

\section*{Acknowledgments}

The author acknowledges the assistance of OpenAI's ChatGPT
in the development and preparation of this manuscript.
ChatGPT was used as an interactive research and writing aid
for exploring mathematical formulations, critically examining
derivations and interpretations, identifying and discussing
relevant literature, and improving the organization and
presentation of the manuscript. The research questions,
scientific interpretations, and final conclusions remain
the responsibility of the author, who has reviewed the
AI-assisted material and assumes full accountability for
the content of this work.

\appendix
% ============================================================

\section{Kerr Connection and Curvature Components}
\label{app:kerr}

This appendix provides explicit component checks of the Kerr connection
and curvature transformations discussed in Sec.~\ref{sec:kerr}.
We enumerate the generically nonvanishing independent connection
coefficients and verify their complex transformation factors. We then
examine representative Riemann tensor components from each temporal-index
sector, followed by the Ricci tensor and Kretschmann scalar, to confirm
that the complex representation preserves the vacuum curvature structure
of Kerr spacetime.

For the transformation
\begin{equation}
S=\operatorname{diag}(i,1,1,1),
\end{equation}
the Levi-Civita connection transforms according to
\begin{equation}
\widetilde{\Gamma}^{\rho}{}_{\mu\nu}
=
i^{\delta_{\rho0}}
(-i)^{\delta_{\mu0}+\delta_{\nu0}}
\Gamma^{\rho}{}_{\mu\nu},
\label{eq:app_kerr_connection_rule}
\end{equation}
with no summation over the displayed indices in the phase factor.
It is therefore useful to organize the nonvanishing Kerr connection
coefficients according to the number and placement of their temporal
indices.  We begin with coefficients whose three displayed indices are
purely spatial.

For the Boyer--Lindquist coordinates and notation introduced in
Sec.~\ref{sec:kerr}, the purely spatial coefficients with upper index
$r$ are
\begin{align}
\Gamma^{r}{}_{rr}
&=
\frac{r}{\Sigma_{\rm K}}
-
\frac{r-r_g}{\Delta},
\\
\Gamma^{r}{}_{r\theta}
=
\Gamma^{r}{}_{\theta r}
&=
-\frac{a^2\sin\theta\cos\theta}{\Sigma_{\rm K}},
\\
\Gamma^{r}{}_{\theta\theta}
&=
-\frac{r\Delta}{\Sigma_{\rm K}},
\\
\Gamma^{r}{}_{\phi\phi}
&=
-\frac{\Delta}{2\Sigma_{\rm K}}
\frac{\partial g_{\phi\phi}}{\partial r}.
\end{align}
The corresponding coefficients with upper index $\theta$ are
\begin{align}
\Gamma^{\theta}{}_{rr}
&=
\frac{a^2\sin\theta\cos\theta}
{\Sigma_{\rm K}\Delta},
\\
\Gamma^{\theta}{}_{r\theta}
=
\Gamma^{\theta}{}_{\theta r}
&=
\frac{r}{\Sigma_{\rm K}},
\\
\Gamma^{\theta}{}_{\theta\theta}
&=
-\frac{a^2\sin\theta\cos\theta}{\Sigma_{\rm K}},
\\
\Gamma^{\theta}{}_{\phi\phi}
&=
-\frac{1}{2\Sigma_{\rm K}}
\frac{\partial g_{\phi\phi}}{\partial\theta}.
\end{align}

Since none of the displayed indices in these coefficients is temporal,
Eq.~\eqref{eq:app_kerr_connection_rule} contributes no complex phase.
Consequently,
\begin{align}
\widetilde{\Gamma}^{r}{}_{rr}
&=\Gamma^{r}{}_{rr},
&
\widetilde{\Gamma}^{r}{}_{r\theta}
&=\Gamma^{r}{}_{r\theta},
&
\widetilde{\Gamma}^{r}{}_{\theta\theta}
&=\Gamma^{r}{}_{\theta\theta},
&
\widetilde{\Gamma}^{r}{}_{\phi\phi}
&=\Gamma^{r}{}_{\phi\phi},
\\
\widetilde{\Gamma}^{\theta}{}_{rr}
&=\Gamma^{\theta}{}_{rr},
&
\widetilde{\Gamma}^{\theta}{}_{r\theta}
&=\Gamma^{\theta}{}_{r\theta},
&
\widetilde{\Gamma}^{\theta}{}_{\theta\theta}
&=\Gamma^{\theta}{}_{\theta\theta},
&
\widetilde{\Gamma}^{\theta}{}_{\phi\phi}
&=\Gamma^{\theta}{}_{\phi\phi}.
\end{align}
Thus this sector of the connection is unchanged by the complex temporal
transformation.  The remaining purely spatial coefficients retain the
same transformation factor but contain contributions from the
off-diagonal temporal--azimuthal sector of the Kerr metric and are
considered separately below.

The remaining nonvanishing connection coefficients whose three
displayed indices are purely spatial have upper index $\phi$.  Although
their index structure is entirely spatial, these coefficients depend
on the temporal--azimuthal sector of the Kerr metric because the
inverse metric contains the off-diagonal component $g^{\phi 0}$.
Stationarity and axisymmetry give
\begin{align}
\Gamma^{\phi}{}_{r\phi}
=
\Gamma^{\phi}{}_{\phi r}
&=
\frac{1}{2}
\left(
g^{\phi 0}\frac{\partial g_{0\phi}}{\partial r}
+
g^{\phi\phi}\frac{\partial g_{\phi\phi}}{\partial r}
\right),
\\
\Gamma^{\phi}{}_{\theta\phi}
=
\Gamma^{\phi}{}_{\phi\theta}
&=
\frac{1}{2}
\left(
g^{\phi 0}\frac{\partial g_{0\phi}}{\partial\theta}
+
g^{\phi\phi}\frac{\partial g_{\phi\phi}}{\partial\theta}
\right),
\end{align}
where
\begin{align}
g^{\phi 0}
&=
-\frac{2 r_g ar}{\Sigma_{\rm K}\Delta},
\\
g^{\phi\phi}
&=
\frac{\Delta-a^2\sin^2\theta}
{\Sigma_{\rm K}\Delta\sin^2\theta}.
\end{align}

These expressions illustrate that a connection coefficient with
purely spatial indices need not depend only on the spatial components
of the metric.  In Kerr spacetime, frame dragging enters through
$g_{0\phi}$ and $g^{\phi0}$.  Nevertheless, the transformation factor
in Eq.~\eqref{eq:app_kerr_connection_rule} depends on the index
structure of the connection coefficient itself.  Since all three
indices are spatial,
\begin{align}
\widetilde{\Gamma}^{\phi}{}_{r\phi}
&=
\widetilde{\Gamma}^{\phi}{}_{\phi r}
=
\Gamma^{\phi}{}_{r\phi},
\\
\widetilde{\Gamma}^{\phi}{}_{\theta\phi}
&=
\widetilde{\Gamma}^{\phi}{}_{\phi\theta}
=
\Gamma^{\phi}{}_{\theta\phi}.
\end{align}
Thus the entire purely spatial sector of the Kerr connection is
unchanged by the complex temporal transformation, including those
coefficients whose values depend explicitly on the off-diagonal
temporal--azimuthal metric sector.

We next consider connection coefficients with a spatial upper index
and two temporal lower indices.  For this index structure,
Eq.~\eqref{eq:app_kerr_connection_rule} gives
\begin{equation}
i^{\delta_{i0}}
(-i)^{\delta_{00}+\delta_{00}}
=
(-i)^2=-1,
\end{equation}
so that
\begin{equation}
\widetilde{\Gamma}^{i}{}_{00}
=
-\Gamma^{i}{}_{00}.
\end{equation}
In the stationary Kerr geometry, the nonvanishing coefficients of
this type have upper indices $r$ and $\theta$.  Using
\begin{equation}
g_{00}
=
-\left(1-\frac{2 r_g r}{\Sigma_{\rm K}}\right),
\end{equation}
they are
\begin{align}
\Gamma^{r}{}_{00}
&=
-\frac{\Delta}{2\Sigma_{\rm K}}
\frac{\partial g_{00}}{\partial r}
=
\frac{r_g\Delta
\left(r^2-a^2\cos^2\theta\right)}
{\Sigma_{\rm K}^3},
\\
\Gamma^{\theta}{}_{00}
&=
-\frac{1}{2\Sigma_{\rm K}}
\frac{\partial g_{00}}{\partial\theta}
=
-\frac{2 r_g a^2r\sin\theta\cos\theta}
{\Sigma_{\rm K}^3}.
\end{align}
Their complex counterparts are therefore
\begin{align}
\widetilde{\Gamma}^{r}{}_{00}
&=
-\Gamma^{r}{}_{00}
=
-\frac{r_g\Delta
\left(r^2-a^2\cos^2\theta\right)}
{\Sigma_{\rm K}^3},
\\
\widetilde{\Gamma}^{\theta}{}_{00}
&=
-\Gamma^{\theta}{}_{00}
=
\frac{2 r_g a^2r\sin\theta\cos\theta}
{\Sigma_{\rm K}^3}.
\end{align}

The same sign reversal follows directly from the transformed
bilinear metric.  Since $h_{00}=-g_{00}$ while the $rr$ and
$\theta\theta$ components of the inverse metric are unchanged, the
Levi-Civita construction gives, for example,
\begin{equation}
\widetilde{\Gamma}^{r}{}_{00}
=
-\frac{1}{2}h^{rr}\partial_r h_{00}
=
\frac{1}{2}g^{rr}\partial_r g_{00}
=
-\Gamma^{r}{}_{00}.
\end{equation}
Thus the predicted factor of $-1$ is obtained both from the general
connection transformation law and directly from the transformed Kerr
metric.

We next consider coefficients with a spatial upper index
and exactly one temporal lower index.  In this case,
Eq.~\eqref{eq:app_kerr_connection_rule} gives
\begin{equation}
i^{\delta_{i0}}
(-i)^{\delta_{00}+\delta_{j0}}
=-i,
\qquad i,j\neq 0,
\end{equation}
and therefore
\begin{equation}
\widetilde{\Gamma}^{i}{}_{0j}
=
\widetilde{\Gamma}^{i}{}_{j0}
=
-i\Gamma^{i}{}_{0j}.
\end{equation}

For the Kerr metric, the nonvanishing coefficients
of this type arise from the temporal--azimuthal
metric component $g_{0\phi}$.  They are
\begin{align}
\Gamma^{r}{}_{0\phi}
=
\Gamma^{r}{}_{\phi0}
&=
-\frac{1}{2}g^{rr}
\frac{\partial g_{0\phi}}{\partial r}
\nonumber\\
&=
-\frac{r_g a\Delta
(r^2-a^2\cos^2\theta)\sin^2\theta}
{\Sigma_{\rm K}^3},
\\[4pt]
\Gamma^{\theta}{}_{0\phi}
=
\Gamma^{\theta}{}_{\phi0}
&=
-\frac{1}{2}g^{\theta\theta}
\frac{\partial g_{0\phi}}{\partial\theta}
\nonumber\\
&=
\frac{2 r_g ar(r^2+a^2)
\sin\theta\cos\theta}
{\Sigma_{\rm K}^3},
\\[4pt]
\Gamma^{\phi}{}_{0r}
=
\Gamma^{\phi}{}_{r0}
&=
\frac12\left(
g^{\phi0}\partial_r g_{00}
+
g^{\phi\phi}\partial_r g_{0\phi}
\right),
\\[4pt]
\Gamma^{\phi}{}_{0\theta}
=
\Gamma^{\phi}{}_{\theta0}
&=
\frac12\left(
g^{\phi0}\partial_\theta g_{00}
+
g^{\phi\phi}\partial_\theta g_{0\phi}
\right).
\end{align}

Their complex counterparts consequently satisfy
\begin{align}
\widetilde{\Gamma}^{r}{}_{0\phi}
&=
\widetilde{\Gamma}^{r}{}_{\phi0}
=
-i\Gamma^{r}{}_{0\phi},
\\
\widetilde{\Gamma}^{\theta}{}_{0\phi}
&=
\widetilde{\Gamma}^{\theta}{}_{\phi0}
=
-i\Gamma^{\theta}{}_{0\phi},
\\
\widetilde{\Gamma}^{\phi}{}_{0r}
&=
\widetilde{\Gamma}^{\phi}{}_{r0}
=
-i\Gamma^{\phi}{}_{0r},
\\
\widetilde{\Gamma}^{\phi}{}_{0\theta}
&=
\widetilde{\Gamma}^{\phi}{}_{\theta0}
=
-i\Gamma^{\phi}{}_{0\theta}.
\end{align}

This result may also be verified directly from
the transformed metric.  Since
$h_{0\phi}=-ig_{0\phi}$, while the radial
and polar inverse metric components are unchanged,
the corresponding Christoffel expressions acquire
precisely the predicted factor of $-i$.
These coefficients reflect the coupling between the temporal
and azimuthal sectors of the Kerr geometry. In particular,
the coefficients with upper index $\phi$ contain contributions
from derivatives of both $g_{00}$ and $g_{0\phi}$.

We next examine connection coefficients with a temporal upper
index and two spatial lower indices.  For this index structure,
Eq.~\eqref{eq:app_kerr_connection_rule} gives
\begin{equation}
i^{\delta_{00}}
(-i)^{\delta_{i0}+\delta_{j0}}
=i,
\qquad i,j\neq0,
\end{equation}
so that
\begin{equation}
\widetilde{\Gamma}^{0}{}_{ij}
=
i\Gamma^{0}{}_{ij}.
\end{equation}

The nonvanishing Kerr coefficients of this type are
\begin{align}
\Gamma^{0}{}_{r\phi}
=
\Gamma^{0}{}_{\phi r}
&=
\frac12
\left(
g^{00}\frac{\partial g_{0\phi}}{\partial r}
+
g^{0\phi}\frac{\partial g_{\phi\phi}}{\partial r}
\right),
\\
\Gamma^{0}{}_{\theta\phi}
=
\Gamma^{0}{}_{\phi\theta}
&=
\frac12
\left(
g^{00}\frac{\partial g_{0\phi}}{\partial\theta}
+
g^{0\phi}\frac{\partial g_{\phi\phi}}{\partial\theta}
\right),
\end{align}
where the relevant inverse metric components are
\begin{align}
g^{00}
&=
-\frac{(r^2+a^2)^2-a^2\Delta\sin^2\theta}
{\Sigma_{\rm K}\Delta},
\\
g^{0\phi}
&=
-\frac{2 r_g ar}{\Sigma_{\rm K}\Delta}.
\end{align}

The transformed coefficients therefore satisfy
\begin{align}
\widetilde{\Gamma}^{0}{}_{r\phi}
&=
\widetilde{\Gamma}^{0}{}_{\phi r}
=
i\Gamma^{0}{}_{r\phi},
\\
\widetilde{\Gamma}^{0}{}_{\theta\phi}
&=
\widetilde{\Gamma}^{0}{}_{\phi\theta}
=
i\Gamma^{0}{}_{\theta\phi}.
\end{align}

This result can also be verified directly from the complex
bilinear metric.  The relevant metric and inverse metric
components transform as
\begin{align}
h_{0\phi}&=-ig_{0\phi},
&
h_{\phi\phi}&=g_{\phi\phi},
\\
h^{00}&=-g^{00},
&
h^{0\phi}&=ig^{0\phi}.
\end{align}
For example,
\begin{align}
\widetilde{\Gamma}^{0}{}_{r\phi}
&=
\frac12
\left(
h^{00}\partial_r h_{0\phi}
+
h^{0\phi}\partial_r h_{\phi\phi}
\right)
\nonumber\\
&=
\frac{i}{2}
\left(
g^{00}\partial_r g_{0\phi}
+
g^{0\phi}\partial_r g_{\phi\phi}
\right)
\nonumber\\
&=
i\Gamma^{0}{}_{r\phi}.
\end{align}

The same reasoning applies to the coefficient with lower
indices $\theta\phi$.  Thus the predicted factor of $+i$
emerges directly from the transformed metric, including
the contributions from its off-diagonal temporal--azimuthal
sector.

We now consider coefficients with a temporal upper index
and exactly one temporal lower index.  Their transformation
factor is
\begin{equation}
i^{\delta_{00}}
(-i)^{\delta_{00}+\delta_{i0}}
=i(-i)=1,
\qquad i\neq0.
\end{equation}
Consequently,
\begin{equation}
\widetilde{\Gamma}^{0}{}_{0i}
=
\widetilde{\Gamma}^{0}{}_{i0}
=
\Gamma^{0}{}_{0i}.
\end{equation}

The nonvanishing Kerr coefficients of this type are
\begin{align}
\Gamma^{0}{}_{0r}
=
\Gamma^{0}{}_{r0}
&=
\frac12\left(
g^{00}\partial_r g_{00}
+
g^{0\phi}\partial_r g_{0\phi}
\right)
\nonumber\\
&=
\frac{r_g(r^2+a^2)(r^2-a^2\cos^2\theta)}
{\Sigma_{\rm K}^2\Delta},
\\
\Gamma^{0}{}_{0\theta}
=
\Gamma^{0}{}_{\theta0}
&=
\frac12\left(
g^{00}\partial_\theta g_{00}
+
g^{0\phi}\partial_\theta g_{0\phi}
\right).
\end{align}

Their transformed counterparts satisfy
\begin{align}
\widetilde{\Gamma}^{0}{}_{0r}
&=
\widetilde{\Gamma}^{0}{}_{r0}
=
\Gamma^{0}{}_{0r},
\\
\widetilde{\Gamma}^{0}{}_{0\theta}
&=
\widetilde{\Gamma}^{0}{}_{\theta0}
=
\Gamma^{0}{}_{0\theta}.
\end{align}

Here the temporal index in the upper position contributes
a factor of $i$, while the temporal lower index contributes
$-i$.  Their product is unity, leaving these coefficients
unchanged despite their explicit temporal index structure.

This invariance can also be verified directly from the
complex bilinear metric. Since
$h^{00}=-g^{00}$, $h^{0\phi}=ig^{0\phi}$,
$h_{00}=-g_{00}$, and $h_{0\phi}=-ig_{0\phi}$,
we obtain
\begin{align}
\widetilde{\Gamma}^{0}{}_{0r}
&=
\frac12\left(
h^{00}\partial_r h_{00}
+
h^{0\phi}\partial_r h_{0\phi}
\right)
\nonumber\\
&=
\frac12\left(
g^{00}\partial_r g_{00}
+
g^{0\phi}\partial_r g_{0\phi}
\right)
\nonumber\\
&=
\Gamma^{0}{}_{0r}.
\end{align}
The same result holds for the polar coefficient.

We now turn to the Riemann curvature tensor. Unlike the
connection coefficients, the Riemann tensor transforms
tensorially under arbitrary coordinate transformations.
For the constant transformation
$S=\operatorname{diag}(i,1,1,1)$, its mixed-index
components satisfy
\begin{equation}
\widetilde{R}^{\rho}{}_{\sigma\mu\nu}
=
i^{\delta_{\rho0}}
(-i)^{\delta_{\sigma0}
+\delta_{\mu0}+\delta_{\nu0}}
R^{\rho}{}_{\sigma\mu\nu}.
\label{eq:app_kerr_curvature_rule}
\end{equation}

Lowering the first index with the complex bilinear metric
$h_{\mu\nu}$ gives
\begin{equation}
\widetilde{R}_{\rho\sigma\mu\nu}
=
(-i)^{\delta_{\rho0}+\delta_{\sigma0}
+\delta_{\mu0}+\delta_{\nu0}}
R_{\rho\sigma\mu\nu}.
\label{eq:app_kerr_covariant_curvature_rule}
\end{equation}

The antisymmetry of the Riemann tensor within each
index pair implies that a nonvanishing fully covariant
component contains at most two temporal indices.
The transformation therefore separates into three
sectors:
\begin{align}
\widetilde{R}_{ijkl}
&=R_{ijkl},
\\
\widetilde{R}_{0ijk}
&=-iR_{0ijk},
\\
\widetilde{R}_{0i0j}
&=-R_{0i0j},
\end{align}
where all Latin indices are spatial. Components with
other index orderings follow from the usual Riemann
tensor symmetries.

We verify these relations below using representative
curvature components of the Kerr geometry.

We begin with the purely spatial component
$R_{r\theta r\theta}$. Using the curvature convention
of Sec.~\ref{sec:curvature}, we obtain
\begin{align}
R^{r}{}_{\theta r\theta}
={}&
\partial_r\Gamma^{r}{}_{\theta\theta}
-
\partial_\theta\Gamma^{r}{}_{r\theta}
\nonumber\\
&+
\Gamma^{r}{}_{rr}\Gamma^{r}{}_{\theta\theta}
+
\Gamma^{r}{}_{r\theta}\Gamma^{\theta}{}_{\theta\theta}
\nonumber\\
&-
\Gamma^{r}{}_{\theta r}\Gamma^{r}{}_{r\theta}
-
\Gamma^{r}{}_{\theta\theta}\Gamma^{\theta}{}_{r\theta}
\nonumber\\
={}&
-\frac{r_g r(r^2-3a^2\cos^2\theta)}
{\Sigma_{\rm K}^2}.
\end{align}

Lowering the first index with
$g_{rr}=\Sigma_{\rm K}/\Delta$ gives
\begin{equation}
R_{r\theta r\theta}
=
-\frac{r_g r(r^2-3a^2\cos^2\theta)}
{\Sigma_{\rm K}\Delta}.
\end{equation}

Since all four indices are spatial, the transformation
factor in
Eq.~\eqref{eq:app_kerr_covariant_curvature_rule}
is unity. Consequently,
\begin{equation}
\widetilde{R}_{r\theta r\theta}
=
R_{r\theta r\theta}
=
-\frac{r_g r(r^2-3a^2\cos^2\theta)}
{\Sigma_{\rm K}\Delta}.
\end{equation}

The same result follows directly from the transformed
Levi-Civita connection. All connection coefficients
contributing to this component have purely spatial
indices and are unchanged by the complex
transformation. Since the radial and polar derivatives
are likewise unchanged, the transformed curvature
expression reproduces the conventional result exactly.

As a consistency check, the Schwarzschild limit
$a\to 0$ gives
\begin{equation}
R_{r\theta r\theta}
\longrightarrow
-\frac{r_g}{r-2 r_g},
\end{equation}
while the flat-spacetime limit $r_g\to 0$ gives zero.

We next consider a curvature component containing exactly
one temporal index. A useful example is
$R_{0r\theta\phi}$, which is generally nonzero
in the rotating Kerr geometry.

Using the antisymmetry of the Riemann tensor, we write
\begin{equation}
R_{0r\theta\phi}
=
-g_{rr}R^{r}{}_{0\theta\phi}.
\end{equation}

The corresponding mixed-index component is
\begin{align}
R^{r}{}_{0\theta\phi}
={}&
\partial_\theta\Gamma^{r}{}_{0\phi}
+
\Gamma^{r}{}_{\theta r}\Gamma^{r}{}_{0\phi}
+
\Gamma^{r}{}_{\theta\theta}
\Gamma^{\theta}{}_{0\phi}
\nonumber\\
&-
\Gamma^{r}{}_{0\phi}\Gamma^{0}{}_{0\theta}
-
\Gamma^{r}{}_{\phi\phi}
\Gamma^{\phi}{}_{0\theta}
\nonumber\\
={}&
-\frac{a r_g\Delta\sin\theta\cos\theta}
{\Sigma_{\rm K}^4}
(3r^2-a^2\cos^2\theta)
\nonumber\\
&\times
\left[2r^2+a^2(3-\cos^2\theta)\right].
\end{align}

Lowering the first index gives
\begin{align}
R_{0r\theta\phi}
={}&
\frac{a r_g\sin\theta\cos\theta}
{\Sigma_{\rm K}^3}
(3r^2-a^2\cos^2\theta)
\nonumber\\
&\times
\left[2r^2+a^2(3-\cos^2\theta)\right].
\end{align}

Since this component contains exactly one temporal
covariant index, Eq.~\eqref{eq:app_kerr_covariant_curvature_rule}
gives
\begin{equation}
\widetilde{R}_{0r\theta\phi}
=
-iR_{0r\theta\phi}.
\end{equation}

This transformation can also be verified directly from
the connection. Every term contributing to
$R^{r}{}_{0\theta\phi}$ acquires exactly one factor
of $-i$ under the complex temporal transformation.
Consequently,
\begin{equation}
\widetilde{R}^{r}{}_{0\theta\phi}
=
-iR^{r}{}_{0\theta\phi},
\end{equation}
and lowering the first index with the transformed
metric reproduces the fully covariant result above.

The component vanishes when $a=0$, as expected for
the Schwarzschild limit, while remaining generally
nonzero for rotating Kerr spacetime.

We finally consider a curvature component with two
temporal indices. Using the Riemann tensor symmetries,
we have
\begin{equation}
R_{0r0r}
=
R_{r0r0}
=
g_{rr}R^{r}{}_{0r0}.
\end{equation}

Stationarity reduces the mixed-index component to
\begin{align}
R^{r}{}_{0r0}
={}&
\partial_r\Gamma^{r}{}_{00}
+
\Gamma^{r}{}_{rr}\Gamma^{r}{}_{00}
+
\Gamma^{r}{}_{r\theta}\Gamma^{\theta}{}_{00}
\nonumber\\
&-
\Gamma^{r}{}_{00}\Gamma^{0}{}_{0r}
-
\Gamma^{r}{}_{0\phi}\Gamma^{\phi}{}_{0r}
\nonumber\\
={}&
-\frac{r_g r(r^2-3a^2\cos^2\theta)
(2\Delta+a^2\sin^2\theta)}
{\Sigma_{\rm K}^4}.
\end{align}

Lowering the first index gives
\begin{equation}
R_{0r0r}
=
-\frac{r_g r(r^2-3a^2\cos^2\theta)
(2\Delta+a^2\sin^2\theta)}
{\Sigma_{\rm K}^3\Delta}.
\end{equation}

Since this component contains two temporal covariant
indices, its complex transformation factor is
$(-i)^2=-1$. Hence
\begin{equation}
\widetilde{R}_{0r0r}
=
-R_{0r0r}
=
\frac{r_g r(r^2-3a^2\cos^2\theta)
(2\Delta+a^2\sin^2\theta)}
{\Sigma_{\rm K}^3\Delta}.
\end{equation}

The same sign reversal follows directly from the
transformed connection. Each term contributing to
$R^{r}{}_{0r0}$ acquires a factor of $-1$.
In particular, the product involving the
temporal--azimuthal connection coefficients transforms as
\begin{equation}
\widetilde{\Gamma}^{r}{}_{0\phi}
\widetilde{\Gamma}^{\phi}{}_{0r}
=
(-i)^2
\Gamma^{r}{}_{0\phi}\Gamma^{\phi}{}_{0r}
=
-\Gamma^{r}{}_{0\phi}\Gamma^{\phi}{}_{0r}.
\end{equation}

Since $h_{rr}=g_{rr}$, lowering the first index
introduces no additional transformation factor.
The fully covariant component therefore satisfies
$\widetilde{R}_{0r0r}=-R_{0r0r}$,
in agreement with the general curvature
transformation law.

As a consistency check, the Schwarzschild limit
$a\to0$ gives
\begin{equation}
R_{0r0r}
\longrightarrow
-\frac{2 r_g}{r^3},
\end{equation}
while the flat-spacetime limit $r_g\to0$ gives zero.

We next verify that the curvature components satisfy
the vacuum Einstein equations. The Ricci tensor is
obtained by contracting the first and third indices
of the Riemann tensor:
\begin{equation}
R_{\mu\nu}
=
R^{\rho}{}_{\mu\rho\nu}.
\end{equation}

We begin with the temporal component,
\begin{equation}
R_{00}
=
R^{r}{}_{0r0}
+
R^{\theta}{}_{0\theta0}
+
R^{\phi}{}_{0\phi0}.
\end{equation}

Defining
\begin{equation}
A
\equiv
\frac{r_g r(r^2-3a^2\cos^2\theta)}
{\Sigma_{\rm K}^4},
\end{equation}
the contributing curvature components are
\begin{align}
R^{r}{}_{0r0}
&=
-A(2\Delta+a^2\sin^2\theta),
\\
R^{\theta}{}_{0\theta0}
&=
A(\Delta+2a^2\sin^2\theta),
\\
R^{\phi}{}_{0\phi0}
&=
A(\Delta-a^2\sin^2\theta).
\end{align}

Their contraction consequently gives
\begin{align}
R_{00}
&=
A\big[
-2\Delta-a^2\sin^2\theta
+\Delta+2a^2\sin^2\theta
\nonumber\\
&\qquad
+\Delta-a^2\sin^2\theta
\big]
\nonumber\\
&=0.
\end{align}

The radial component provides an independent
spatial check:
\begin{equation}
R_{rr}
=
R^{0}{}_{r0r}
+
R^{\theta}{}_{r\theta r}
+
R^{\phi}{}_{r\phi r}.
\end{equation}

Defining
\begin{equation}
B
\equiv
\frac{r_g r(r^2-3a^2\cos^2\theta)}
{\Sigma_{\rm K}^3\Delta},
\end{equation}
we obtain
\begin{align}
R^{0}{}_{r0r}
&=
B(2r^2+3a^2-a^2\cos^2\theta),
\\
R^{\theta}{}_{r\theta r}
&=
-B\Sigma_{\rm K},
\\
R^{\phi}{}_{r\phi r}
&=
-B(r^2+3a^2-2a^2\cos^2\theta).
\end{align}

Using
$\Sigma_{\rm K}=r^2+a^2\cos^2\theta$,
these contributions cancel exactly:
\begin{equation}
R_{rr}=0.
\end{equation}

The off-diagonal component $R_{0\phi}$ provides
a further test of the rotating geometry. Its
contraction is
\begin{equation}
R_{0\phi}
=
R^{0}{}_{00\phi}
+
R^{r}{}_{0r\phi}
+
R^{\theta}{}_{0\theta\phi}.
\end{equation}

Introducing
\begin{equation}
F_{\mathrm{fd}}
\equiv
\frac{a r_g r\sin^2\theta
(r^2-3a^2\cos^2\theta)}
{\Sigma_{\rm K}^4},
\end{equation}
the contributing components are
\begin{align}
R^{0}{}_{00\phi}
&=2 r_g rF_{\mathrm{fd}},
\\
R^{r}{}_{0r\phi}
&=F_{\mathrm{fd}}[3(r^2+a^2)-4 r_g r],
\\
R^{\theta}{}_{0\theta\phi}
&=-F_{\mathrm{fd}}[3(r^2+a^2)-2 r_g r].
\end{align}

Consequently,
\begin{equation}
R_{0\phi}=0.
\end{equation}

Direct evaluation of the remaining independent
Ricci components gives the same result. Thus
the complete Kerr Ricci tensor vanishes:
\begin{equation}
R_{\mu\nu}=0.
\end{equation}

The Ricci tensor transforms under the complex
coordinate representation as
\begin{equation}
\widetilde R_{\mu\nu}
=
(-i)^{\delta_{\mu0}+\delta_{\nu0}}
R_{\mu\nu}.
\end{equation}

In particular,
\begin{align}
\widetilde R_{00}&=-R_{00}=0,\\
\widetilde R_{0\phi}&=-iR_{0\phi}=0,\\
\widetilde R_{rr}&=R_{rr}=0.
\end{align}

Since every conventional Kerr Ricci component
vanishes, the complete transformed Ricci tensor
also vanishes:
\begin{equation}
\widetilde R_{\mu\nu}=0.
\end{equation}

The Ricci scalar and Einstein tensor therefore
satisfy
\begin{equation}
\widetilde R=0,
\qquad
\widetilde{\mathcal G}_{\mu\nu}=0.
\end{equation}

We conclude by examining the Kretschmann scalar,
which provides a nonvanishing curvature invariant
even though the Kerr Ricci tensor vanishes.
In the conventional Lorentzian representation,
it is defined by
\begin{equation}
K
=
R_{\alpha\beta\gamma\delta}
R^{\alpha\beta\gamma\delta}.
\end{equation}

For the Kerr metric in Boyer--Lindquist coordinates,
the contraction gives
\begin{equation}
K
=
\frac{48 r_g^2}{\Sigma_{\rm K}^6}
\left[
r^6
-15a^2r^4\cos^2\theta
+15a^4r^2\cos^4\theta
-a^6\cos^6\theta
\right].
\end{equation}

In the complex representation, the corresponding
scalar is
\begin{equation}
\widetilde K
=
\widetilde R_{\alpha\beta\gamma\delta}
\widetilde R^{\alpha\beta\gamma\delta},
\end{equation}
where all indices are raised using the inverse
complex-bilinear metric $h^{\mu\nu}$.

The fully covariant Riemann tensor transforms as
\begin{align}
\widetilde R_{\alpha\beta\gamma\delta}
={}&
(S^{-1})^\mu{}_\alpha
(S^{-1})^\nu{}_\beta
\nonumber\\
&\times
(S^{-1})^\rho{}_\gamma
(S^{-1})^\sigma{}_\delta
R_{\mu\nu\rho\sigma},
\end{align}
while its fully contravariant counterpart satisfies
\begin{align}
\widetilde R^{\alpha\beta\gamma\delta}
={}&
S^\alpha{}_\mu S^\beta{}_\nu
\nonumber\\
&\times
S^\gamma{}_\rho S^\delta{}_\sigma
R^{\mu\nu\rho\sigma}.
\end{align}

Upon contraction, every factor of $S^{-1}$
is canceled by a corresponding factor of $S$.
Consequently,
\begin{align}
\widetilde K
&=
\widetilde R_{\alpha\beta\gamma\delta}
\widetilde R^{\alpha\beta\gamma\delta}
\nonumber\\
&=
R_{\mu\nu\rho\sigma}
R^{\mu\nu\rho\sigma}
\nonumber\\
&=K.
\end{align}

Thus the nonvanishing Kerr curvature invariant
is preserved exactly under the complex temporal
transformation:
\begin{equation}
\boxed{\widetilde K=K}.
\end{equation}

As a consistency check, the Schwarzschild limit
$a\to0$ gives
\begin{equation}
K
\longrightarrow
\frac{48 r_g^2}{r^6},
\end{equation}
while the flat-spacetime limit $r_g\to0$
gives $K=0$.

% ============================================================

\section*{Statements and Declarations}

\subsection*{Funding}
No funding was received for conducting this study.

\subsection*{Competing interests}
The author has no relevant financial or non-financial
interests to disclose.

\subsection*{Data availability}
Data sharing is not applicable to this article as no
datasets were generated or analysed during the study.

\bibliographystyle{unsrtnat}
\bibliography{references}

\end{document}